\documentclass[twocolumn,floatfix,superscriptaddress,nofootinbib,prb]{revtex4-1}

\usepackage{xcolor}

\usepackage{bm}
\usepackage{mathtools}
\newcommand{\tr}{\operatorname{Tr}}
\newcommand{\ketbra}[2]{\ket {#1} \hskip -0.8ex \bra {#2}}

\newcommand{\expct}[1]{\left\langle #1 \right\rangle}
\newcommand{\expcth}[1]{\left\langle #1 \right\rangle_{\mathrm{Haar}} }

\newcommand{\eff}{\mathrm{eff}}

\usepackage{amssymb}
\usepackage{amsmath}
\usepackage{amsthm}
\usepackage{amsfonts}
\usepackage{braket}
\usepackage{graphicx}
\usepackage{epstopdf}
\usepackage{epsfig}
\usepackage[toc,page,title,titletoc,header]{appendix}
\usepackage{dsfont,amsthm,amsbsy}
\usepackage{hyperref} 
\usepackage[xspace]{ellipsis}
\usepackage{cancel}

\definecolor{dkblue}{rgb}{0,0,0.5} 
\newtheorem{prop}{Proposition}
\newtheorem{cor}{Corollary}

\DeclareMathOperator{\erf}{erf}

\begin{document}

\title{ Mana and thermalization: probing the feasibility of near-Clifford Hamiltonian simulation }

\author{Troy J. Sewell}
\affiliation{Joint Center for Quantum Information and Computer Science, University of Maryland, College Park, Md, 20742}

\author{Christopher David White}
\affiliation{Joint Center for Quantum Information and Computer Science, University of Maryland, College Park, Md, 20742}
\affiliation{Condensed Matter Theory Center, University of Maryland, College Park, Md, 20742}

\begin{abstract}
  Quantum hydrodynamics is the emergent classical dynamics governing transport of conserved quantities in generic strongly-interacting quantum systems.
  Recent matrix product operator methods \cite{white_quantum_2017,rakovszky_dissipation-assisted_2020} have made simulations of quantum hydrodynamics in 1+1d tractable,
  but they do not naturally generalize to 2+1d or higher,
  and they offer limited guidance as to the difficulty of simulations on quantum computers.
  Near-Clifford simulation algorithms are not limited to one dimension,
  and  future error-corrected quantum computers will likely be bottlenecked by non-Clifford operations.
  We therefore investigate the non-Clifford resource requirements for simulation of quantum hydrodynamics
  using ``mana'', a resource theory of non-Clifford operations.
  For infinite-temperature starting states 
  we find that the mana of subsystems quickly approaches zero,
  while for starting states with energy above some threshold 
  the mana approaches a nonzero value.
  Surprisingly, in each case the finite-time mana is governed by the subsystem entropy,
  not the thermal state mana;
  we argue that this is because mana is a sensitive diagnostic of finite-time deviations from canonical typicality.
\end{abstract}

\maketitle

\section{Introduction}

Quantum hydrodynamics---the long-wavelength, long-time dynamics governing transport of conserved quantities---is believed to be efficiently simulable on classical computers, even for strongly-interacting systems.
If a system's Hamiltonian satisfies the \textit{eigenstate thermalization hypothesis }(ETH) \cite{deutsch_quantum_1991,srednicki_chaos_1994,dalessio_quantum_2016} it will rapidly reach local thermodynamic equilibrium.
After that time local observables are well described by a Gibbs state with spatially varying thermodynamic potentials.
Since hydrodynamics is presumptively local \cite{chaikin_principles_2000,forster_hydrodynamic_1975}
one expects a local approximation to be enough to compute long-time dynamics. References 
\onlinecite{white_effective_2021,von_keyserlingk_operator_2021} offer quantitative arguments that this is the case.

Recent work has built on these conceptual insights to create workable numerical methods in one dimension. 
The ``generalized relaxation time approximation'' \cite{lopez-piqueres_hydrodynamics_2020}
treats integrable models perturbed by small integrability-breaking terms;
it replaces the detailed effect of the perturbation
by a local phenomenological collision integral
with a single parameter, a relaxation time.
Near-equilibrium transport properties are accessible in non-equilibrium steady-state setups. \cite{prosen_matrix_2009,langer_real-time_2009,znidaric_thermalization_2010,prosen_matrix_2015,weimer_simulation_2021}
For unitary quench dynamics far from integrability there is a new generation of
matrix product operator methods,
\textit{density matrix truncation} (DMT) \cite{white_quantum_2017}
and
\textit{dissipation assisted operator evolution} (DAOE).\cite{rakovszky_dissipation-assisted_2020}
These methods assume that non-local information is unimportant and can be discarded,
if one proceeds carefully---but so far the only practical use of either of these methods
has been for a model close to free-fermion integrability.\cite{ye_emergent_2019}
All these approaches share two key assumptions:
that local approximations to the full state ``simple'', in some sense;
and that capturing those local properties is enough to simulate the system's dynamics
(at least in some hydrodynamical regime).
These matrix product operator methods are restricted to one-dimensional systems.

\begin{figure}[t]
  \begin{minipage}{0.45\textwidth}
    \includegraphics[width=\linewidth]{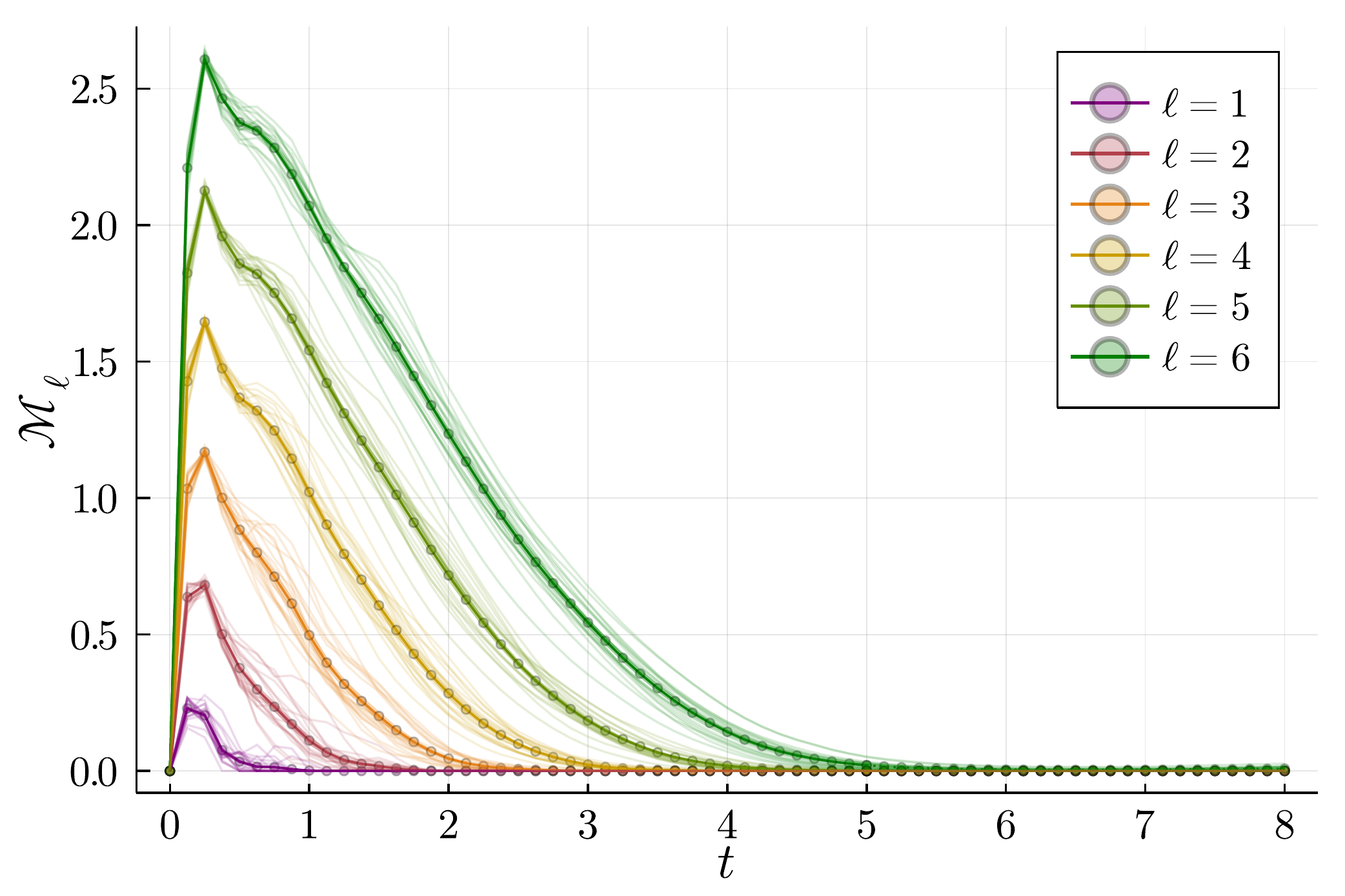}
  \end{minipage}
  \caption{
    \textbf{Subsystem mana in real-time evolution} of infinite-temperature starting states on a chain of 50 sites.
    Bold lines show the average over 20 random starting states (cf Sec.~\ref{ss:initial}),
    faint lines show trajectories of individual states,
    and dot mark points at which we measure the mana with time step $dt = 1/8$.
    We use TEBD (cf Sec.~\ref{ss:method}).
    In each case we see a rapid rise to approximately the average mana of a Haar state (black line in bottom plot),
    due to local thermalization of the subsystem,
    followed by a slow decay as the subsystem entangles with the rest of the system.
     }
  \label{fig:submana-inftemp-t}
\end{figure}

But the insights that led to these methods
are not limited to one-dimensional physics.
In two (or more) spatial dimensions,
ETH Hamiltonians will still locally thermalize,
and one expects that Gibbs states will still have compact representations.
The questions, then, are---what data structures and algorithms are suitable for higher-dimensional mixed-state dynamics?
How computationally intensive is the early-time complexity ``hump''?
And how computationally intensive is the long-time dynamics?

The effective model of Ref. \onlinecite{white_effective_2021} offers one route to higher-dimensional simulations.
In that model one explicitly represents the small-diameter components of an operator (or density matrix), and replaces dynamics in the large-diameter subspace by decay into a vacuum.
This effective model generalizes naturally to more than one dimension,
and has tractable Hilbert space dimension.
But checking convergence is challenging,
and the effective model requires nontrivial physics---the rate of long-operator decay rate---as input.
(The bulk of Ref. \onlinecite{white_effective_2021} was devoted to computing that decay rate.)

We propose that near-Clifford algorithms and data structures offer a promising avenue for simulations of hydrodynamics in more than one dimension.
Clifford circuits are efficiently simulable on classical computers,\cite{gottesman_heisenberg_1998,aaronson_improved_2004,anders_fast_2006}
because they map each Pauli operator to a single other Pauli operator.
Pure Clifford circuits have been used to construct analytically tractable ETH Hamiltonians,\cite{gopalakrishnan_facilitated_2018} and related quantum cellular automaton models have been used to study hydrodynamics.\cite{alba_operator_2019,iaconis_multipole_2020}
Circuits with few non-Clifford gates
(or many gates that are nearly Clifford)
are also classically simulable.\cite{pashayan_estimating_2015,bravyi_trading_2016,bravyi_improved_2016,bennink_unbiased_2017,howard_application_2017,bravyi_simulation_2018,huang_approximate_2019,bu_efficient_2019}
But we wish to do more than build model systems:
we wish to simulate any given (ETH) Hamiltonian, in any dimension.

At first sight, Clifford or near-Clifford circuits are ill-suited to simulating hydrodynamics.
They can map any particular input state to at most a finite number of states.
(In particular, they map the computational basis state only to stabilizer states.)
By contrast the orbit of an initial state under evolution generated by an ETH Hamiltonian
traces out a continuous manifold.
Moreover capturing chaotic growth in OTOCs requires many non-Clifford gates \cite{leone_quantum_2021}.
But if we satisfy ourselves with non-unitary local approximations to the system's dynamics---as we do in using DMT, DAOE, or the relaxation-time approximation---a path opens up.
Stochastic Clifford circuits (that is, averages over an ensemble of Clifford circuits) can simulate many mixed states,
in fact every state in the convex hull of the stabilizer states;
one might construct such an ensemble by using randomized Trotter decompositions or similar techniques.%
\cite{campbell_random_2019,childs_faster_2019,childs_theory_2021,tran_faster_2021,faehrmann_randomizing_2021}

Each algorithm for near-Clifford simulation has classical computational complexity exponential in some measure of the distance of the circuit from a pure Clifford circuit.
One can estimate these circuit measures
by computing so-called \textit{magic monotones} \cite{veitch_resource_2014} for the states produced by the circuits.
A magic monotone is a function on a quantum state that is non-decreasing under Clifford gates and certain other reasonable operations;
consequently, it lower-bounds the number of non-Clifford operations required to produce the state.
Many magic monotones exist.\cite{veitch_resource_2014,howard_application_2017,wang_efficiently_2018,regula_convex_2018,beverland_lower_2019,leone_stabilizer_2022}

We ask:
under what circumstances are local approximations to hydrodynamics accessible to near-Clifford simulations?
We use the \textit{mana} \cite{veitch_resource_2014} of subsystem reduced density matrices
as a proxy for that accessibility.
We choose it in part because it is closely related to the quantity controlling the difficulty of the Monte Carlo method of Ref. \onlinecite{pashayan_estimating_2015},
and in part because it is computable without solving a minimization problem.
We consider time evolution of a stabilizer state,
and measure the mana of local reduced density matrices as a function of time. 
We find that local reduced density matrices display a clear ``complexity hump'' (Fig.~\ref{fig:submana-inftemp-t}):
for times $t \lesssim \varepsilon^{-1}$ the local energy scale%
\footnote{
  In the model \eqref{eq:ham} an appropriate local energy scale is
  \[ \varepsilon = \sqrt{2(J^2 + h_x^2 + h_z^2)} \approx 2.4 \]
  for our parameter values $J = h_x = h_z = 1$.
}
these local subsystems rise to nearly maximal mana
while for $t \gtrsim \varepsilon^{-1}$ the mana decreases,
broadly following the mana of a Haar state with the appropriate entropy.
For finite-temperature states, we additionally notice that
the subsystem mana deviates from the Gibbs state mana (which is zero for sufficiently small inverse temperature\cite{sarkar_characterization_2020}).
We attribute this to mana's sensitivity to small deviations from canonical typicality.

The paper is organized as follows.
In Sec.~\ref{s:model-init} we describe our model (a variant of the $q = 3$ Potts model),
our procedure for choosing initial states,
and our numerical methods,
and we briefly describe mana. 
In Sec.~\ref{s:inf} we treat the evolution of mana for infinite-temperature states,
while in Sec.~\ref{s:finite} we treat the evolution of mana for finite-temperature states.

\section{Model, initial state, and methods}\label{s:model-init}

\subsection{Model}

\begin{figure}
  \centering
  \includegraphics[width=0.45\textwidth]{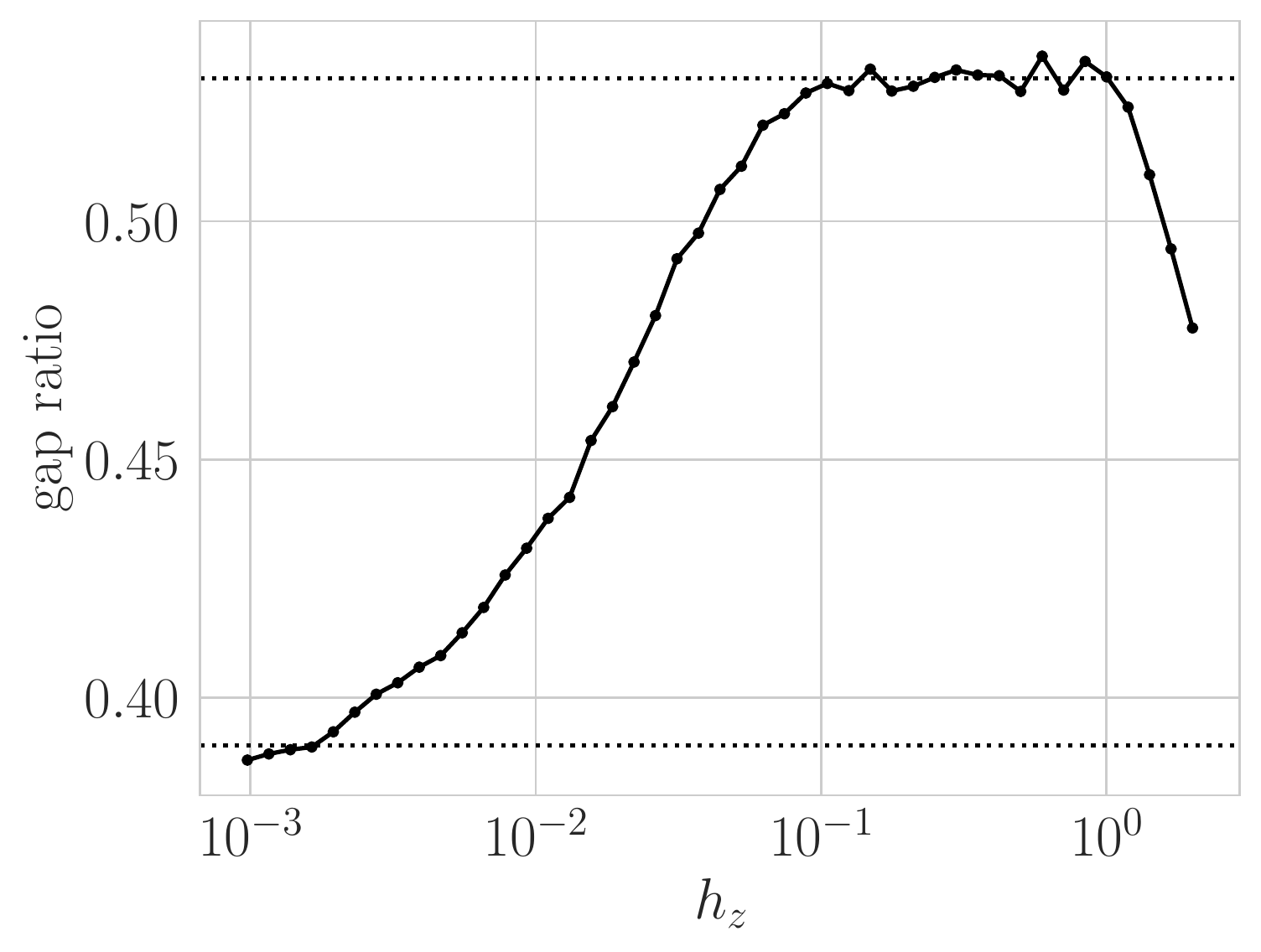}
  \caption{
    \textbf{Gap ratio} of Eq.~\eqref{eq:gapratio} for the model Eq.~\eqref{eq:ham} with $h_x = J = 1$, as a function of $h_z$,  on a system of $L = 9$ sites. For $0.1 \lesssim h_z \lesssim 1.0$ the model has the GOE gap ratio $r \approx 0.53$ (upper dotted line); for $h_z \ll 0.1$ it drops below the Poisson gap ratio $r \approx 0.39$ (lower dotted line)
  }
  \label{fig:gap-ratio}
\end{figure}

Studies of thermalization and hydrodynamics
typically use a transverse-field Ising model with additional longitudinal field.
Because mana is only defined for qudits of odd dimension,%
\footnote{
  We performed much of this work before the R\'enyi entropy of magic\cite{leone_stabilizer_2022} was defined for mixed states.
  Going forward we suggest others use that monotone.
}
we cannot use that model.
Instead we use the analogous qutrit model
\begin{align}
  \label{eq:ham}
  \begin{split}
    H &=   - J \sum_j [Z^\dagger_j Z_j + h.c.]  - h_x \sum_j [X_j + X_j^\dagger] \\
    &\qquad - h_z \sum_j [Z_j + Z_j^\dagger]
  \end{split}
\end{align}
where $X = \sum_{m = 0}^2 \ketbra{m + 1 \mod 3}{m}$
and $Z = \sum_{m = 0}^2 e^{-2\pi i m / 3} \ketbra{m}{m}$
are the clock and shift operators.
We take $J = h_x = h_z = 1$ and system size $L = 50$, 
except where otherwise specified.

This model is ETH for most parameter values.
Even with $h_z = 0$, where this model becomes the $\mathbb Z_3$ Potts model, it is only integrable at the critical point $J = h_x$.%
\cite{fateev_self-dual_1982,von_gehlen_zn-symmetric_1985,au-yang_commuting_1987,fendley_parafermionic_2012}
\footnote{
  This is in contrast to the transverse-field Ising model, which requires the longitudinal field to break integrability.
  The $\mathbb Z_3$ Potts model can be written as a quadratic of $\mathbb Z_3$ parafermion operators via the Fradkin-Kadanoff transformation,\cite{fradkin_disorder_1980}
  just as the transverse-field Ising model can be written in terms of fermions via the Jordan-Wigner and Bogoliubov transformations.
  But the (abelian) anyonic commutation relations of the parafermion operators
  mean that the resulting Hamiltonian cannot be trivially diagonalized.
  \onlinecite{fendley_free_2014} constructs integrable ``free-parafermion'' models, but these models are not Hermitian.
  I am grateful to Aaron Chew for a helpful conversation about these matters.
}
We add the longitudinal-field term to robustly break integrability,
even at $J = h_x$,
and the $\mathbb Z_3$ onsite rotation symmetry,
which will lead to more complicated hydrodynamics.

To check that this Hamiltonian is in fact ETH,
we measure the eigenstate gap ratio.
The eigenstate gap ratio is
\begin{equation}
  \label{eq:gapratio}
  r := \expct{
  \frac{\min(\delta_\alpha, \delta_{\alpha + 1})}
  {\max(\delta_\alpha, \delta_{\alpha + 1})}}\;,
\end{equation}
$\delta_\alpha := E_{\alpha + 1} - E_{\alpha}$, where the average is over eigenstates.
We additionally average over symmetry sectors.
We plot $r$ as a function of $h_z$ 
for a system of $L = 9$ sites
in Fig.~\ref{fig:gap-ratio};
we find that the system has GOE level-spacing statistics for $0.1 \lesssim h_z \lesssim 1.0$,
indicating that it satisfies the ETH.
(In the limit $h_z \gg 1$ the uniform field term dominates; this will lead to a gap ratio $r = 0$ due to degeneracies.
Similarly, in the limit $h_z \ll 1$ the model regains its $\mathbb Z_3$ symmetry,
again leading to degeneracies and a gap ratio $r = 0$.
In each case the model presumably remains ETH for sufficiently large system sizes.)

Despite the fact that we ultimately seek two-dimensional algorithms,
we use the one-dimensional model \eqref{eq:ham}.
We do so precisely because there already exist data structures and algorithms%
---matrix product states (MPS) and time-evolving block decimation (TEBD)---%
for one-dimensional systems.
To use an effective model like \onlinecite{white_effective_2021} risks assuming our conclusion.

\subsection{Initial state}\label{ss:initial}

\begin{figure}
  \centering
\begin{minipage}{.45\textwidth}
  \includegraphics[width=\linewidth]{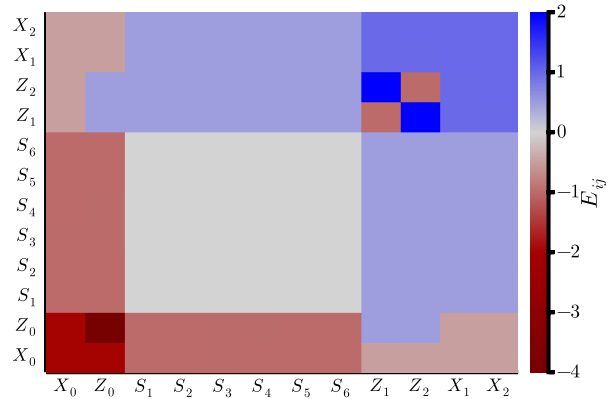}
\end{minipage}
\caption{
  \textbf{Energy of pairs of single-qutrit stabilizer states} in the Hamiltonian \eqref{eq:ham} with $J = h_x = h_z = 1$. The energy for the stabilizer product pair is defined as the bond term for the pair plus half the single site terms from each qutrit.
  $X_\alpha$ and $Z_\alpha$ label eigenstates of the shift operator $X$ and the clock operator $Z$, respectively, while $S_\alpha$ are eigenstates of the other onsite generalized Pauli operators $ZX, ZX^2$.
}
\label{fig:energy-density-possibilities}
\end{figure}

We wish to study the effect of local thermalization on subsystem mana.
To do so cleanly,
we choose each of our initial states to be the product of onsite stabilizer states
(eigenstates of the single-site generalized Pauli matrices)
picked to give constant energy density.
We use a product state 
because we expect that the trajectory of a subsystem's mana
will be intimately tied to the way it entangles with the rest of the system.
(Additionally, choosing our initial state to be a product state
keeps our matrix product state bond dimensions tractable for slightly longer.)
We use stabilizer states, which have zero mana,
so we can watch the initial growth of the mana,
as well as its transfer from short-range degrees of freedom
to long-range degrees of freedom.
We choose constant initial energy density
because we do not wish to confuse the effects of local thermalization with the those of long-time hydrodynamical relaxation
(which will drive the system's dynamics after the initial thermalization).

Our requirement that the energy density be spatially homogeneous
strongly constrains our initial state,
and indeed the energy densities we can choose.
Fig.~\ref{fig:energy-density-possibilities} illustrates the energy densities we can achieve with the product of two stabilizer states.
The energy density is a two-site operator---call the energy density on sites $j, j+1$ by
\begin{align}
  \begin{split}
    \varepsilon_{j,j+1} \equiv &-J [Z^\dagger_j Z_{j+1} + h.c.]\\
    &- \frac 1 2 h_x [(X_j + X_{j+1}) + h.c.] \\
    &+ \frac 1 2 h_z  [(Z_j + Z_{j+1}) + h.c.]
  \end{split}
\end{align}
for $1 < j < L-1$.
(One must take care at the ends of an open chain, i.e. $j = 1, L-1$.)
We can construct a state at a particular constant energy density 
by choosing the states $\ket {\psi_1}, \ket{\psi_2}$ on sites $1$ and $2$ to have that energy density,
and then choosing the site on each successive site $j > 2$ such that $\varepsilon_{j-1, j} = \varepsilon_{1,2}$.
This state selection fixes the energy density on the $L-1$ pairs of neighboring sites, but fails to account for the extra contribution of half the on-site field therms for the first and last site of the chain.
The initial state is only accepted if the total energy is $L$ times the specified local energy density.
This allows for some fluctuations in energy density at the ends of the chain arising from the extra on-site field contributions as long as these fluctuations do not alter the desired total energy. 
Only a discrete set of energy densities is possible,
and many possible energy densities admit only one state.
If there is only one state in a manifold,
we cannot average to avoid non-generic effects.
We therefore restrict ourselves to energy density manifolds with at least two states per bond;
this gives a manifold with $\mathcal N > 2^{L-1}$ possible states.
Fig.~\ref{fig:energy-density-possibilities}
illustrates the possibilities for $J = h_t = h_l = -1$.

We give some further details of the initial state configuration in Appendix \ref{section: init}

\subsection{Method}\label{ss:method}

The majority of our simulations use time evolution with TEBD \cite{vidal_efficient_2003,vidal_efficient_2004} with a second-order Trotter decomposition.
We find that Trotter step $dt = 1/16$ and bond dimension $\chi = 512$ give good convergence for $t \lesssim 6$
as measured by the half-chain entanglement entropy;
the subsystem mana is converged for $t \lesssim 8$.
We indicate the regime $6 \le t \le 8$ by dotted lines and lower color saturation.
App.~\ref{app:conv} gives details of our convergence testing.

\subsection{Quantity of interest}
We measure \textit{mana} $\mathcal M$.\cite{veitch_resource_2014}
In this section we give a very brief pr\'ecis of the relevant properties of mana.
In App.~\ref{app:mana} we describe how to calculate it.
For slightly less brief pr\'ecis from a similar point of view see Refs. \onlinecite{white_mana_2020,white_conformal_2020};
for more details see Refs. \onlinecite{veitch_resource_2014,gross_hudsons_2006,gross_non-negative_2007}.

Mana is a \textit{magic monotone}, meaning that it is non-increasing under Clifford unitaries,
partial traces, and (on average) stabilizer measurements.
For a pure state, the mana is zero if and only if the state is a stabilizer state.\cite{gross_hudsons_2006,gross_non-negative_2007}
Classical statistical mixtures of stabilizer states likewise have zero mana,
but some zero-mana mixed states are not statistical mixtures of stabilizer states.

Mana is \textit{multiplicative},
in the sense that for two density matrices $\rho_1, \rho_2$
\begin{align}
  \label{eq:mana-multiplicative}
  \mathcal M(\rho_1\otimes \rho_2) = \mathcal M(\rho_1) + \mathcal M(\rho_2)\;.
\end{align}
One therefore expects it to be extensive for states with short range correlations.
In fact on a system of $\ell$ qudits each with dimension $d$ one can bound
\begin{align}
  \label{eq:mana-bound}
  \mathcal M(\rho) \le \frac 1 2 (\ell \ln d - S_2)\;,
\end{align}
$S_2$ is the second R\'enyi entropy of the state $\rho$,
and the mana of a Haar-random state is extensive with subextensive corrections.\cite{white_mana_2020}

Mana is computed in terms of the \textit{Wigner norm} $\mathcal W$:
\[ \mathcal M(\rho) = \ln \mathcal W(\rho)\;.\]
Sometimes it is convenient to work in terms of this Wigner norm,
which shares the properties of mana, suitably translated.


\section{Mana at infinite temperature}\label{s:inf}

\begin{figure}
  \begin{minipage}{0.45\textwidth}
    \includegraphics[width=\linewidth]{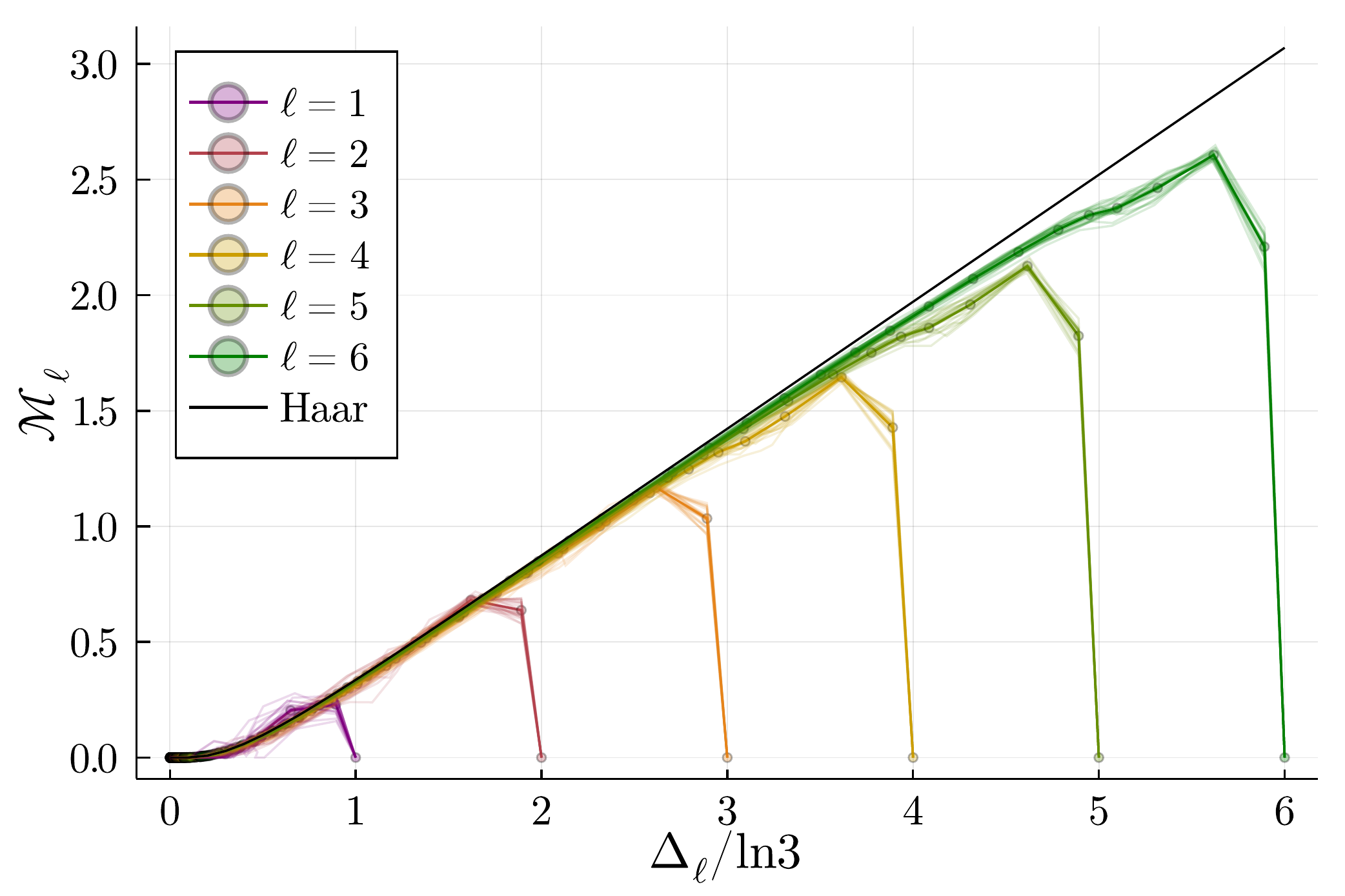}
  \end{minipage}
  \caption{\textbf{Subsystem mana in real-time evolution of infinite-temperature starting states}
    as a function of entropy deficit $\Delta(t) = \ell \ln 3 - S_2(t)$, with $S_2$ the subsystem second R\'enyi entropy
    (cf Fig.~\ref{fig:submana-inftemp-t}).
    Bold lines show the average over 20 random starting states (cf Sec.~\ref{ss:initial}),
    faint lines show trajectories of individual states,
    and dot mark points at which we measure the mana with time step $dt = 1/8$.
    In each case we see a rapid rise to approximately the average mana of a Haar state,
    due to local thermalization of the subsystem,
    followed by a slow decay as the subsystem entangles with the rest of the system.
 }
 \label{fig:submana-inftemp-delta}
\end{figure}

Consider a length-$\ell$ subsystem of our chain (we consider the $\ell$ most central sites).
The evolution of the subsystem mana is governed by the competition between two effects.
The subsystem's internal dynamics locally randomize the state, for a fast rise in the mana density. 
At the same time the coupling between the subsystem and its complement
steadily increases the subsystem's entropy:
since the mana is bounded by Eq.~\eqref{eq:mana-bound}
 this ongoing increase of entanglement must decrease the mana.
For short times the first effect dominates and the mana rapidly rises;
for longer times the second is dominates and the mana must slowly decay.
In the long-time limit $\ell \ln 3 - S_2 = 0$ (up to a Page correction,
which will be small for $\ell \ll L/2$),
so the subsystem mana must be
\[ \lim_{t \to \infty} \mathcal M = 0 \;.\]
The system as a whole, by contrast, is in a pure state.
One expects this pure state to be essentially a random state in the microcanonical ensemble,
and so to have Haar-like magic
\begin{align}
  \mathcal M \approx \frac 1 2 (\ell \ln 3 - \ln \pi/2)\;.
\end{align}
This whole-system saturation was observed for a related model in Ref. \onlinecite{goto_chaos_2021}.

In Fig.~\ref{fig:submana-inftemp-t}
we show the subsystem mana as a function of time.
The subsystems consist of the $\ell$ central sites of the MPS state.
We clearly see both effects---rapid initial rise due to local randomization,
followed by decay to the infinite temperature value $\mathcal M = 0$.

 \begin{figure}
  \centering
  \begin{minipage}{0.23\textwidth}
  \includegraphics[width=\linewidth]{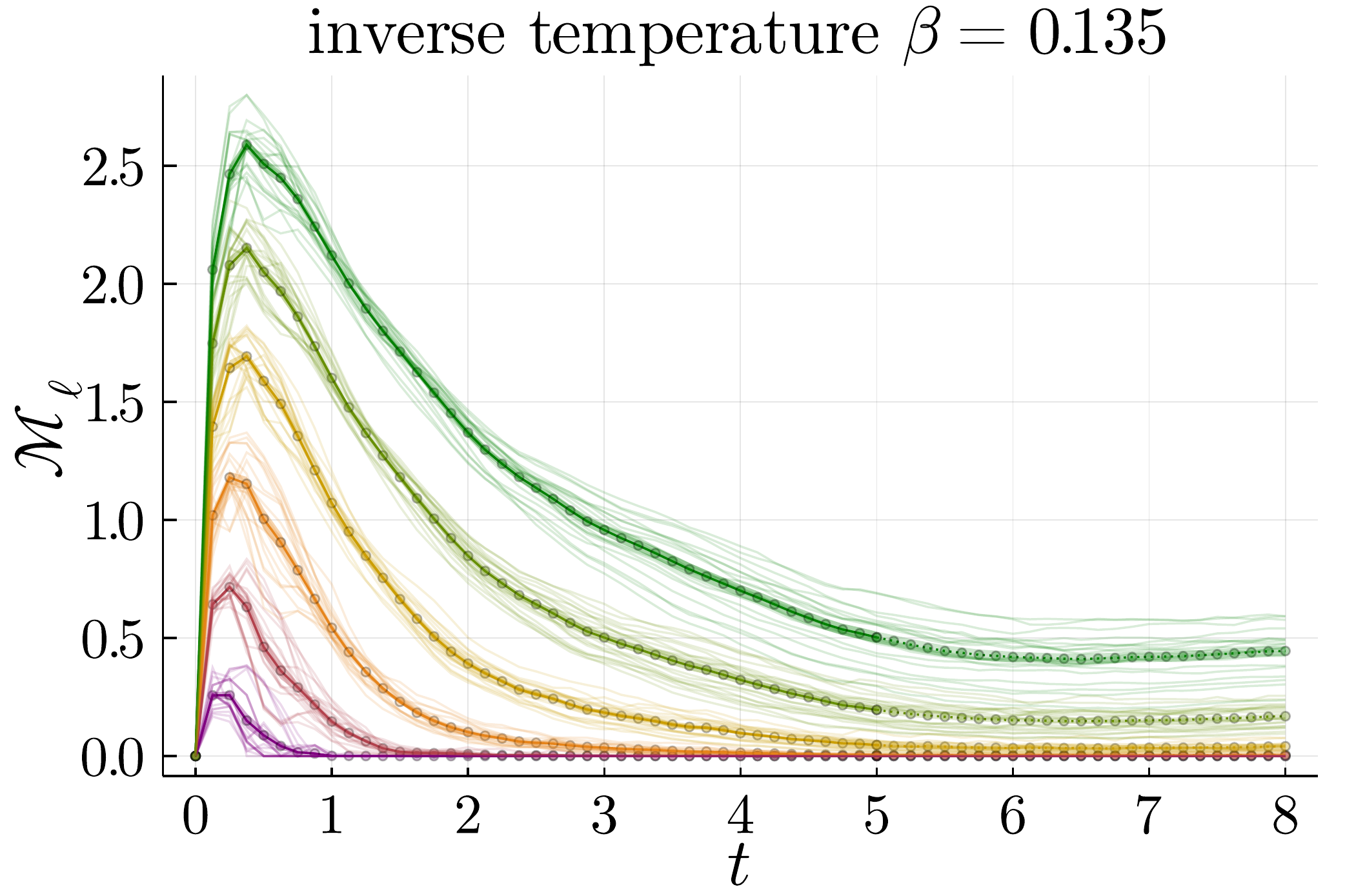}
 \end{minipage}
   \begin{minipage}{0.23\textwidth}
  \includegraphics[width=\linewidth]{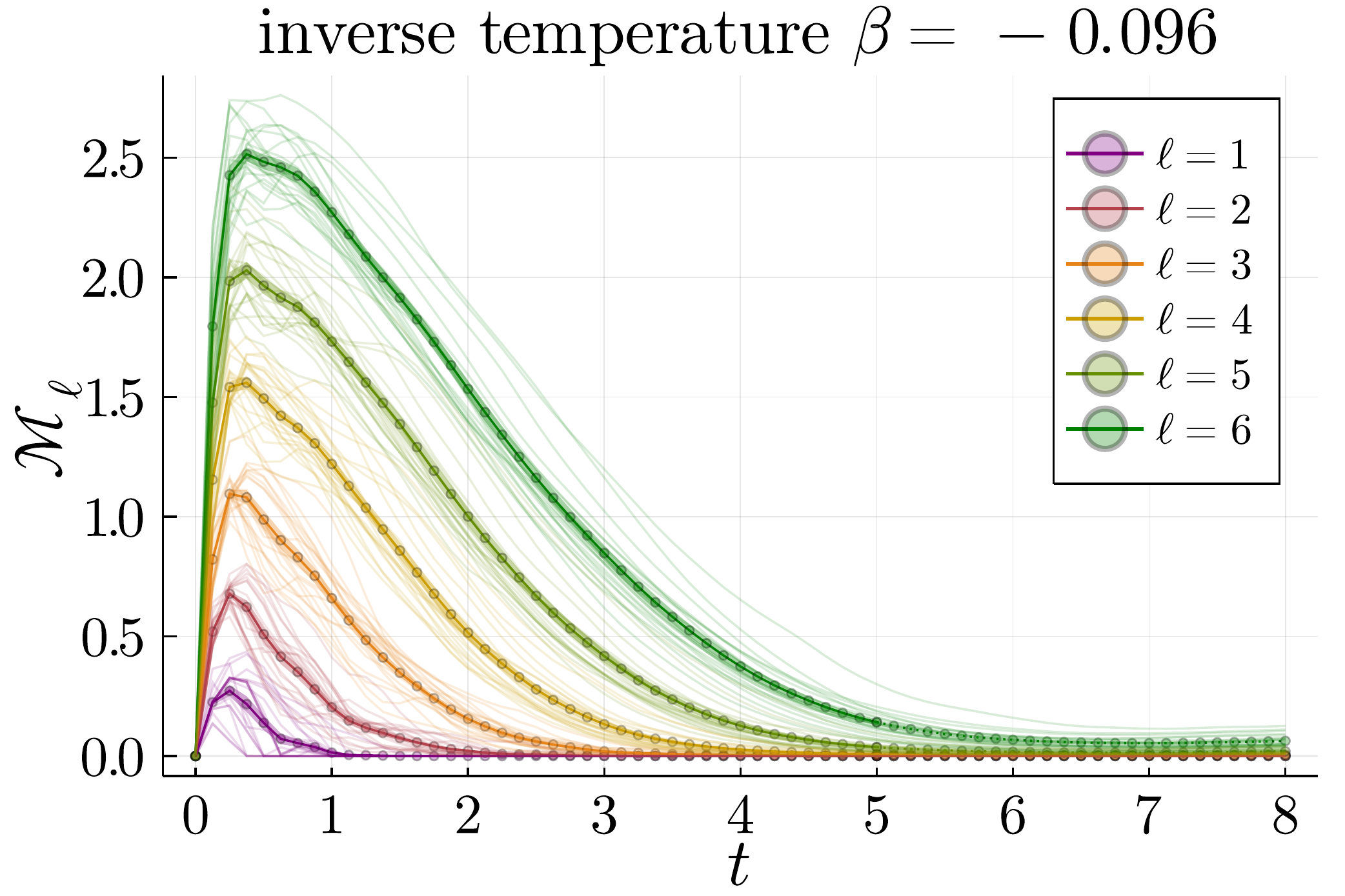}
 \end{minipage}
 \\
   \begin{minipage}{0.23\textwidth}
  \includegraphics[width=\linewidth]{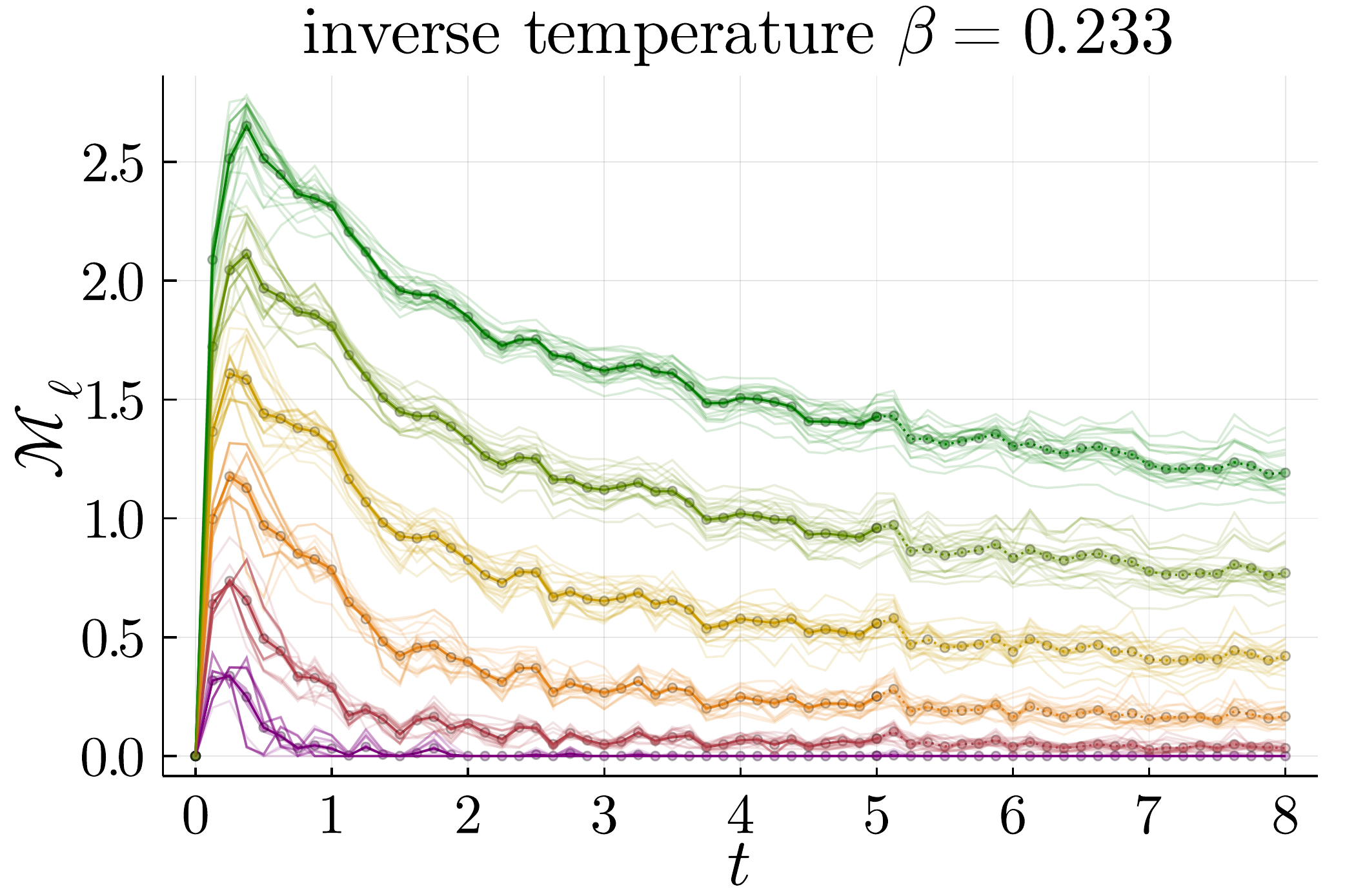}
 \end{minipage}
   \begin{minipage}{0.23\textwidth}
  \includegraphics[width=\linewidth]{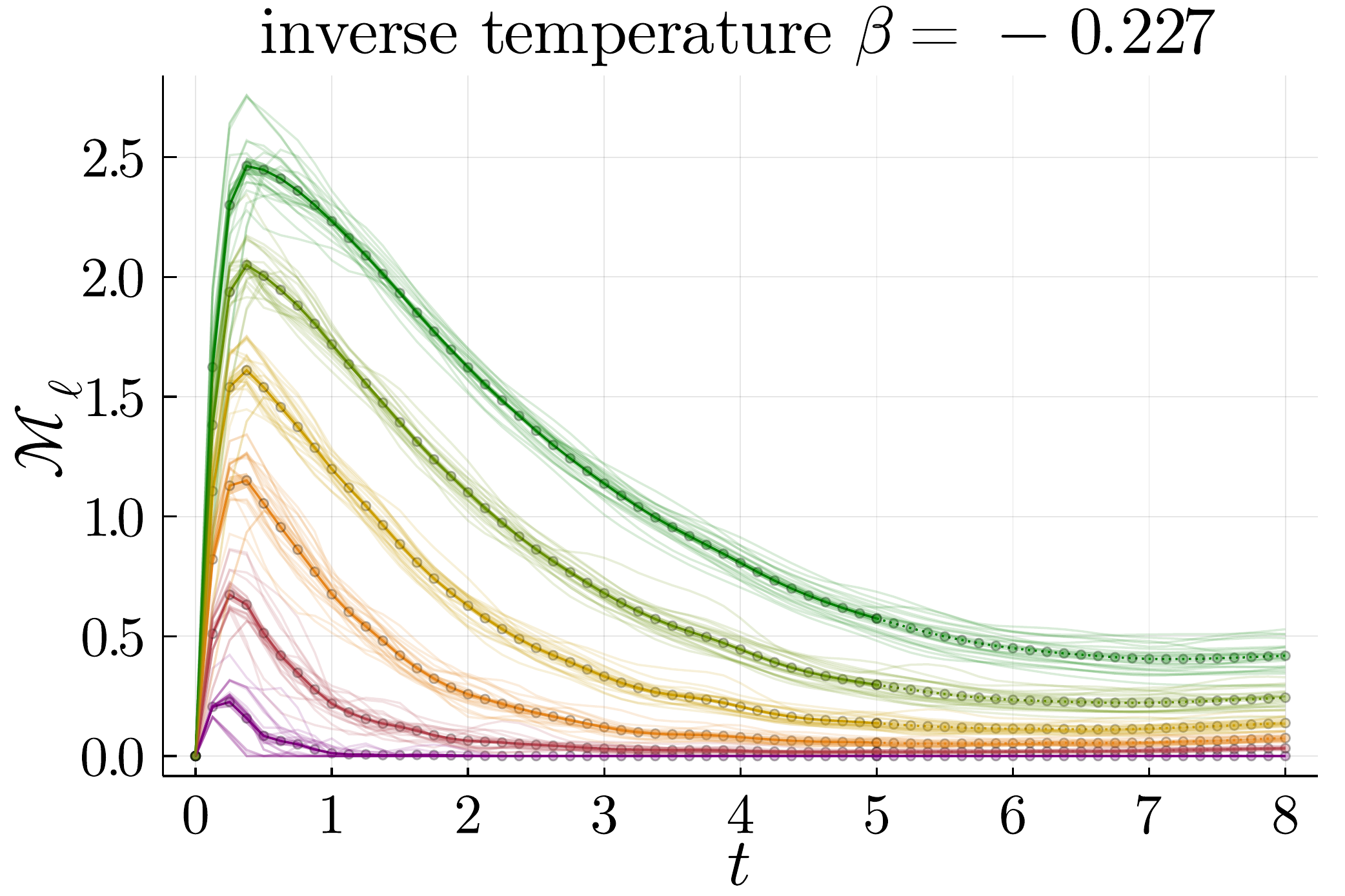}
\end{minipage}

  \caption{
    \textbf{Subsystem mana in real-time evolution of finite-temperature starting states} on a chain of 50 sites.
    Bold lines show the average over 20 random starting states (cf Sec.~\ref{ss:initial}),
    faint lines show trajectories of individual states,
    and dot mark points at which we measure the mana with time step $dt = 1/8$.
    As in the infinite-temperature case we see a rapid rise,
    due to local thermalization of the subsystem,
    followed by a slow decay as the subsystem entangles with the rest of the system---but the decay no longer continues to mana $\mathcal M_\ell = 0$ at accessible times.
     }
  \label{fig:submana-fintemp}

 \end{figure}

The bound Eq. \eqref{eq:mana-bound} depends solely on the ``entropy deficit''
\begin{equation} \Delta = \ell \ln d - S_2\;; \end{equation}
one is entitled to ask how close the subsystem mana comes to saturating that bound.
Moreover one might expect the decay of the subsystem mana
to track the mana of a subsystem of a Haar-random state.
For large Hilbert space dimension the Wigner norm of a subsystem of a Haar state
is controlled solely by the entropy deficit,
and given by\cite{white_mana_2020}
\begin{align}
  \label{eq:haar-wigner}
  \begin{split}
    \expcth{\mathcal W} &= \sqrt{2/\pi}\;(\sigma/\mu) e^{-\mu^2/2\sigma^2} + \erf(\mu/\sigma\sqrt{2})
  \end{split}
\end{align}
where
\[  \frac {\sigma^2}{\mu^2} = e^{\Delta} - 1\;.\]
For $\Delta \gtrsim \ln \pi / 2$ the mana resulting from this expression becomes\cite{white_mana_2020}
\begin{align}
  \label{eq:haar-mana}
 \mathcal M \approx \frac 1 2 [\Delta - \ln \pi / 2]\;.
\end{align}
In Fig.~\ref{fig:submana-inftemp-delta} we plot the subsystem mana against the entropy deficit $\Delta$.
Each subsystem starts at zero entropy $S_2$,
hence large entropy deficit $\Delta$,
and moves right to left to smaller entropy deficit.
(We plot the entropy as a function of time in App.~\ref{s:subent}.)
Dot mark points where we measure the mana,
every time step $dt = 1/8$.
We see again a fast early rise and a long-time decay,
matching the Haar prediction based on \eqref{eq:haar-wigner}.

At intermediate times the state mana undershoots the Haar prediction.
This is because different parts of the subsystem are not fully entangled.
Heuristically, the subsystem behaves like a tensor product of Haar states on smaller subsystems.
The mana is the sum of the manas of these smaller subsystems
(cf \eqref{eq:mana-multiplicative}),
each of which comes with a $\ln \pi / 2$ correction from \eqref{eq:haar-mana}.
Once the system is fully (internally) entangled
one can think of it as a single Haar state with a single $\ln \pi / 2$ correction:
this gives the black line in Fig.~\ref{fig:submana-inftemp-delta}.



\section{Mana at finite temperature}\label{s:finite}

\begin{figure}[t]
  \centering
  \begin{minipage}{0.23\textwidth}
    \includegraphics[width=\linewidth]{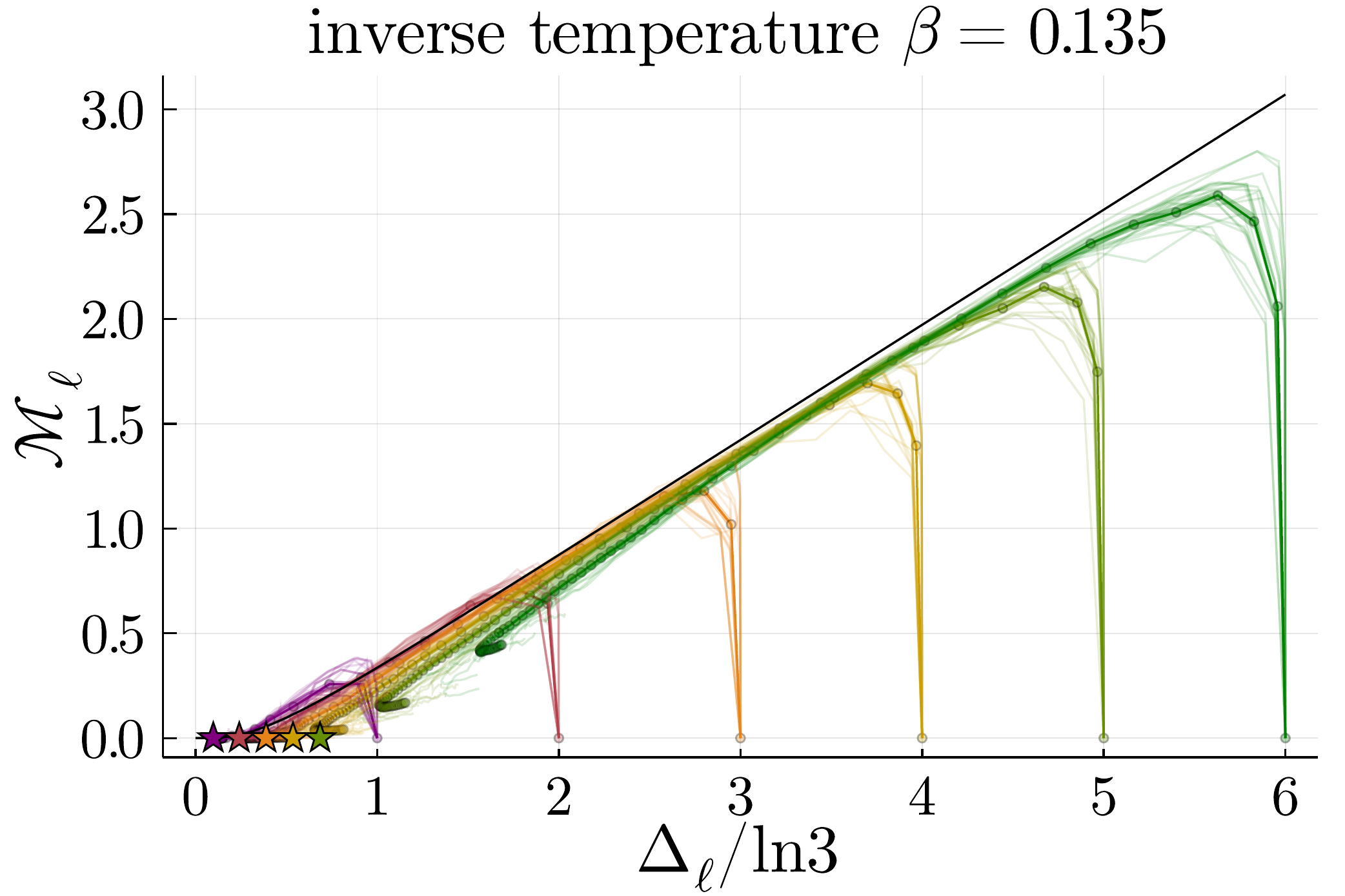}
  \end{minipage}
  \begin{minipage}{0.23\textwidth}
    \includegraphics[width=\linewidth]{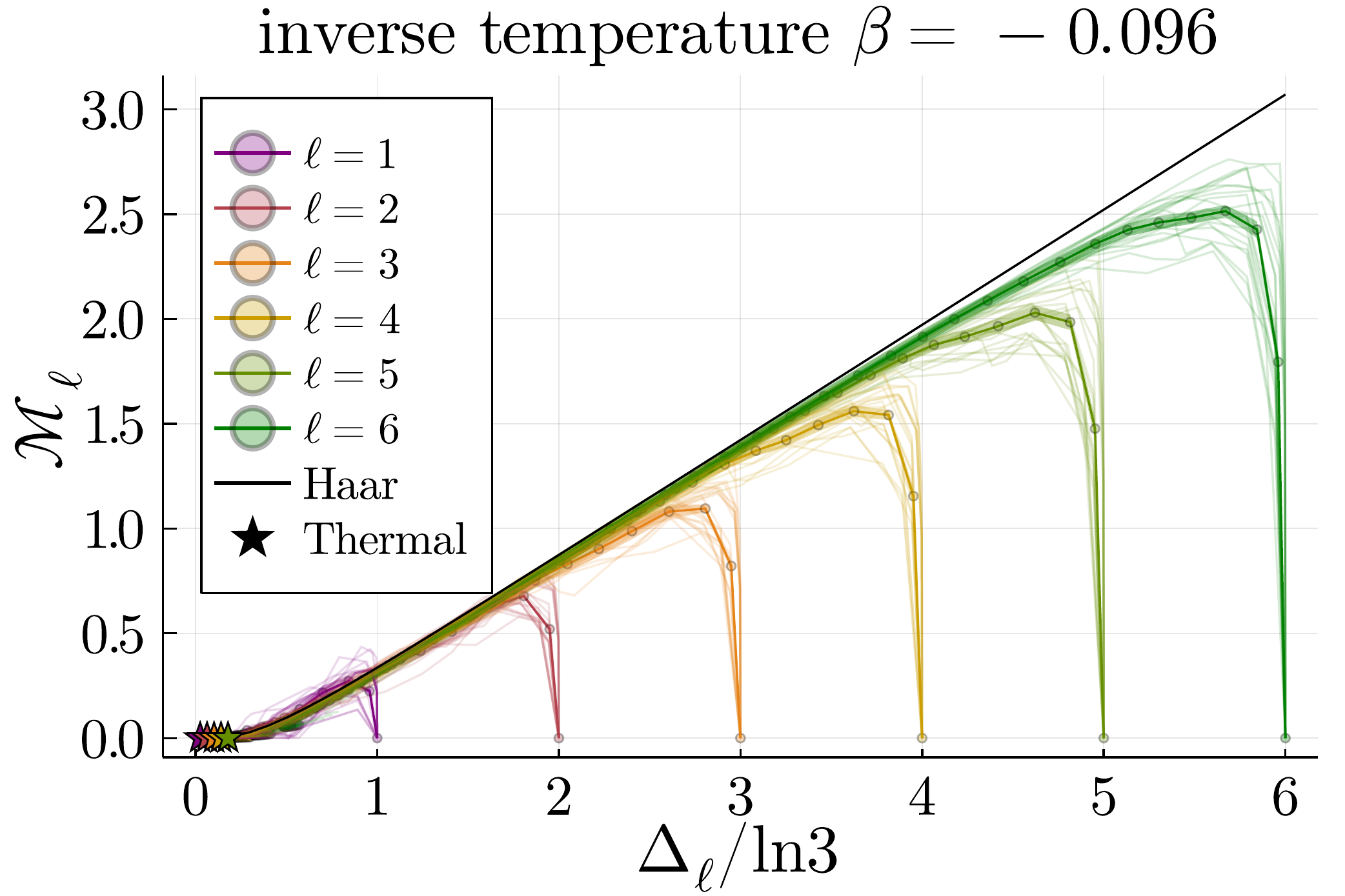}
  \end{minipage}
  \\
  \begin{minipage}{0.23\textwidth}
    \includegraphics[width=\linewidth]{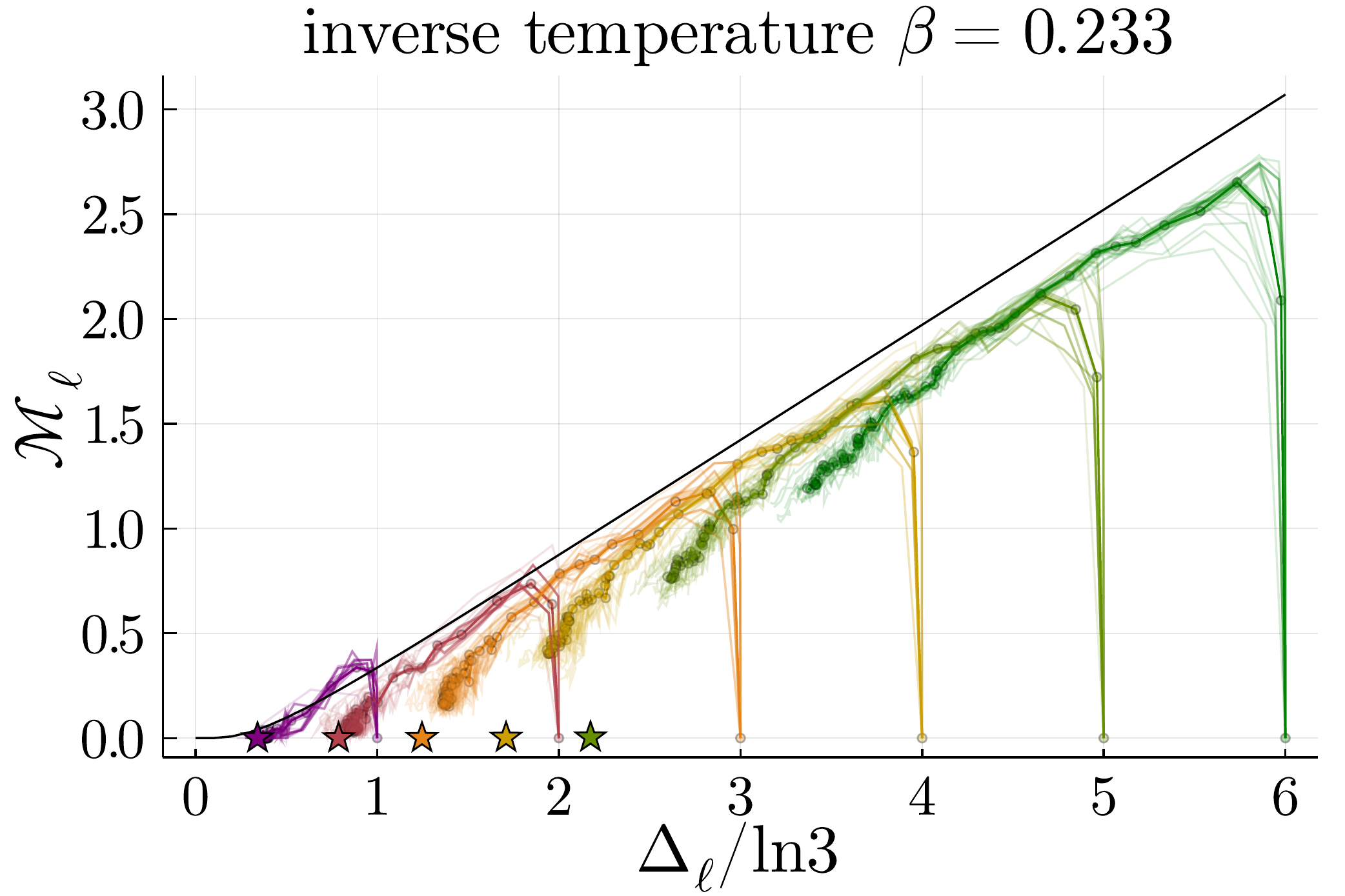}
  \end{minipage}
  \begin{minipage}{0.23\textwidth}
    \includegraphics[width=\linewidth]{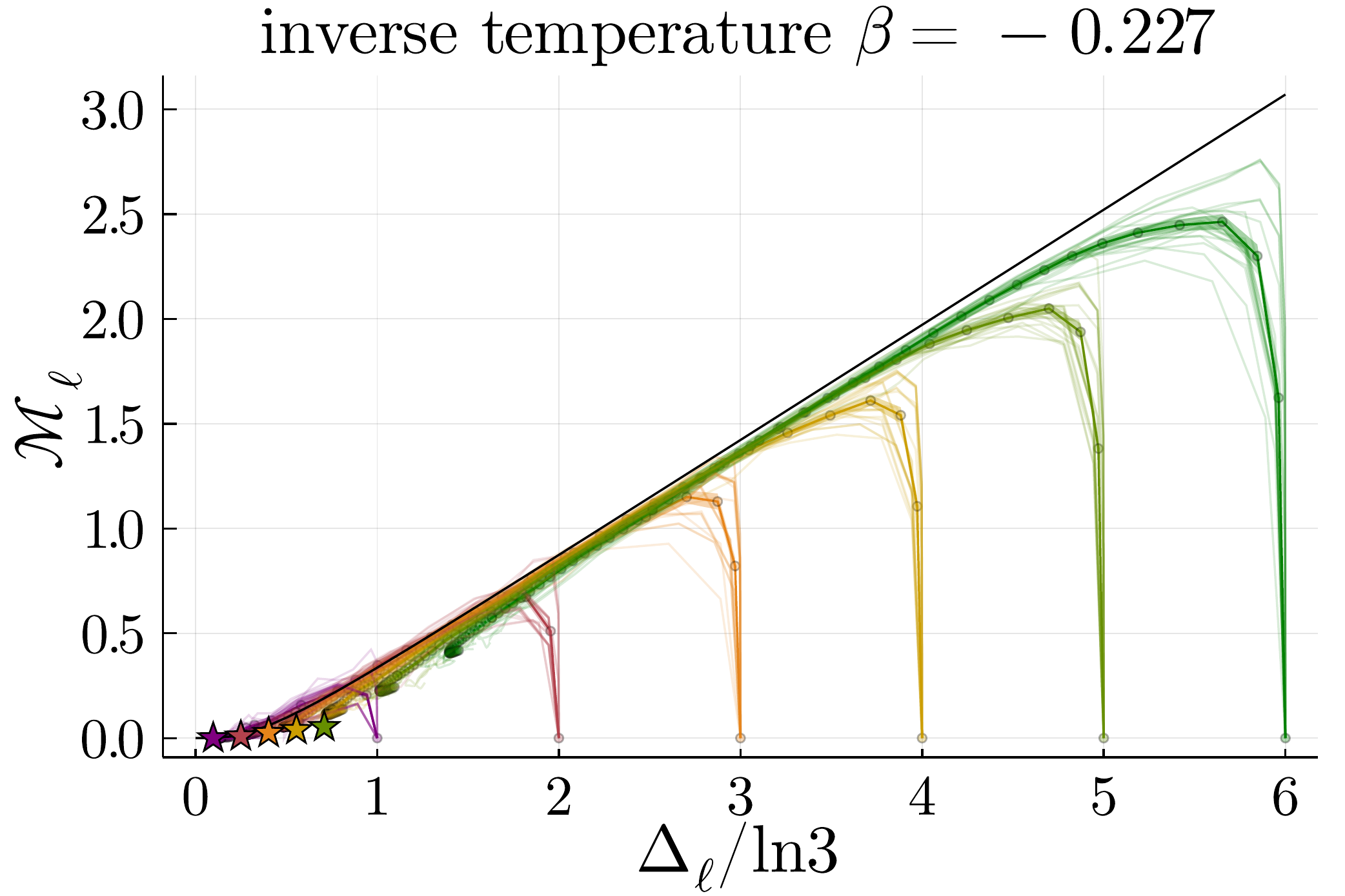}
  \end{minipage}

  \caption{
    \textbf{Subsystem mana in real-time evolution of finite-temperature starting states}
    as a function of entropy deficit $\Delta(t) = \ell \ln 3 - S_2(t)$, $S_2$ the subsystem second R\'enyi entropy
    (cf Fig.~\ref{fig:submana-inftemp-t}).
    Stars mark Gibbs state values.
    Time increases as entropy deficit decreases, i.e. from right to left.
    In each case the long-time endpoint
    ($t = 8$, leftmost on each curve)
    has mana and entropy noticeably larger than the Gibbs value.
  }
  \label{fig:entdef-fintemp}
\end{figure}

How does this picture change when the initial state has nonzero energy?
Fig.~\ref{fig:submana-fintemp} shows subsystem mana as a function of time,
and Fig.~\ref{fig:entdef-fintemp} shows it as a function of the subsystem's entropy deficit.
In each case we show a variety of energies, labeled by equilibrium temperature.%
\footnote{
  The model has one conserved quantity, the energy.
  Every energy $E_0$ corresponds to an equilibrium temperature $\beta(E_0)$ such that
  \[ E_0 = Z^{-1} \tr H e^{-\beta(E_0)H}\;; \]
  if the state after time evolution locally approximates any Gibbs state,
  it must be this Gibbs state.
}

We see an initial rise followed by a slow decay, broadly following the Haar value, as in the infinite-temperature case.
In both Fig.~\ref{fig:submana-fintemp} and Fig.~\ref{fig:entdef-fintemp}, the $\beta = 0.233$ averages display much more variation than other temperatures.
We believe that the variation is due to the tight constraints on initial states at this energy (cf App.~\ref{section: init}).
Although there are exponentially many suitable initial states, they are locally similar.

\begin{figure}
  \centering
  \begin{minipage}{0.45\textwidth}
  \includegraphics[width=\linewidth]{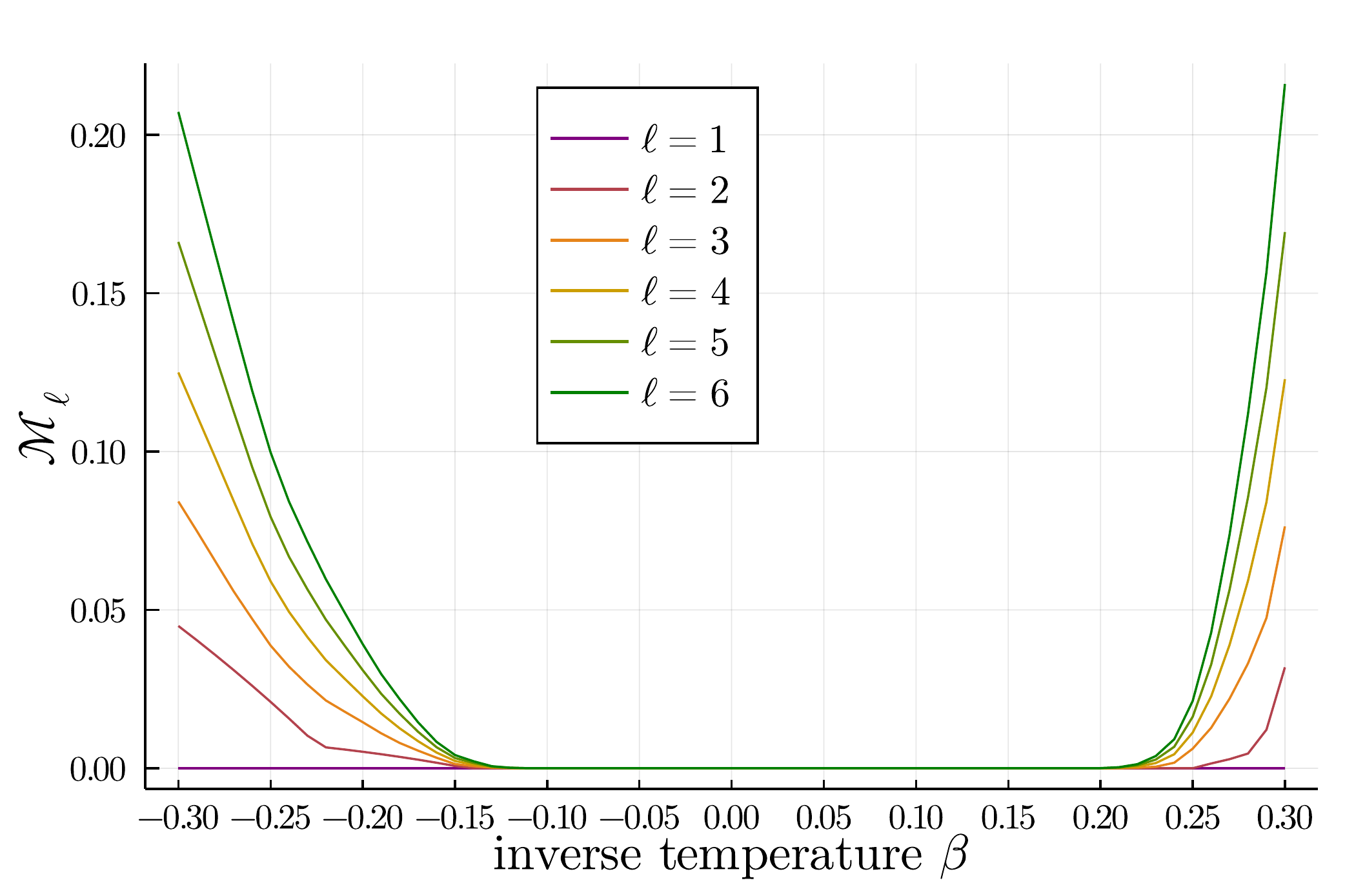}
 \end{minipage}
 \caption{\textbf{Subsystem mana for Gibbs states} of the Hamiltonian \eqref{eq:ham} on 9 qutrits with periodic boundary conditions. We take $J = h_z = h_z = 1$.
  }
 \label{fig:thermal-exact}
 
\end{figure}

But careful examination of 
Fig.~\ref{fig:entdef-fintemp} presents a mystery.
We extract the subsystem mana and entropy of a Gibbs state from exact-diagonalization simulations on small systems (cf Fig.~\ref{fig:thermal-exact} and App.~\ref{s:gibbs});
we mark those values in Fig.~\ref{fig:entdef-fintemp} with a dot.

We attribute the discrepancy to a subtle finite-time effect related to deviations from so-called \textit{canonical typicality}.%
\cite{tasaki_quantum_1998,goldstein_canonical_2006,popescu_entanglement_2006,reimann_typicality_2007,goldstein_approach_2010,goldstein_normal_2010,reimann_generalization_2015}
Essentially, the discrepancy is controlled by the size of the Hilbert space of the region with which our subsystem is entangled.
As that region grows the mana will approach its thermal value---%
but our MPS simulations are limited to times for which the entangled region is small.

To understand this, consider first subsystem mana in finite systems at long times.
Heuristically, one expects long-time states to behave much like
Haar-random states on a microcanonical subspace
(we rehearse the standard intuition behind this statement in App.~\ref{thermalization-intuition}).
Almost all such states have subsystem reduced density matrices near in trace norm to
the subsystem reduced density matrices of the microcanonical density matrix,\cite{popescu_entanglement_2006}
hence to that of the Gibbs state.

But this is not enough:
two density matrices close in trace norm can have widely divergent Wigner norms.
We show in App.~\ref{s:mana-bnd} that if density matrices $\rho,\sigma$ on a dimension-$d$ subspace have
\[ \|\rho - \sigma\|_1 \le \eta\;, \]
then
\begin{equation}
  \label{eq:mana-trace-bound}
  \Big|\ \|\rho \|_W - \| \sigma\|_W \ \Big| \le \min(d^{2}\eta, d^{1/2}\sqrt \eta) \;.
\end{equation}
In App.~\ref{s:mana-typicality} we argue that the Hilbert space dimension factors in \eqref{eq:mana-trace-bound}
mean that it can only give very wide bounds on mana when combined with the result of Popescu.
Moreover a Gibbs state can have zero mana but be near the boundary of the zero-mana region; a nearby thermal state may then have small but nonzero mana.

\begin{figure}[t!]
  \centering
  \begin{minipage}{0.23\textwidth}
  \includegraphics[width=\linewidth]{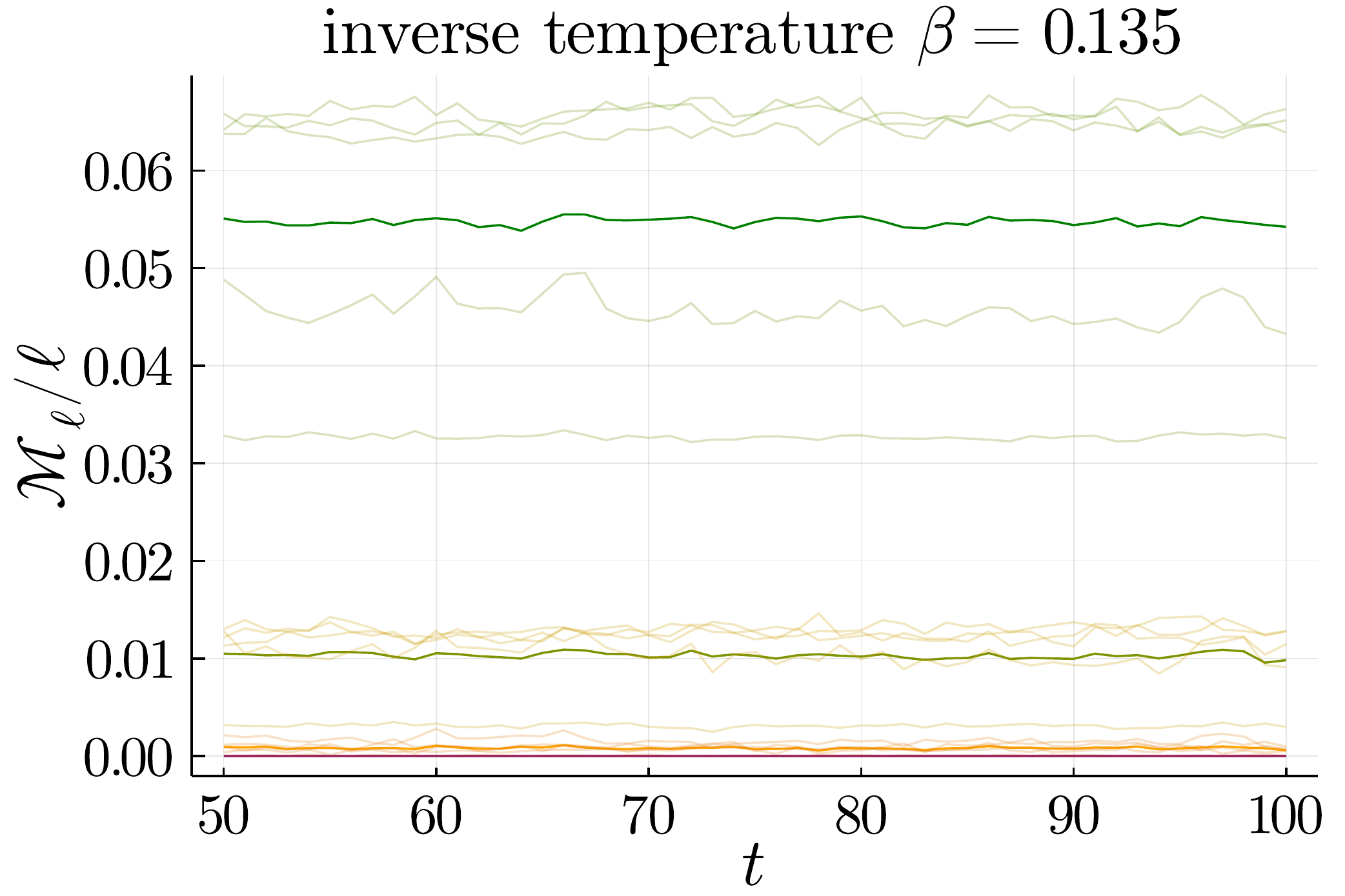}
 \end{minipage}
   \begin{minipage}{0.23\textwidth}
  \includegraphics[width=\linewidth]{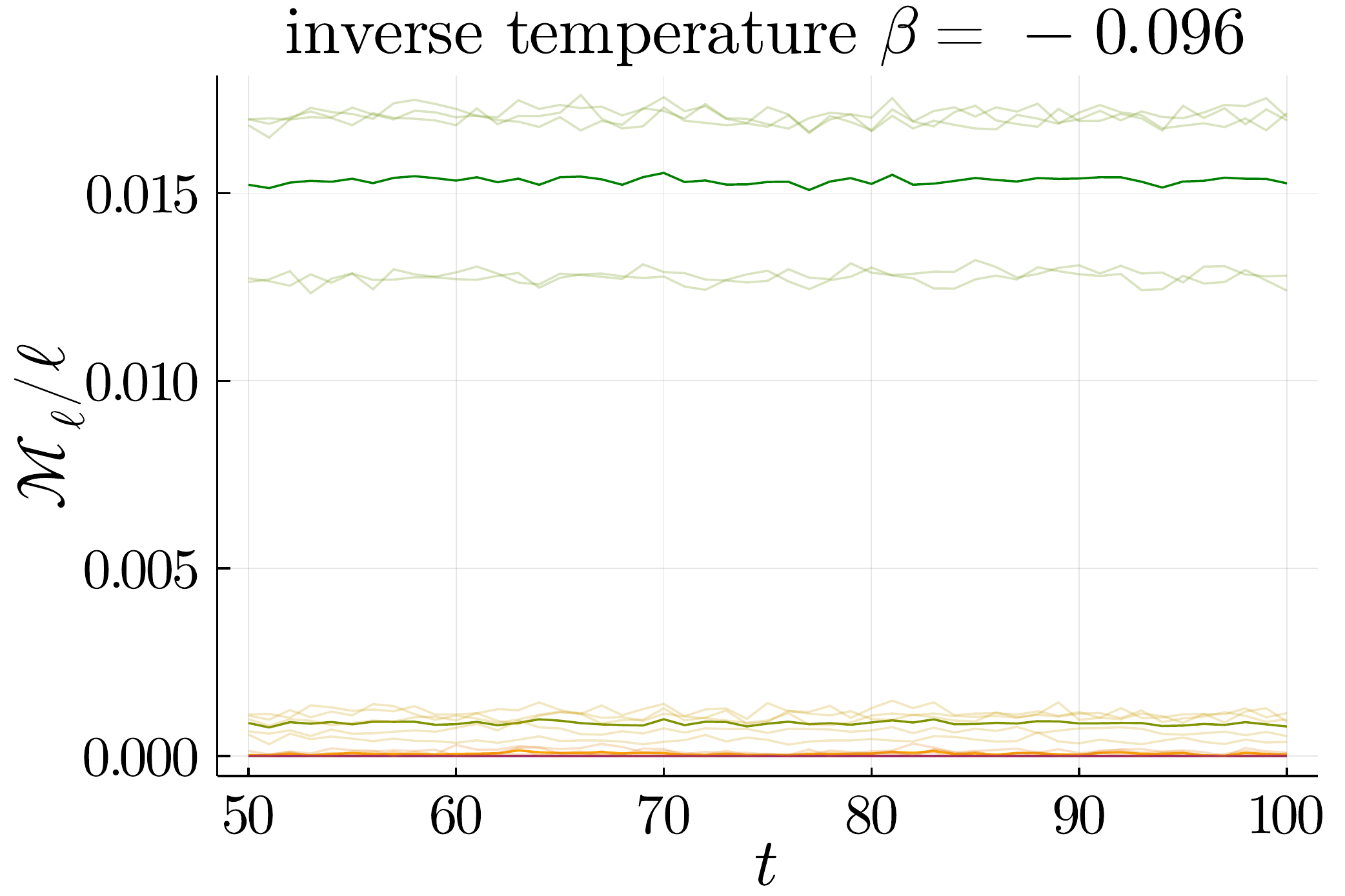}
 \end{minipage}
 \\
   \begin{minipage}{0.23\textwidth}
  \includegraphics[width=\linewidth]{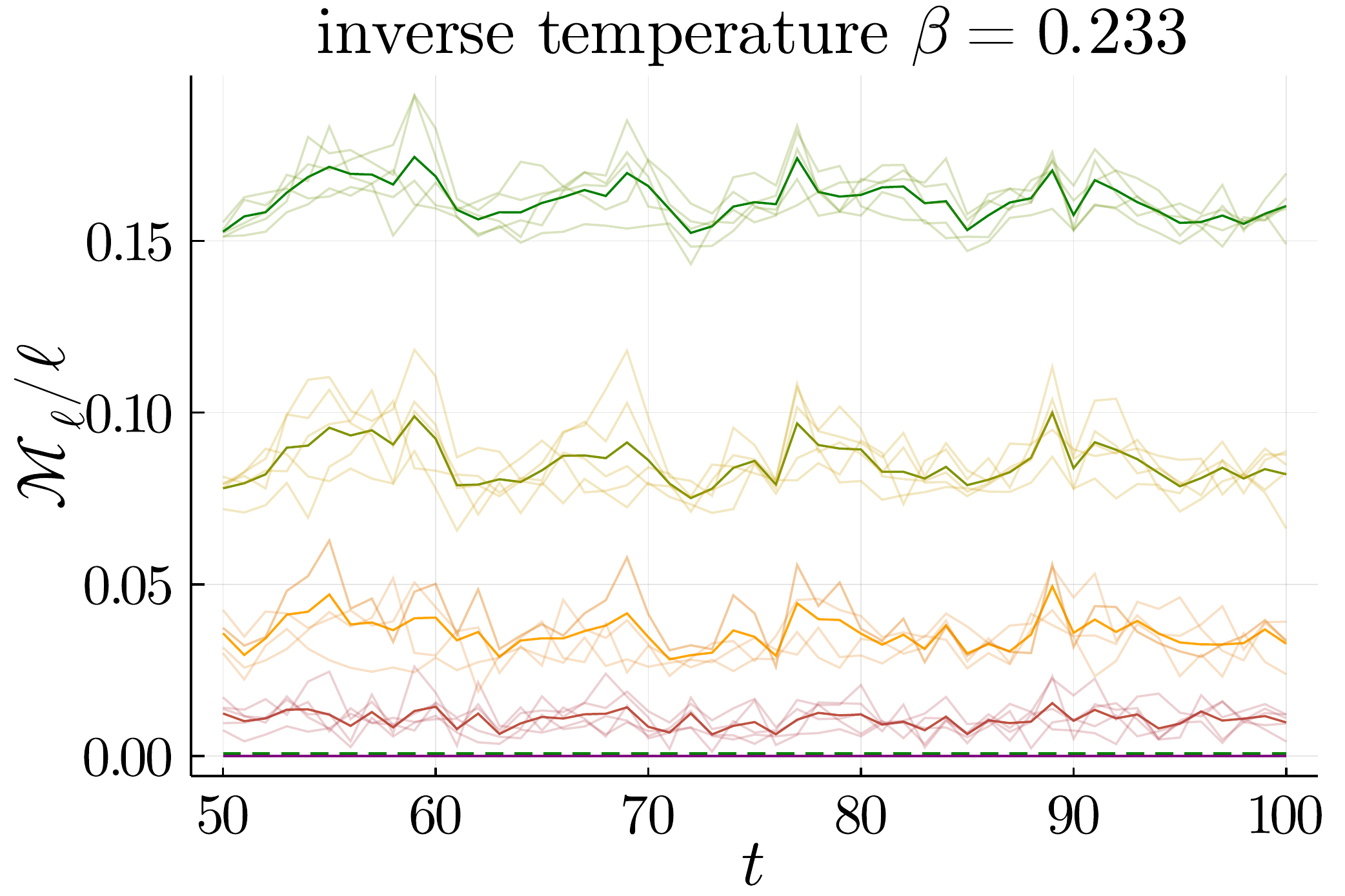}
 \end{minipage}
   \begin{minipage}{0.23\textwidth}
  \includegraphics[width=\linewidth]{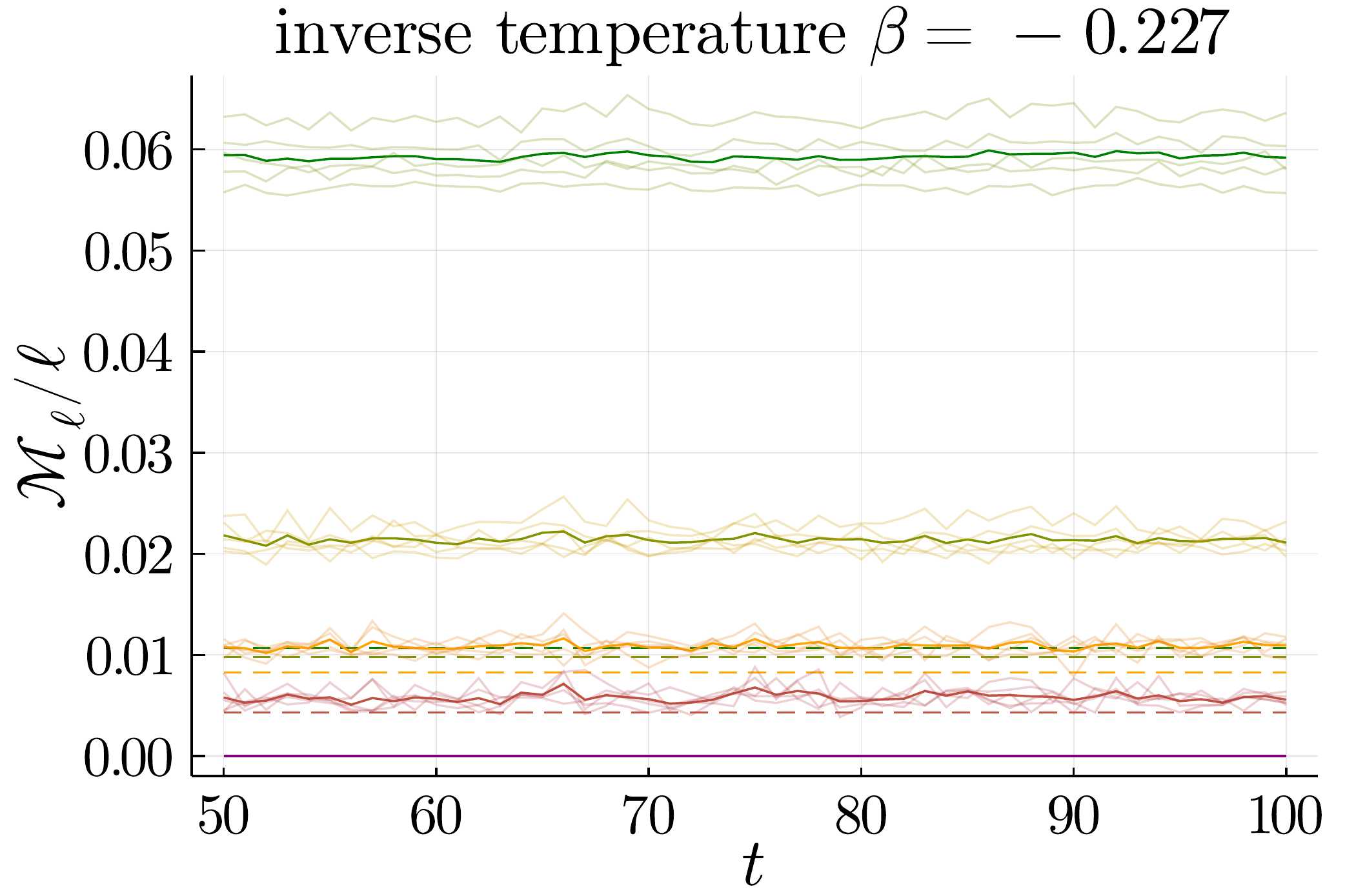}
 \end{minipage}
 \caption{
   \textbf{Long-time subsystem mana per site in Krylov evolution} of an $L = 11$ site chain with thermal values marked with horizontal dashed lines. The average value of five samples is given by a dark solid line with each sample plotted using a light line. Thermal values for all samples except $\beta = -0.227$ are too small to be seen on the plot scale.  }
 \label{fig:submana-fintemp-krylov}
 \end{figure}

 In Fig.~\ref{fig:submana-fintemp-krylov} we plot subsystem mana as a function of time in small (Krylov-accessible) systems at much longer times than are accessible to matrix product states;
we see a steady-state deviation between the mana of the time-evolved state and the Gibbs distribution mana.

So much for finite systems at large times.
What about large systems at finite times?
We have considered fixed-sized subsystems;
as the surrounding system becomes large,
even bounds based on \eqref{eq:mana-trace-bound} and the typicality result of \onlinecite{popescu_entanglement_2006}
will strongly constrain the mana.
But a finite-time state will be very different from most random microcanonical states:
those states have nearly maximal entanglement,
while the state at time $t$ has entanglement entropy $S \approx 2ct$.
That finite-time state, then, can be crudely modeled by a Haar state from the microcanonical subspace on a system of size $l + 2ct$.
(To construct a less crude model one might use random matrix product states.
\cite{garnerone_typicality_2010,garnerone_statistical_2010,collins_matrix_2013}
A random MPS would mimic not only the entanglement of the subsystem with its surroundings,
but also its internal entanglement structure.)
This crude model is broadly consistent with Fig.~\ref{fig:submana-fintemp},
in that the decay times increase linearly with subsystem time.
We leave a more careful comparison of numerics with predictions from canonical typicality to future work.

\section{Discussion}\label{s:discussion}

We find that local subsystems of zero-energy initial states have zero mana after a short local thermalization time proportional to the subsystem size, 
consistent with a characterization as an infinite temperature Gibbs state.
This suggests that infinite-temperature hydrodynamics may be simulated classically methods with low overhead.
In the context of quantum simulation, there may be effective dynamics using mixtures of stabilizer states and Clifford circuits, or circuits with few non-Clifford gates, that also effectively reproduce the infinite-temperature hydrodynamics.
At finite temperature the landscape is more complicated:
for sufficiently high (but still finite) temperature the Gibbs state, hence the long-time thermal state, has zero mana;
while for somewhat lower temperatures, the subsystem mana is small but nonzero.
Additionally, regardless of temperature, the subsystem mana takes a long time to relax to the thermal value, because mana is sensitive to small deviations from canonical typicality.

Because mana is sensitive to small deviations from the Gibbs state, approximate simulations may be able to achieve some desired precision using states with much lower mana than that of the target state.
This could allow approximate simulations to use fewer non-Clifford resources required than would be for exact simulation, which would broaden the scope of physics accessible using near-Clifford simulation techniques and reduce the cost of quantum simulation.

We would be remiss not to outline some limitations of our work.
Most seriously, comparisons between the mana complexity hump of Fig.~\ref{fig:submana-inftemp-t} or \ref{fig:submana-fintemp}
and the entropy complexity hump of e.g. \onlinecite{rakovszky_dissipation-assisted_2020} Fig.~2
can be importantly misleading.
The cost of an MPO calculation is,
broadly speaking,
\begin{equation}
  \label{eq:mpo-cost}
 [\text{MPO cost}] \sim \sum_{\text{bonds}} \exp\left[\alpha S_{\text{bond}}\right] \propto L\;,
\end{equation}
where $S_{\text{bond}}$ is the entanglement entropy across a bond and $\alpha$ is some power.
The simulation cost may be dominated by the peak bond entropy,
but it is still polynomial in system size.
Straightforward near-Clifford simulations, by contrast, will have cost
\begin{align}
  \label{eq:nc-cost}
  \begin{split}
    [\text{near-Clifford cost}] &\sim \exp\left[\gamma\hskip -1.2ex \sum_{\text{subsystems}}\hskip -1.2ex \mathcal M_{\text{subsystem}} \right] \\[1.5ex]
    &\sim e^{\gamma m L}
  \end{split}
\end{align}
for some finite mana density $m$.
That is, even if we take advantage of the insensitivity of hydrodynamics to details of long-range correlations,
the cost is still exponential in system size.
Worse, since the peak mana is close to the maximum mana,
this suggests that there are no cost savings to be had from short-range approximations.

We believe this obstacle is superable.
Because the peak mana occurs at short times,
when the system displays only short-range entanglement,
one should be able to decouple simulation of different subsystems,
effectively exchanging sum and exponential in \eqref{eq:nc-cost} and giving cost polynomial (indeed linear) in system size.
Concretely, imagine dividing the system into subsystems of some length $l$,
separating those subsystems separately for $t \propto l$
and then re-introducing the couplings.
While this dramatically changes the early- to intermediate-time dynamics,
it should not change the long-time hydrodynamics,
and it suggests that more sophisticated schemes are possible.

The nonzero finite-temperature long-time mana is another limitation of our work.
Absent decoupling tricks like those required for early-time simulations,
this nonzero subsystem mana density means that the simulation cost is exponential in system size,
albeit with small exponent.

And looming behind these limitations is the fact that
that while mana controls the difficulty of some classical simulation methods
and lower-bounds the non-Clifford resources required to create a state,
these bounds are not constructive:
knowing that a state has low mana
does not give one a recipe for constructing it.
We leave algorithms---the analogues of DMT or DAOE---to future work.

We have framed this result in terms of classical simulation,
because we expect that to be its immediate application,
but in many ways it is more naturally understood in the context of future error-corrected quantum computers.
In many error-correction architectures Clifford gates are ``easy''.
Non-Clifford gates, by contrast, must be performed by costly magic-state distillation and injection schemes:
magic comes dear.
Our results suggest that good local approximations to long-time states will dramatically reduce requirements for this key resource.

\acknowledgements
We are grateful to Brian Swingle for many helpful conversations.
CDW thanks the U.S. Department of Energy (DOE), Office of Science, Office of Advanced Scientific Computing Research (ASCR) Quantum Computing Application Teams program, for support under fieldwork proposal number ERKJ347. TJS is grateful for support from the Department of Energy under award number DE-SC0019139 and to Alex Bocharov for useful comments.

\bibliography{references}
\appendix

\section{Initial state sampling} 
\label{section: init}

For our MPS time evolution, we choose pure initial states with no mana or entanglement, ie tensor products of the twelve single qutrit stabilizer states. 
We also want to sample from initial states with the same energy density for local subsystems to thermalize to a consistent temperature without hydrodynamic energy transport. 
Looking at the energy for pairs of single qutrit stabilizers we can see what energy densities and therefore temperatures will be accessible to this choice of initial states. 
The energy densities for pairs of stabilizer states are grouped into the X eigenstates (1, 11, 12), the Z eigenstates (2, 9, 10), and other stabilizer states (3-8). 
The energies with highest degeneracy are any of the non X or Z eigenstates, which have zero energy density when paired together, or a slightly positive or negative energy density when paired with one of the X or Z eigenstates. 
When $h_l =0$ states 3-8 only have nonzero energy density when paired with an X eigenstate. 
This means our initial state sequences will be random among states 3-8 for infinite temperature, or alternating between an X or Z eigenstate and one of the other eigenstates. 
The colored tables and graphs show the energy densities for different pairs of these single qutrit stabilizer states, colored so negative energies are red and positive energies are blue. 
Variation of the on-site longitudinal field only affects pairs with a Z eigenstate since all other single qutrit stabilizers have zero expectation value for this term.

\begin{figure}[t]
  \centering
\begin{minipage}{.45\textwidth}
  \includegraphics[width=\linewidth]{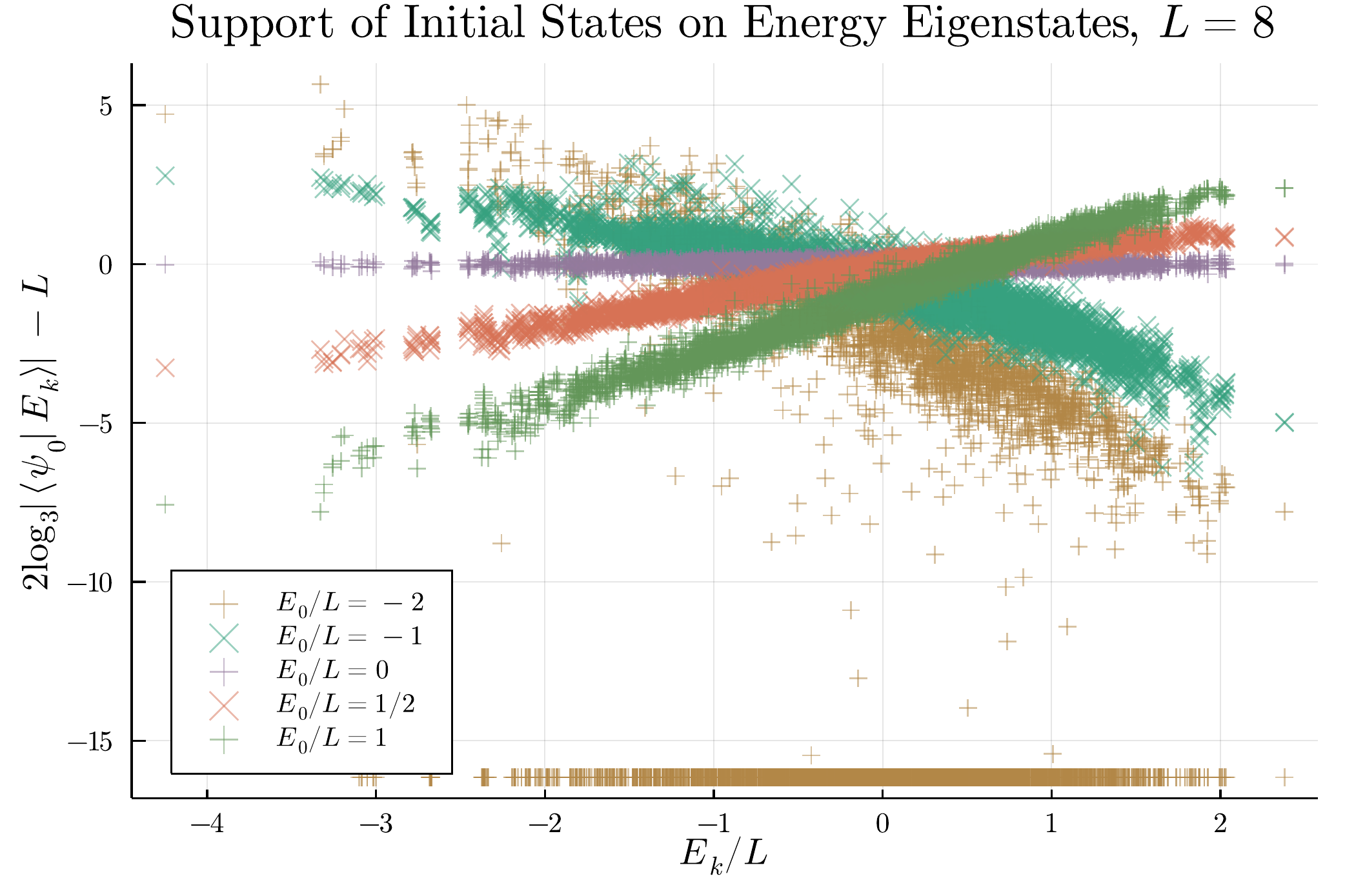}
\end{minipage}
\begin{minipage}{.45\textwidth}
  \includegraphics[width=\linewidth]{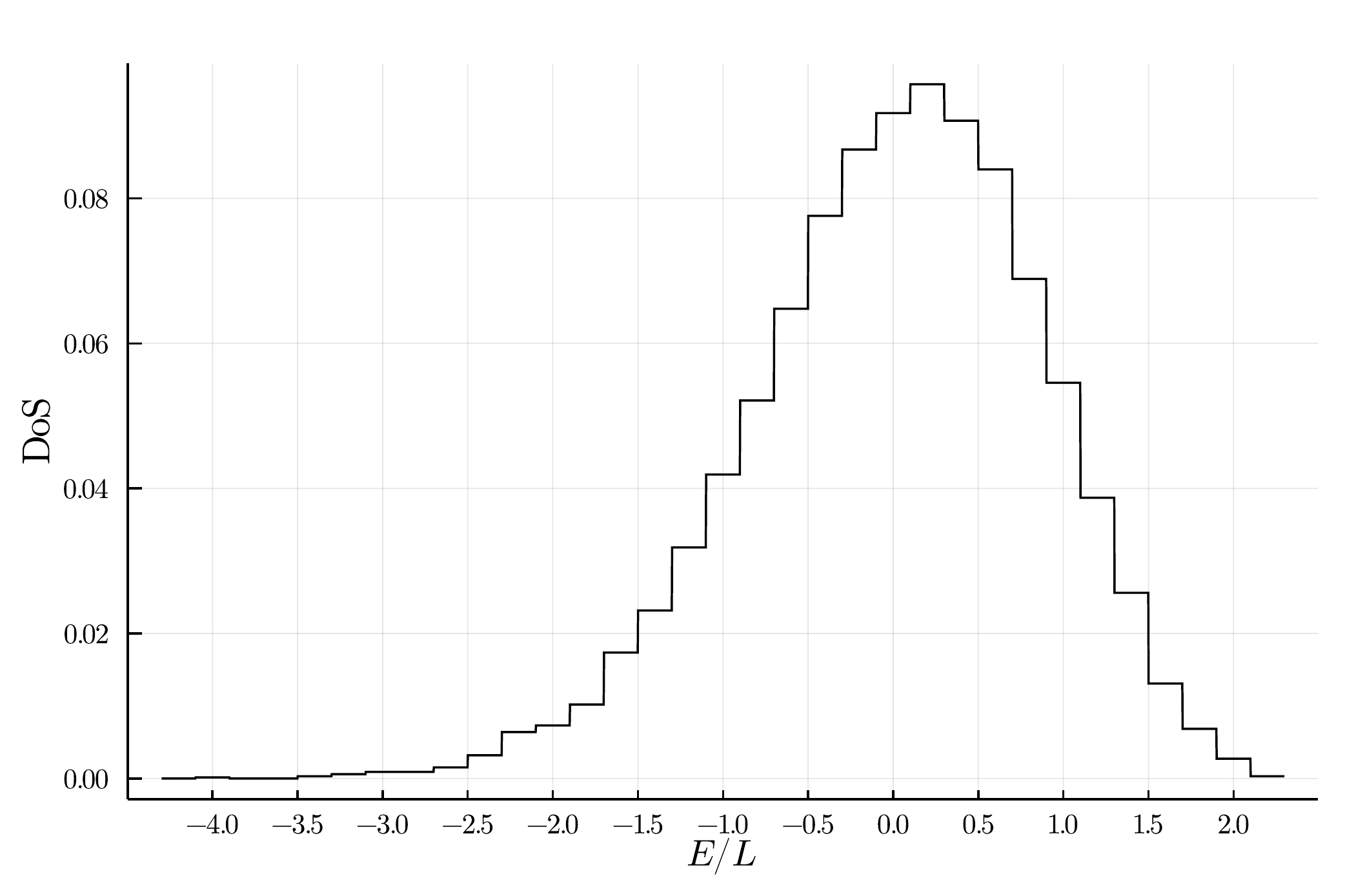}
\end{minipage}
\caption{
  \textbf{Top}: Average support of initial states over energy eigenstates in 8 qutrit system. The quantity $|\braket{\psi_0 |E_k}|$, which measures the support of an initial state $\ket{\psi_0}$ with each each energy eigenstate (labeled by their energy density $E_k$), is averaged over stabilizer product states with the proper energy for each temperature. 
  The relative support of initial states on different energy eigenstates roughly scales as the square root of the Boltzman weights $\sim \exp( -\beta E_k /2)$.  \textbf{Bottom}: Normalized density of states with 0.2 width bins for energy density.}
\label{fig:energy-density-support}
\end{figure}

\begin{figure}[t]
  \centering
  \begin{minipage}{0.48\textwidth}
  \includegraphics[width=\linewidth]{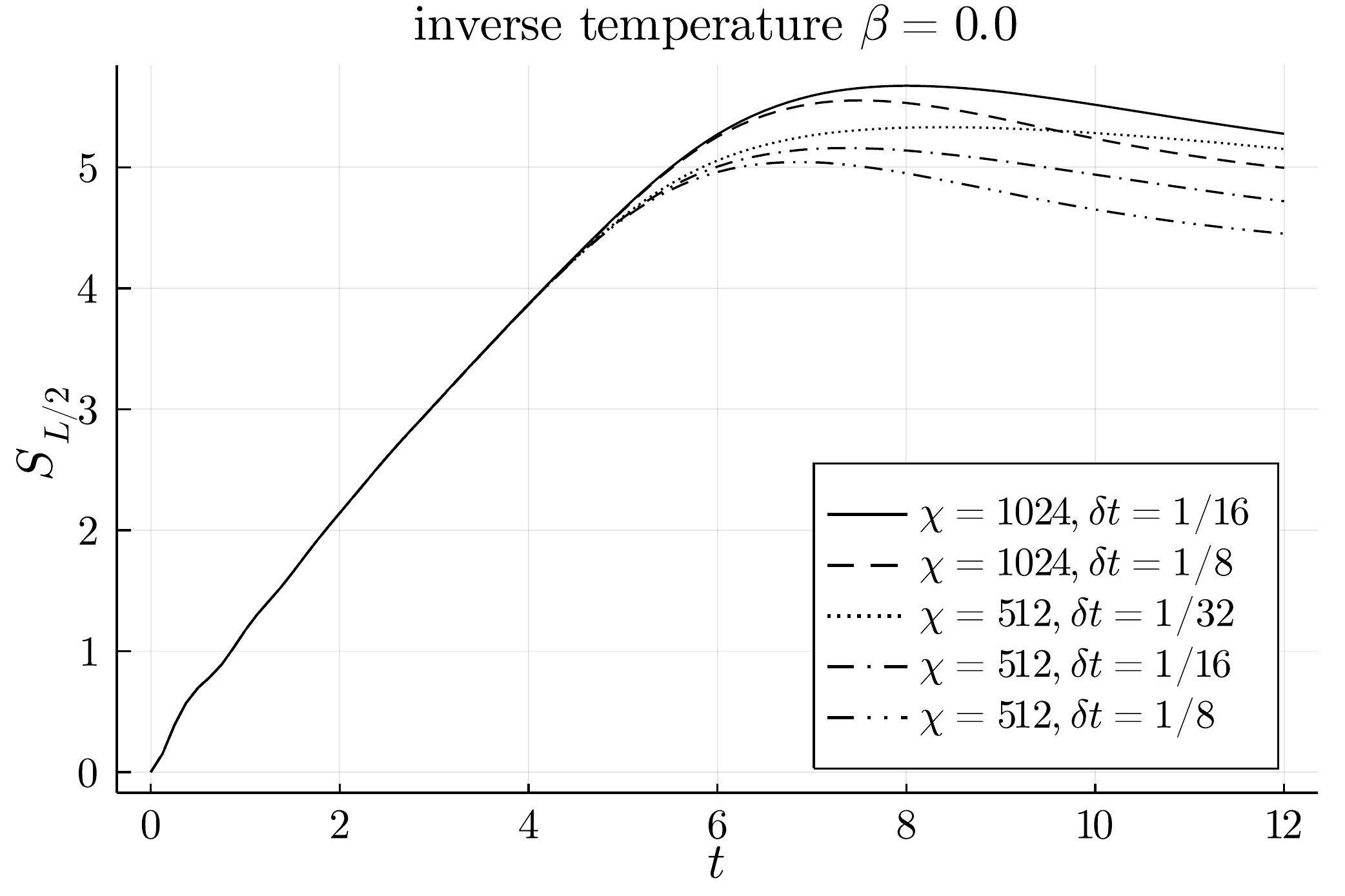}
  \end{minipage}
  \\
  \begin{minipage}{0.23\textwidth}
  \includegraphics[width=\linewidth]{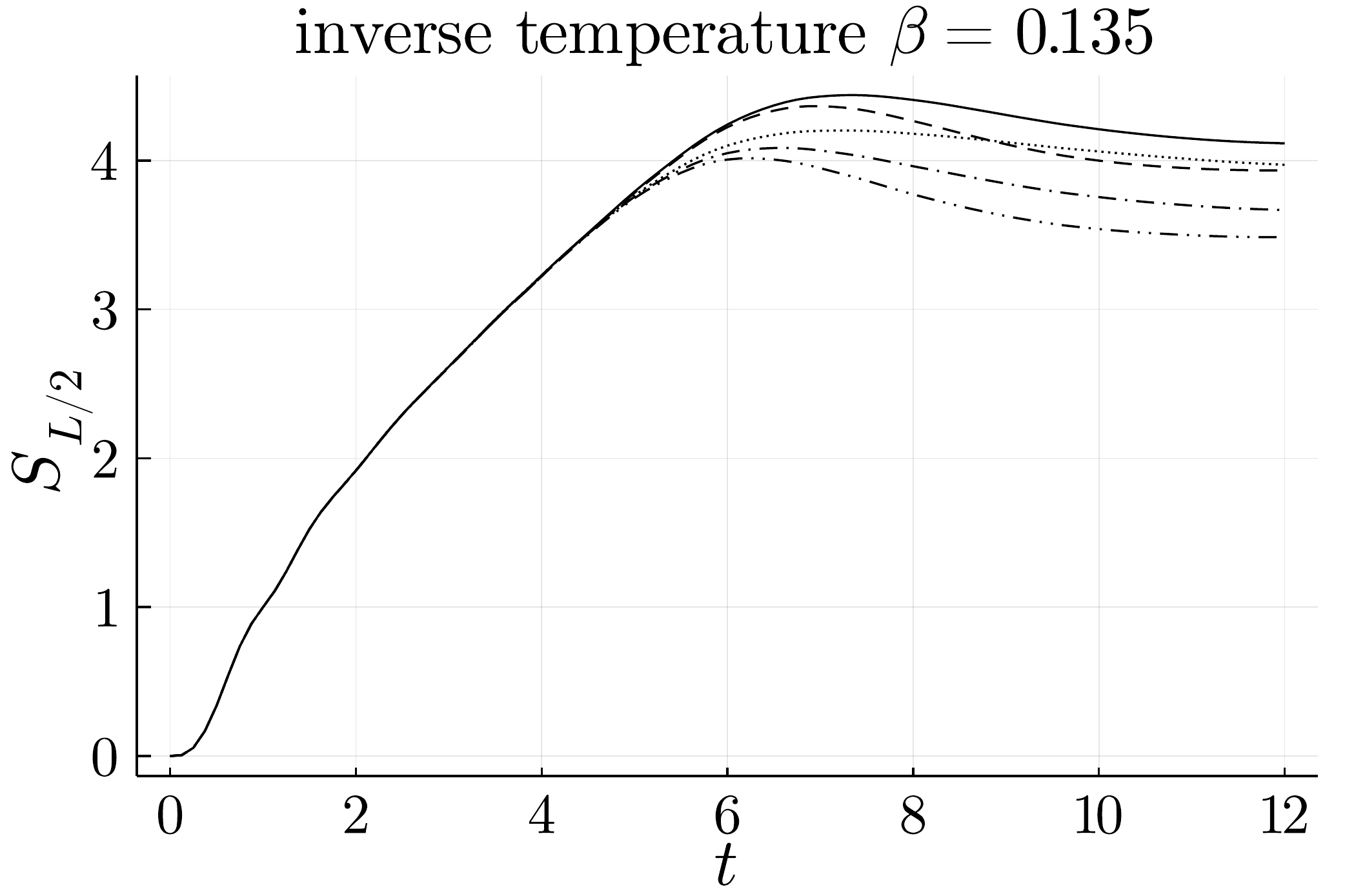}
 \end{minipage}
   \begin{minipage}{0.23\textwidth}
  \includegraphics[width=\linewidth]{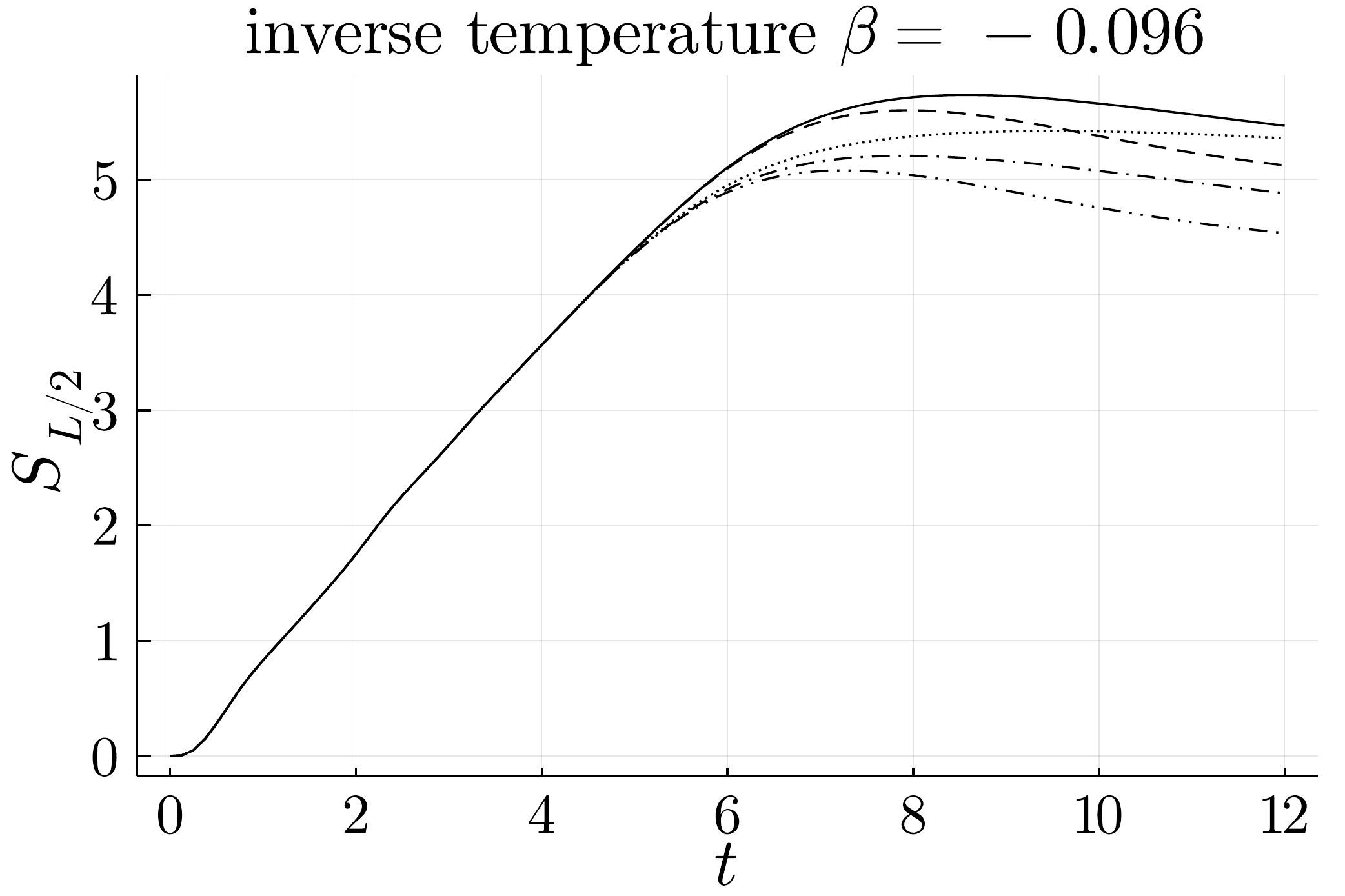}
 \end{minipage}
 \\
   \begin{minipage}{0.23\textwidth}
  \includegraphics[width=\linewidth]{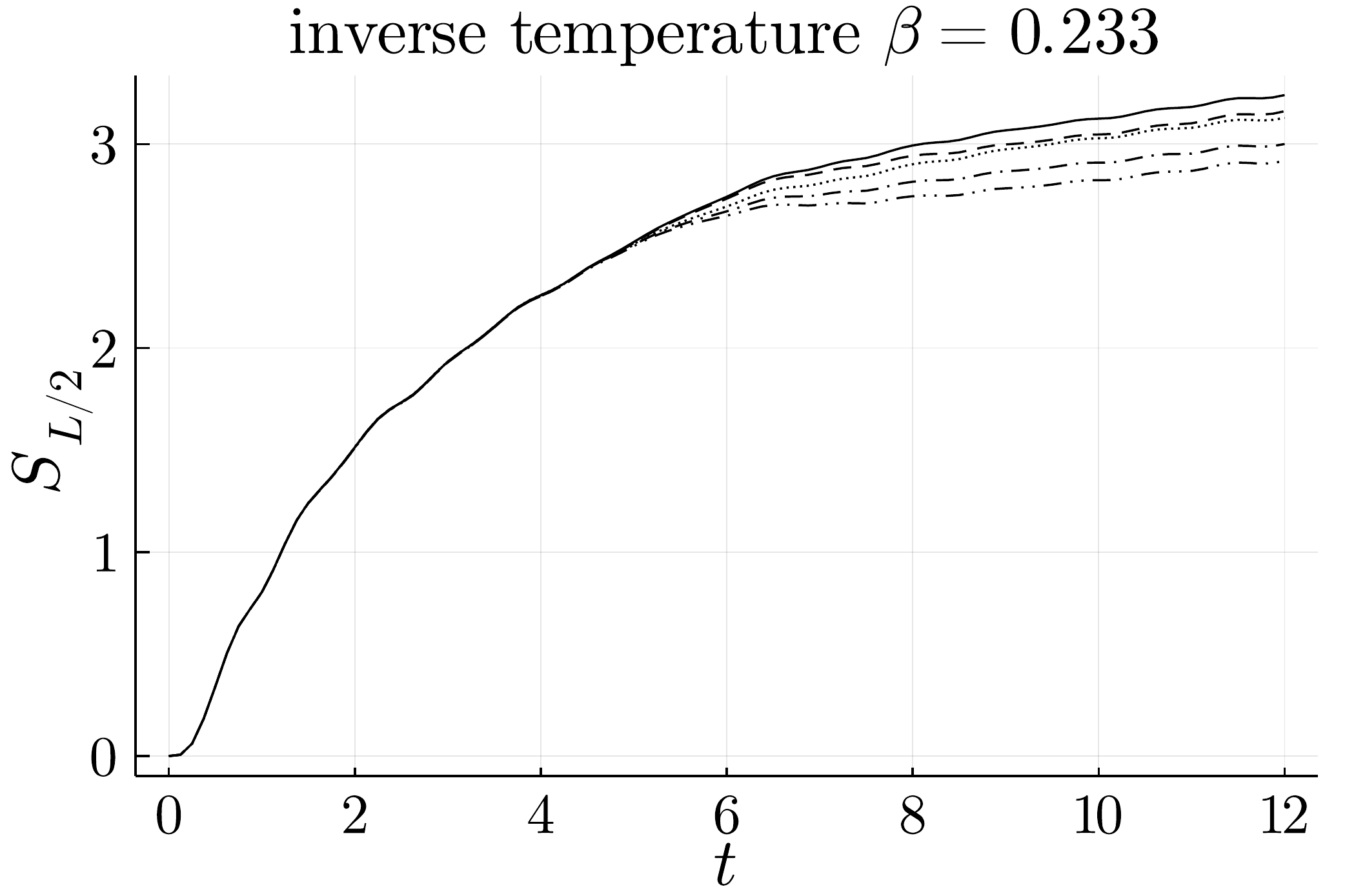}
 \end{minipage}
   \begin{minipage}{0.23\textwidth}
  \includegraphics[width=\linewidth]{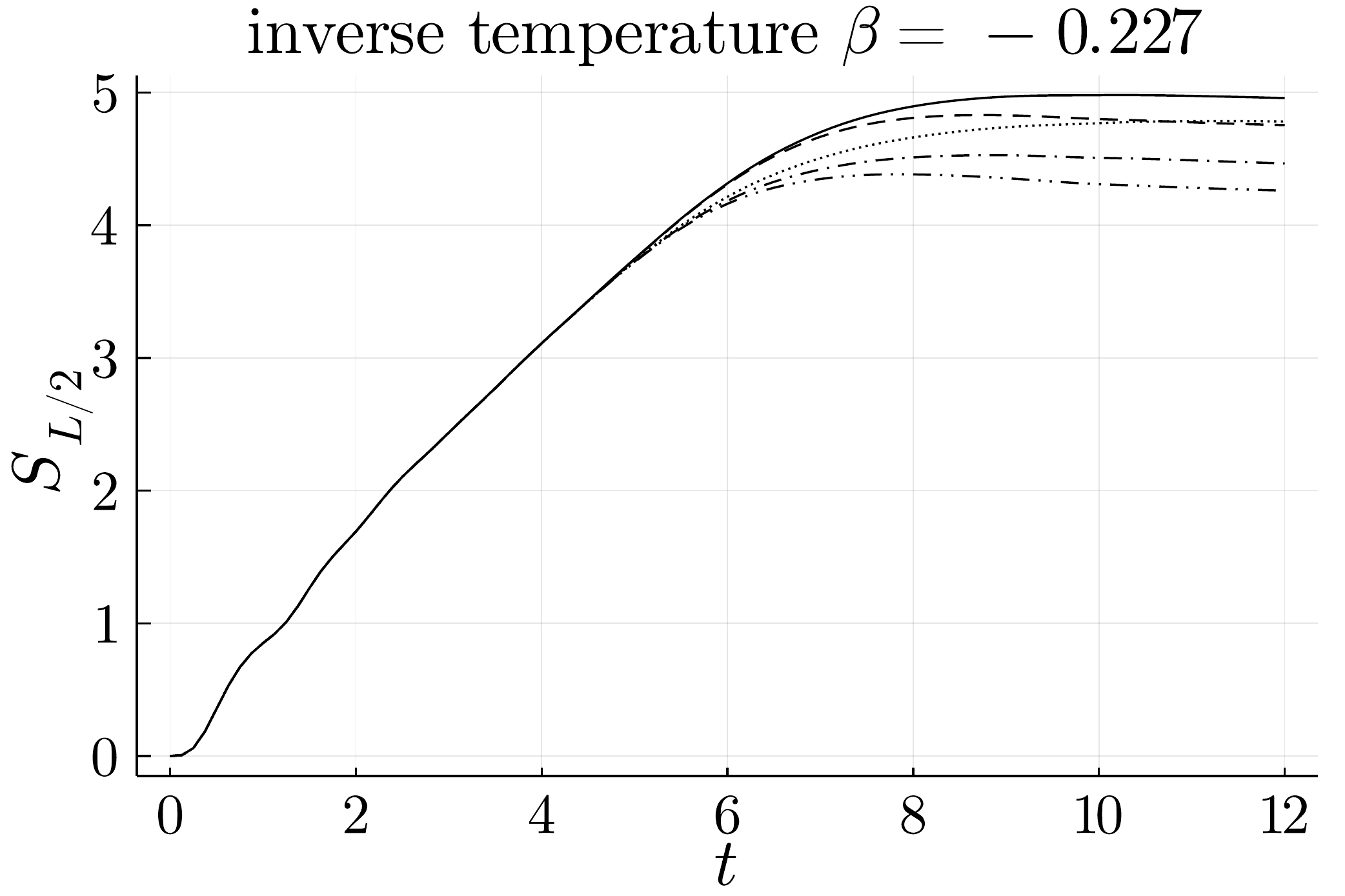}
 \end{minipage}
 \caption{50 qutrit MPS, half-system entropy convergence with bond dimension and trotter step.
 }
 \label{fig:halfent-convergence}
 \end{figure}

\begin{figure}
  \centering
  \begin{minipage}{0.48\textwidth}
  \includegraphics[width=\linewidth]{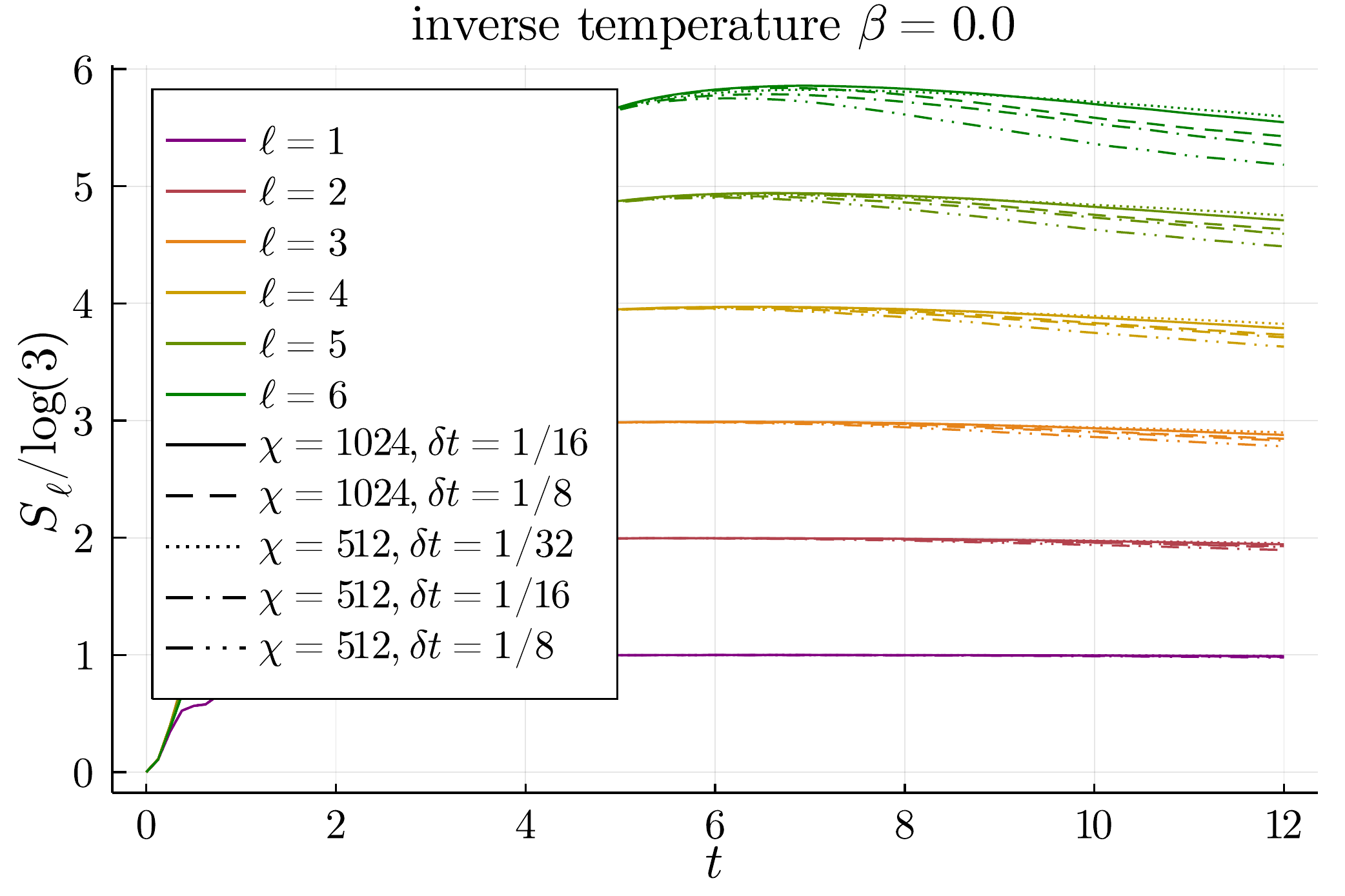}
  \end{minipage}
  \\
  \begin{minipage}{0.23\textwidth}
  \includegraphics[width=\linewidth]{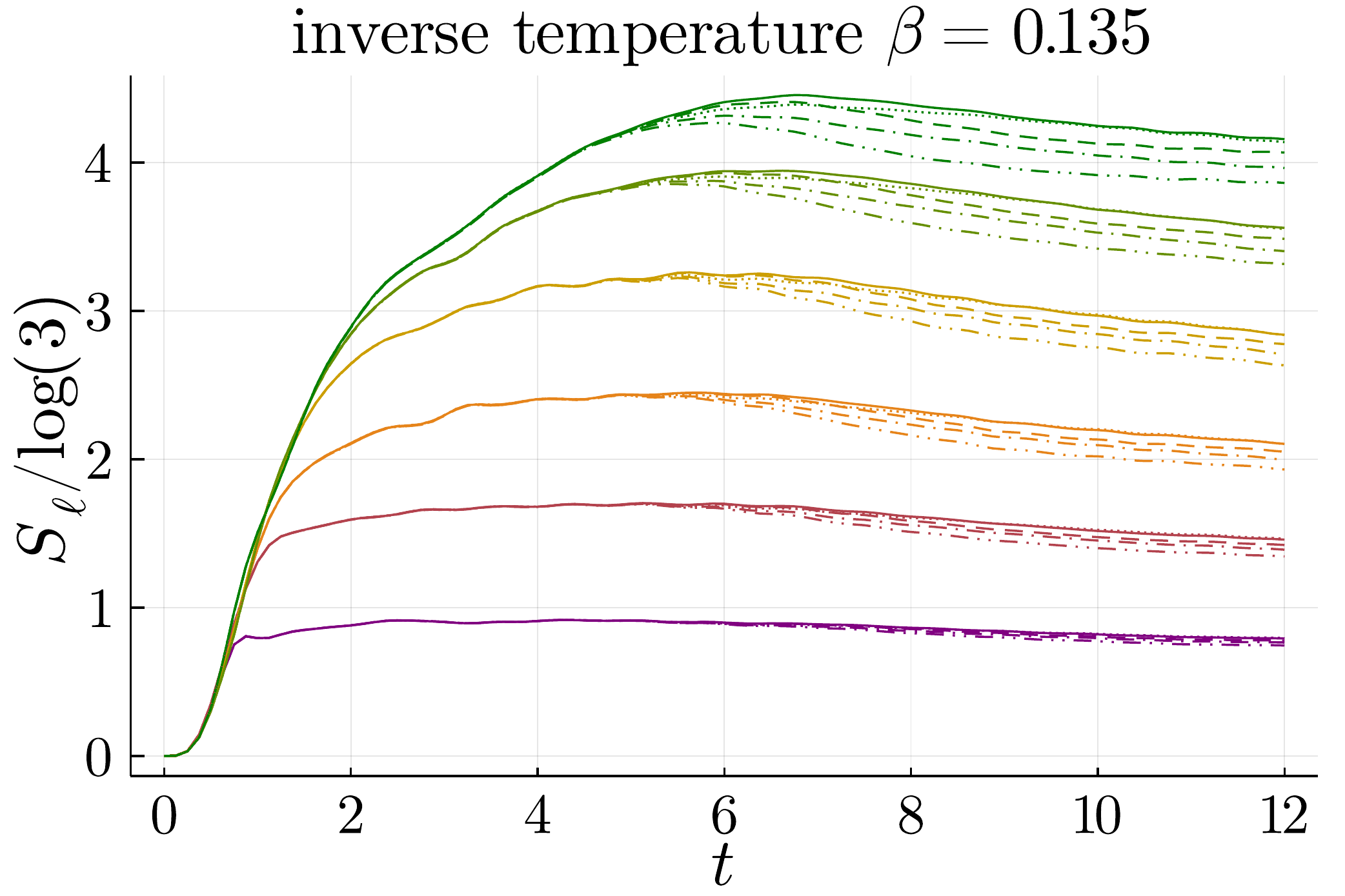}
 \end{minipage}
   \begin{minipage}{0.23\textwidth}
  \includegraphics[width=\linewidth]{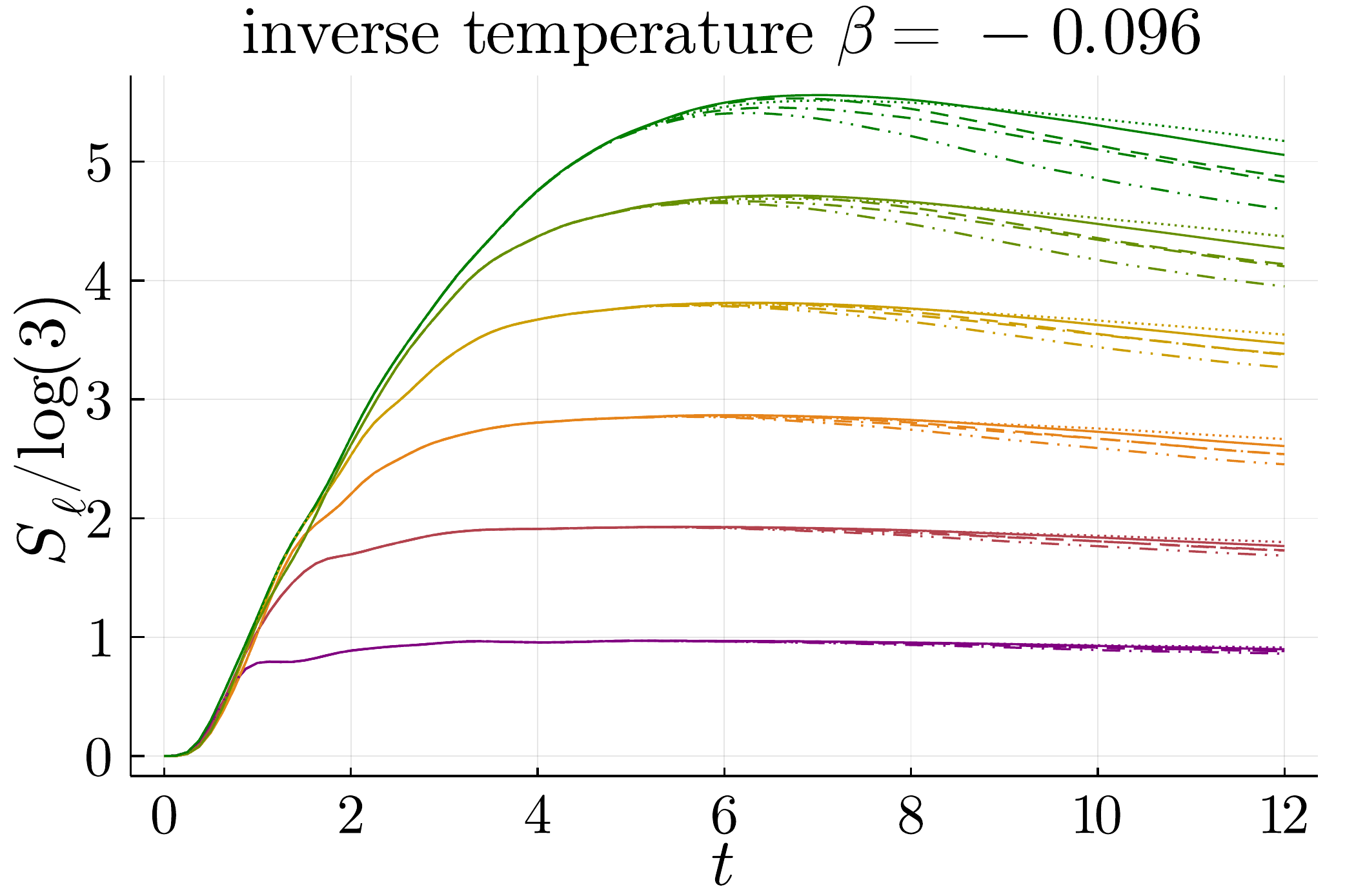}
 \end{minipage}
 \\
   \begin{minipage}{0.23\textwidth}
  \includegraphics[width=\linewidth]{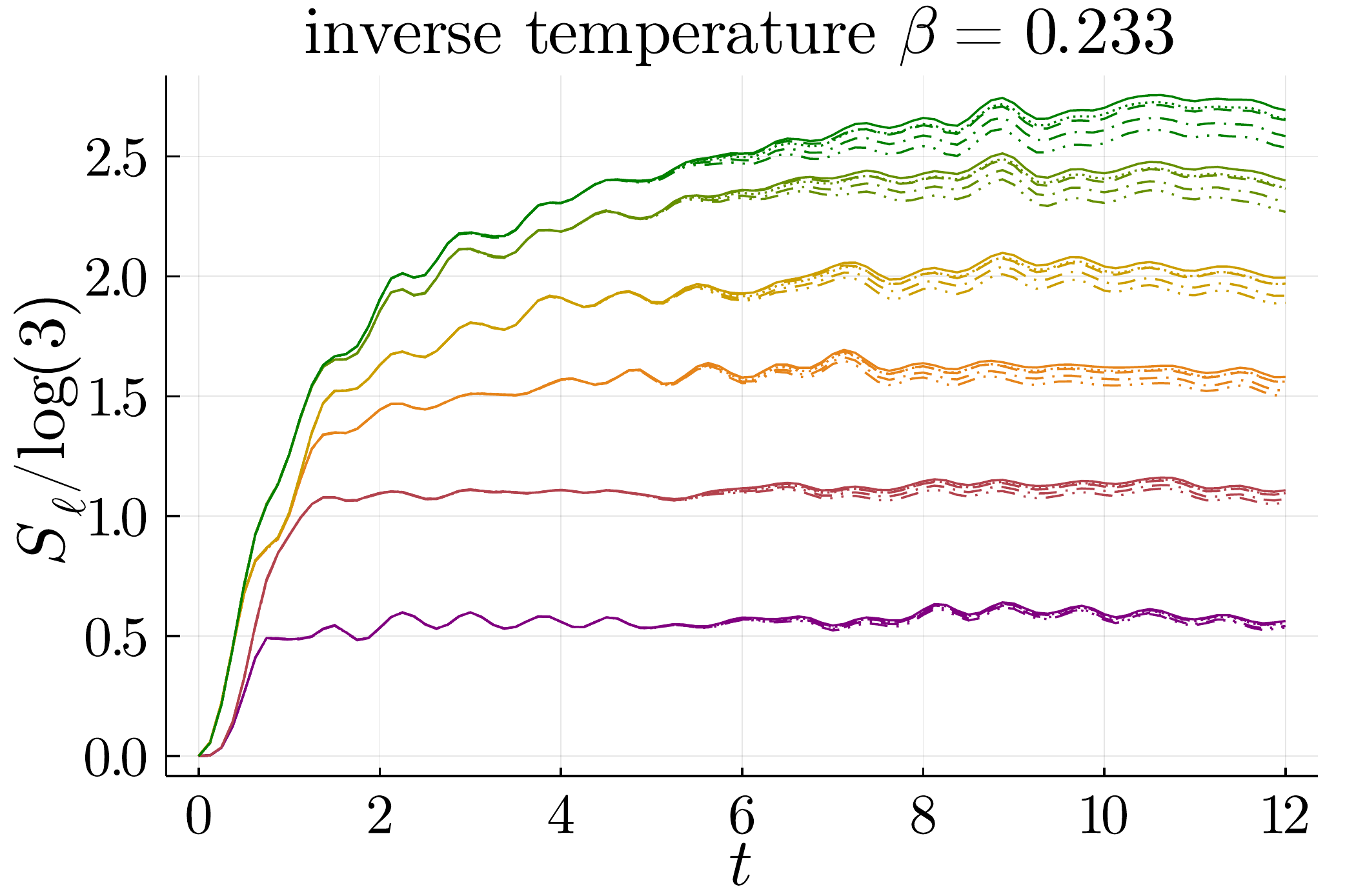}
 \end{minipage}
   \begin{minipage}{0.23\textwidth}
  \includegraphics[width=\linewidth]{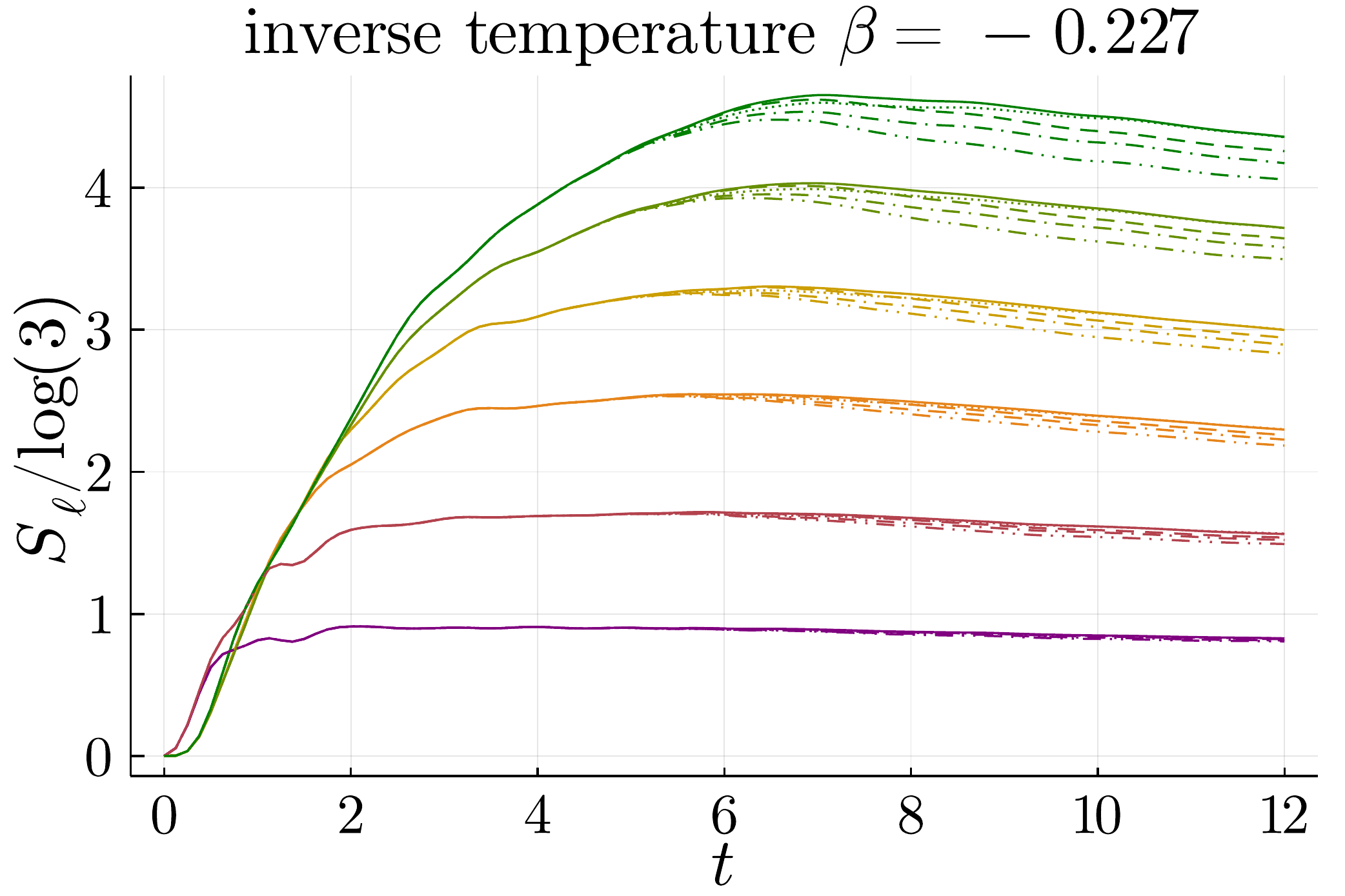}
 \end{minipage}
 \caption{50 qutrit MPS, subsystem entropy convergence with bond dimension and Trotter step for a single initial state at each temperature.
 }
 \label{fig:subent-convergence}
 \end{figure}

The infinite-temperature states---those with $\varepsilon_{j,j+1} = 0$---deserve special consideration,
both for their simplicity and their importance.
Consider the six eigenstates of $ZX$ and $ZX^2$.
call them $\ket{\phi_\alpha}$, and write
\begin{equation}
  \ket{\phi_{\alpha}\phi_\beta} = \ket{\phi_{\alpha}} \otimes \ket{\phi_{\beta}}\;.
\end{equation}
Then term-by-term
\begin{equation}
  \braket{\phi_\alpha\phi_\beta | \epsilon_{j,j+1} | \phi_{\alpha}\phi_\beta } = 0\;,
\end{equation}
for every choice of $\phi_{\alpha}$, $\phi_\beta$, and any state of the form
\begin{equation}
  \label{eq:inf-temp}
  \ket{\phi_{\bm{\alpha}}} = \ket{\phi_{\alpha_1}} \otimes \ket{\phi_{\alpha_2}} \otimes \dots \otimes \ket{\phi_{\alpha L}}
\end{equation}
has zero energy density everywhere.
Our infinite-temperature states, then, are $\ket{\phi_{\bm\alpha}}$ for random strings $\alpha$.

The initial states are chosen to be stabilizer states, so that they have zero initial mana, and product states, so that there is no initial entanglement and all subsystems are pure. The energy of these initial states are determined by all the nearest neighbor pairs of stabilizer states. The only stabilizer states with nonzero expectation value of Hamiltonian terms are the eigenstates of the generalized $Z$ and $X$ operators. 
The transverse and longitudinal field terms each have one low energy eigenvstate, denoted $X_0$ and $Z_0$ respectively. These terms also have two degenerate positive energy eigenstates denoted $X_1$ and $X_2$ for the transverse field and $Z_1$ and $Z_2$ for the longitudinal field. 
The states $Z_i$ also have nonzero expectation value for the bond terms, with negative energy when the nearest neighbor pairs are the same and positive when they are different. 
The other six stabilizer states are denoted $S_i$ and have zero expectation with all Hamiltonian terms.

We consider in Fig.~\ref{fig:energy-density-support} the magnitude of the inner product of initial states with energy eigenstates in an 8-qutrit system, averaged over initial states for a given temperature. 
We find that initial states have support over the full range of energy eigenstates, with the average support scaling roughly as the square root of the appropriate Boltzmann weights at that temperature, i.e. $|\braket{\psi_0 | E_k }| \sim \exp( -\beta E_k /2)$.

 The exception is for the case where $\beta = 0.233$ ($E_0 /L = -2$), where there are many energy eigenstates orthogonal to all of the stabilizer product states at that temperature.
 This makes sense because this temperature has the fewest number of initial states and significant correlations in those initial states imposed by the energy density selection.
 Unlike at other temperatures these constraints for $\beta = 0.233$ has restricted the initial states to have support over only a subspace of the energy eigenstates.
 The the average support within this subspace, however, still scales according to the appropriate Boltzmann weights.

\section{Defining and calculating mana} \label{app:mana}

\begin{figure}
  \centering
  \begin{minipage}{0.48\textwidth}
    \includegraphics[width=\linewidth]{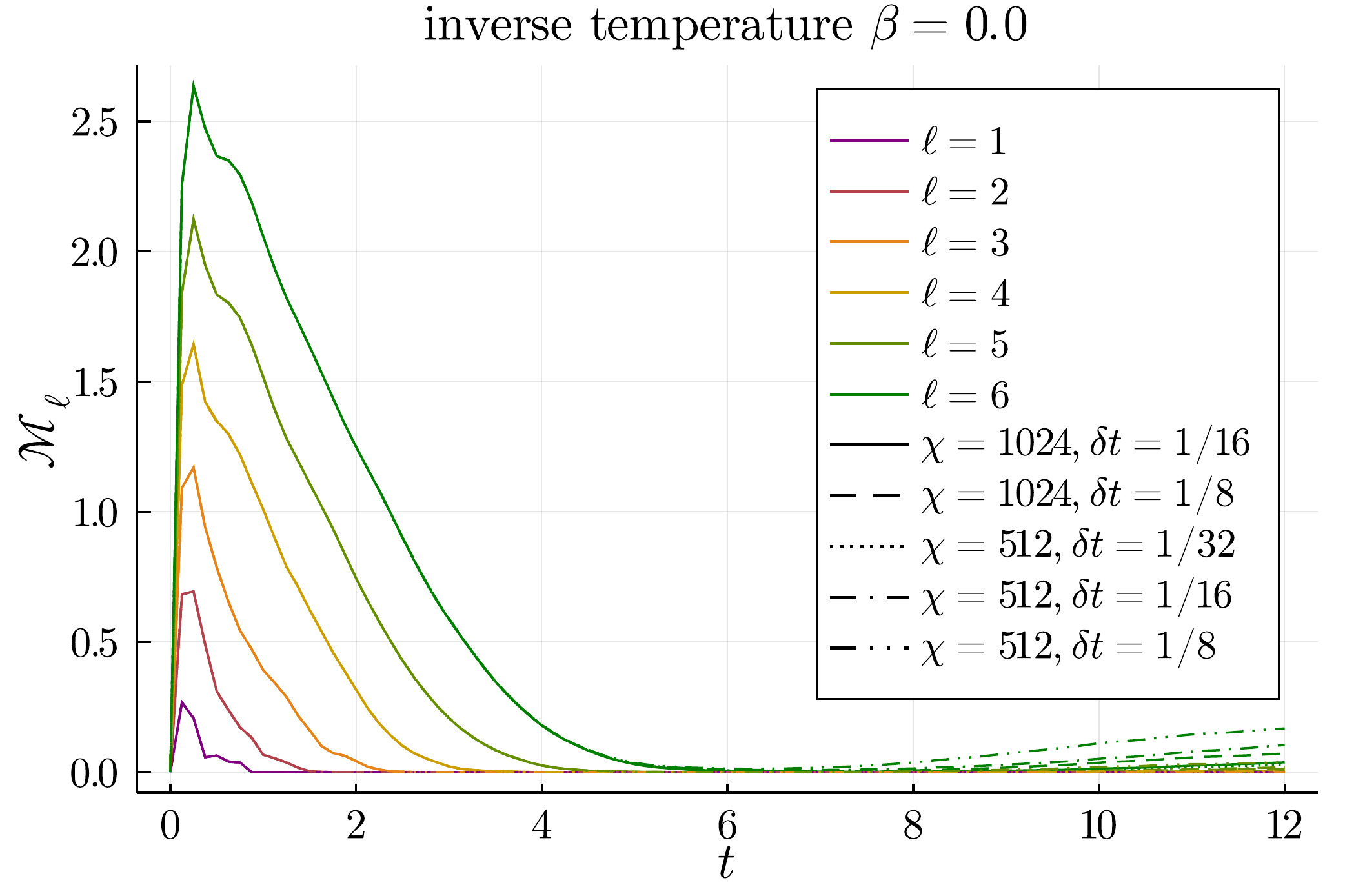}
  \end{minipage}
  \\
  \begin{minipage}{0.23\textwidth}
    \includegraphics[width=\linewidth]{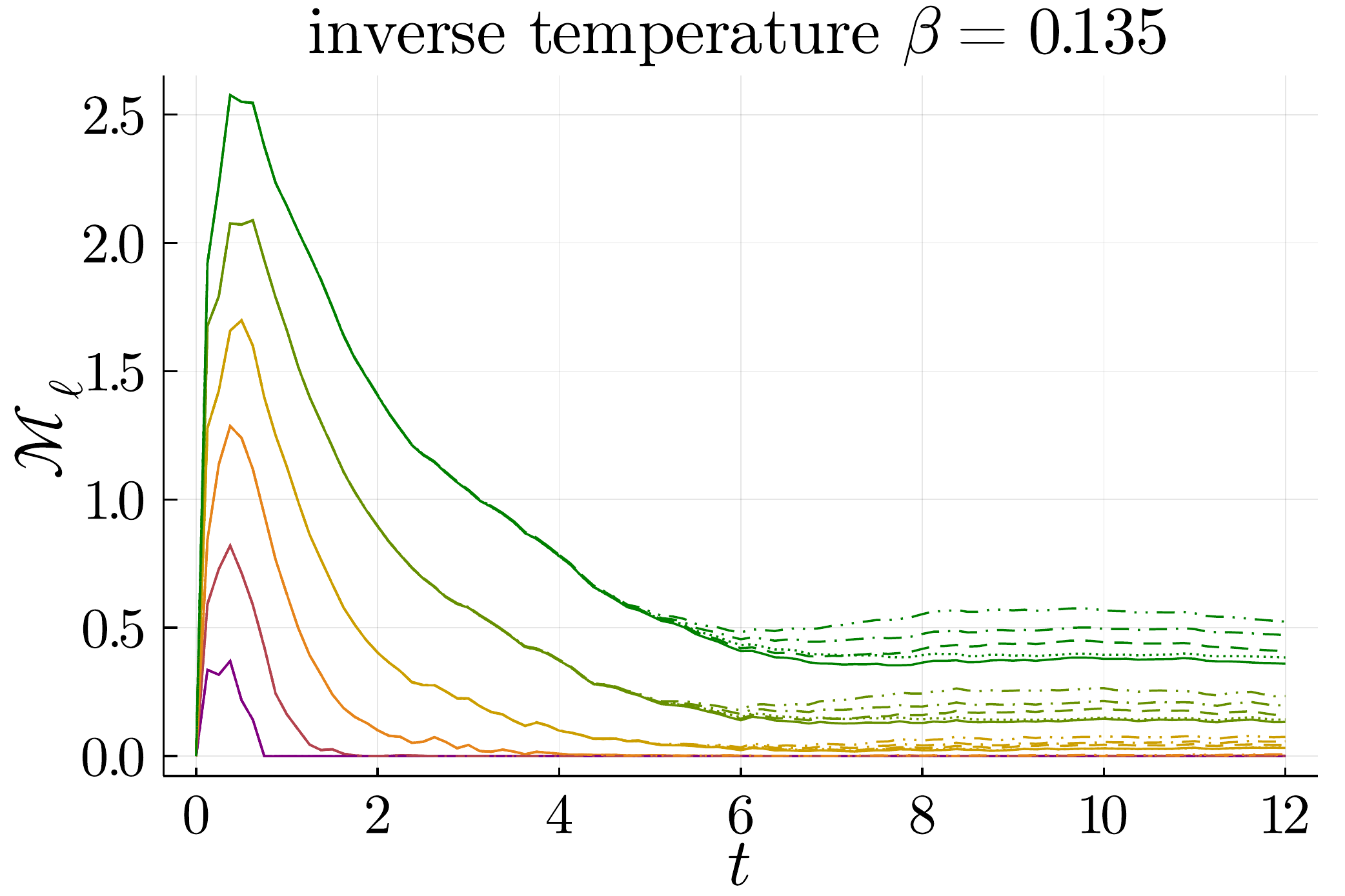}
  \end{minipage}
  \begin{minipage}{0.23\textwidth}
    \includegraphics[width=\linewidth]{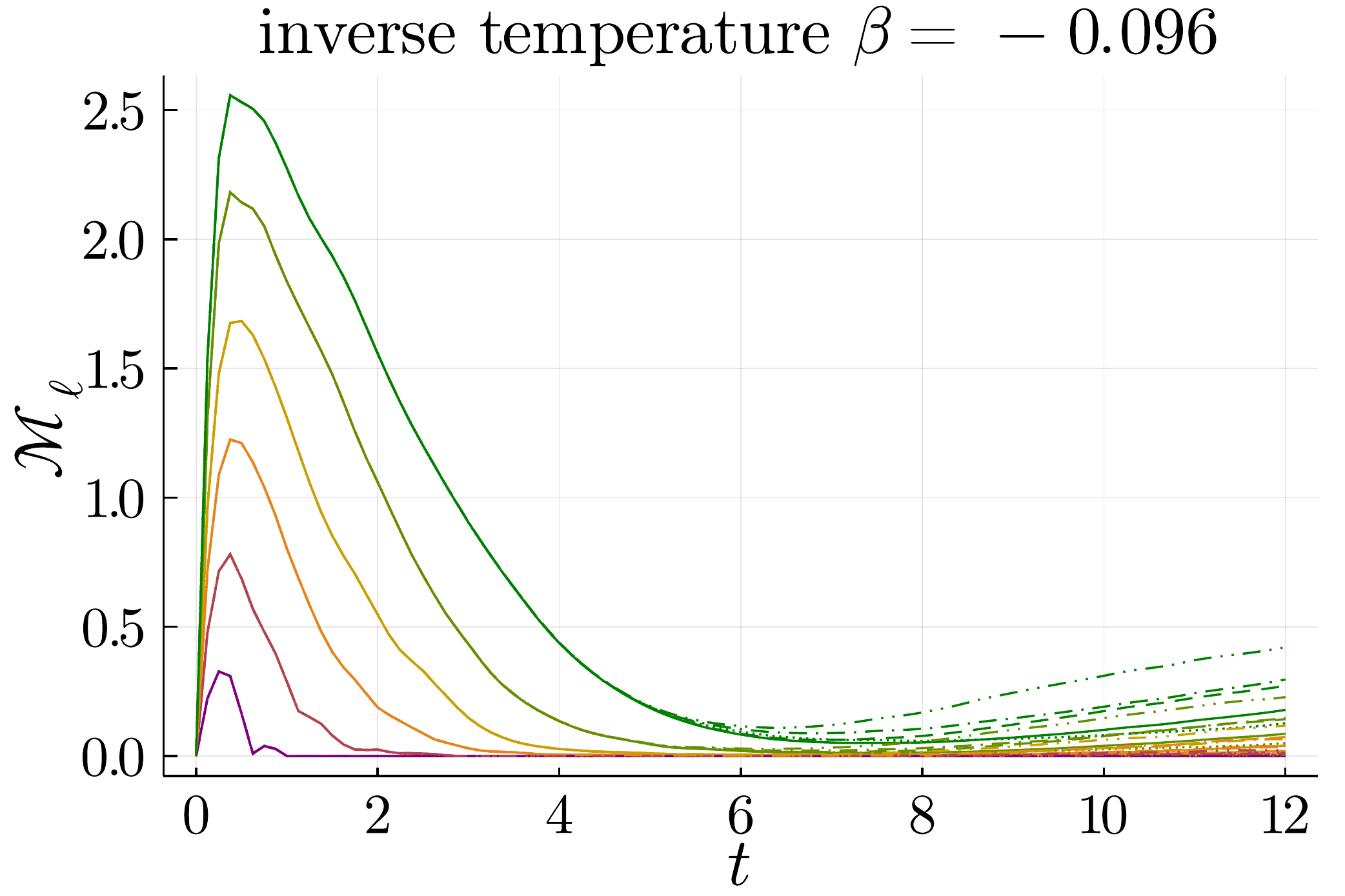}
  \end{minipage}
  \\
  \begin{minipage}{0.23\textwidth}
    \includegraphics[width=\linewidth]{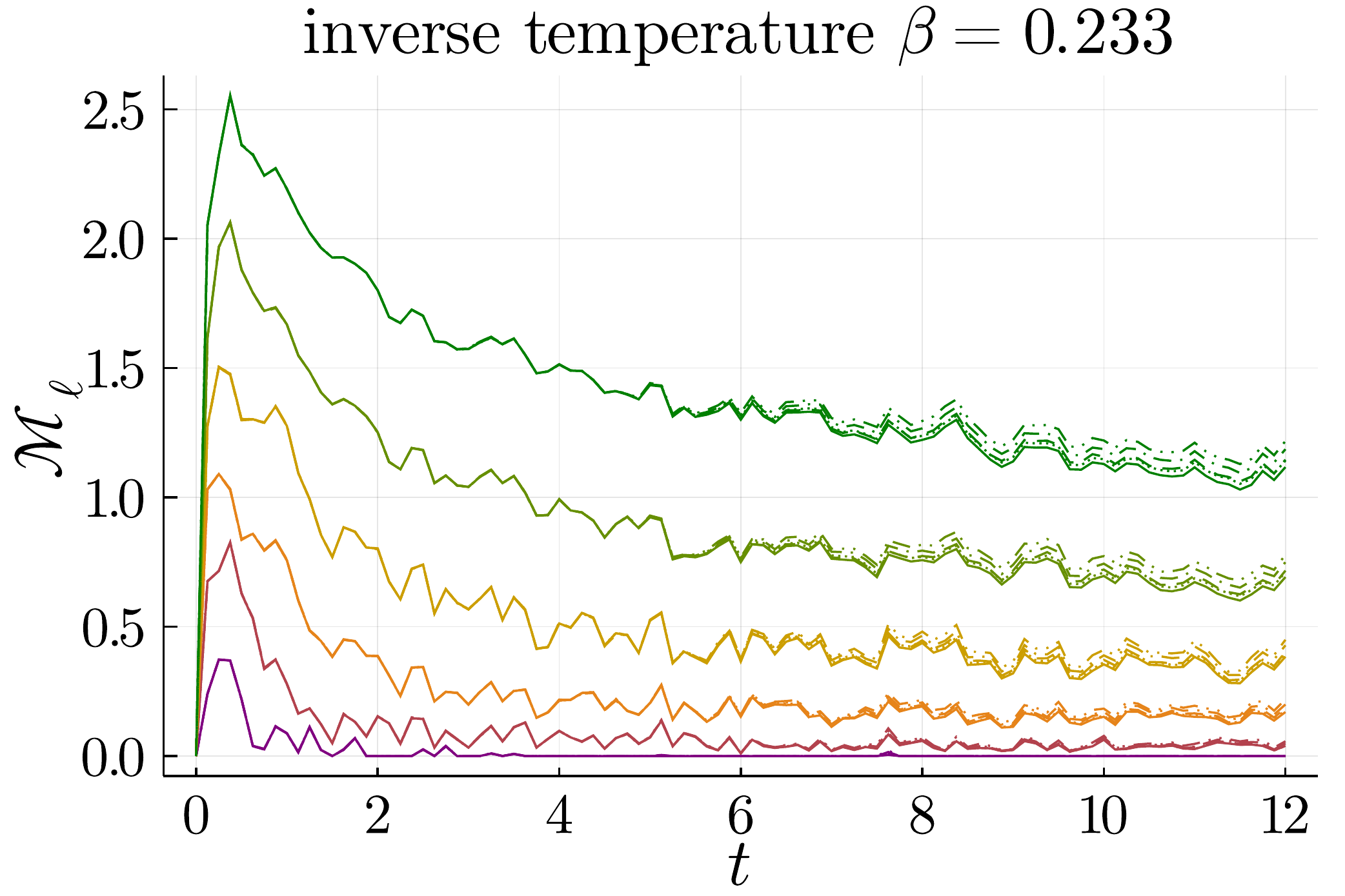}
  \end{minipage}
  \begin{minipage}{0.23\textwidth}
    \includegraphics[width=\linewidth]{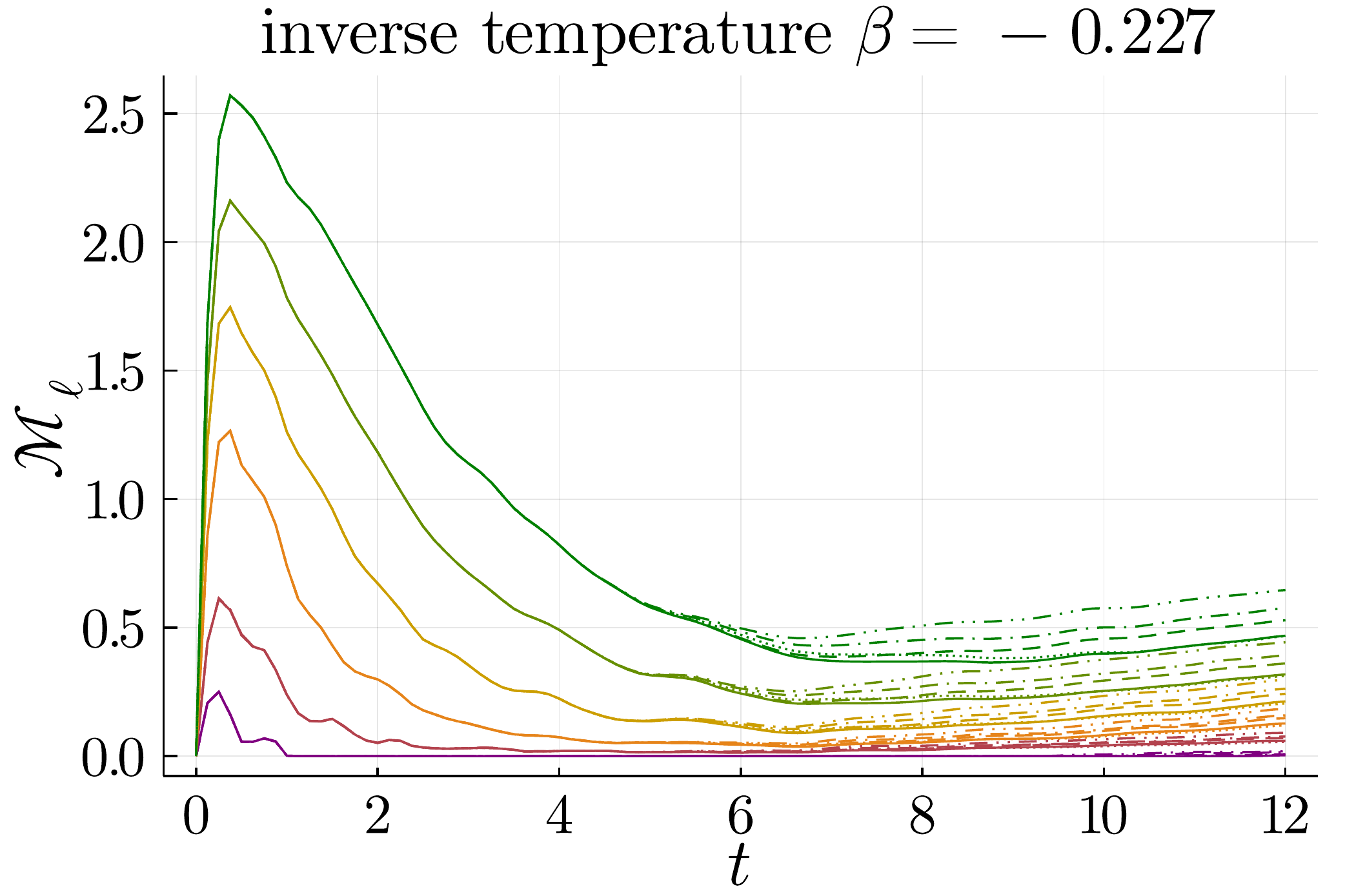}
  \end{minipage}
  \caption{50 qutrit MPS, subsystem mana convergence with bond dimension and Trotter step
    for the same single state at each temperature as in Fig.~\ref{fig:subent-convergence} 
  }
  \label{fig:submana-convergence}
\end{figure}

Consider a system of $\ell$ qudits each of dimension $d$.
(NB this notation is different from that of Sec.~\ref{s:finite} and Appendices \ref{s:mana-bnd}, \ref{s:mana-typicality}.
There we write $d$ for the dimension of a whole Hilbert space;
here we write $d$ for the dimension of the component qubits.)

To calculate the mana, first recall the generalized Pauli strings
\begin{align}
  \label{eq:gen-pauli}
  T_{\bm p \bm q} = \omega^{2^{-1} \bm p \cdot \bm q} \prod_j  Z_j^{p_j}X_j^{q_j}
\end{align}
where $\omega = e^{2\pi i / d}$, $2^{-1} = (d+1)/2$ is the multiplicative inverse of $2$ in the field $\mathds Z_d$, and $Z_j$ and $X_j$ are the clock and shift operators on site $j$.
Just as the Pauli strings generalize continuous single-particle Weyl operators to a product of discrete rings,
we can generalize the Wigner function to a discrete Wigner function
\begin{align}
  W_{\bm p\bm q}(\rho) = d^{-\ell} \sum_{\bm p'\bm q'} \omega^{\bm p\cdot \bm q' - \bm p'\cdot \bm q} \tr \rho T_{\bm p' \bm q'}
\end{align}
{
This may also be written in terms of the \textit{phase-space point operators} $A_{\bm p \bm q}$ as
\begin{align}
  W_{\bm p\bm q}(\rho) = d^{-\ell}  \tr [\rho A_{\bm p \bm q}],
\end{align}
where
\begin{align}
  \label{eq:psp-def}
A_{\bm p \bm q} = T_{\bm p \bm q} A_0 T^\dagger_{\bm p \bm q}  \hspace{5mm}  \mathrm{with} \hspace{5mm} A_0 = d^{-\ell} \sum_{\bm p \bm q} T_{\bm p \bm q}.
\end{align}
}
The discrete Wigner function has $\sum_{\bm p \bm q} W_{\bm p \bm q}(\rho) = \tr \rho = 1$, and%
---for classical mixtures of stabilizer states $\sigma_1, \sigma_2$---
\begin{align}
 W_{\bm p\bm q}(\alpha\sigma_1 + (1-\alpha) \sigma_2) \ge 0.
\end{align}
(Among pure states \textit{only} stabilizer states have $W_{\bm p, \bm q}(\rho) \ge 0$ for all $\bm p, \bm q$.
This property is called the \textit{discrete Hudson's theorem}.\cite{gross_hudsons_2006})
The Wigner norm
\begin{align}
  \| \rho \|_W = \sum_{\bm p \bm q} |W_{\bm p \bm q}(\rho)| 
\end{align}
measures the size of the negative part of the Wigner function, and the mana is
\begin{align}
  \mathcal M(\rho) = \ln \|\rho \|_W\;.
\end{align}

\section{Convergence in Bond Dimension and Trotter Step}\label{app:conv}

The accuracy of simulations of time evolution using matrix product states is limited by both the bond dimension of the MPS and the Trotter step size of the TEBD evolution.
Trotterization of the time-evolution operator to second order with time step $\eta t$ incurs an error of order $\delta t^3$,
but a truncation also occurs at each time step; this truncation projects the state back into the space of MPS of the given bond dimension and causes additional error.

As the initial product state evolves in time entanglement grows in the system but is ultimately limited by the bond dimension of the MPS.
The largest entanglement in the system would occur across a central bipartition,
so we compare the entanglement entropy for this bipartition with several different bond dimensions and Trotter step sizes for a particular unentangled zero energy initial state of a 50 site MPS.
We see in Fig.~\ref{fig:halfent-convergence} that the half system entanglement entropy for MPS evolution of bond dimension $\chi = 512$ and $\chi =1024$ diverge at a time $t\sim 6$,
while Trotter error is not significant (at either bond dimension) until later times $t \sim 8$.
We therefore consider our simulations reliable

While the half system entropy suggests global properties of the MPS begin to diverge around time $t=6$,
entropies and mana for subsystem of size $l \leq 6$ are much better converged as seen in Figures
\ref{fig:subent-convergence}, \ref{fig:submana-convergence}.
 No noticeable difference in these quantities is seen for the entire simulated duration up to $t=12$ for the smallest subsystems, 
 with slight differences noted at later times for the largest subsystems. 
 The spread in entropy of the six qutrit subsystem between these different samples at time $t=12$ is 2.7\%.

 We also see from these examples looking at a single state that while mana of larger systems is always greater or equal to mana of subsystems, the entropy of subsystems can grow at different rates for early times. In our examples at finite energy density the one and six qutrit subsystems have a slower entropy growth than the intermediate subsystems. This is due to the initial state on the edges of our subsystem and how this affects the mixing for initial dynamics. In the case of our zero energy states all sites have single qutrit stabilizers which are neither $X$ nor $Z$ eigenstates and mix at identical rates. For the finite energy density cases our initial states are different eigenstates of the $X$ or $Z$ operators, and so the initial mixing rate depends on the exact pairing of stabilizer states on each edge of the subsystem. For later times once the systems have been sufficiently mixed the larger subsystems have larger entropies as expected.

\section{Subsystem entropy}\label{s:subent}

  \begin{figure}
   \centering
    \begin{minipage}{0.48\textwidth}
    \includegraphics[width=\linewidth]{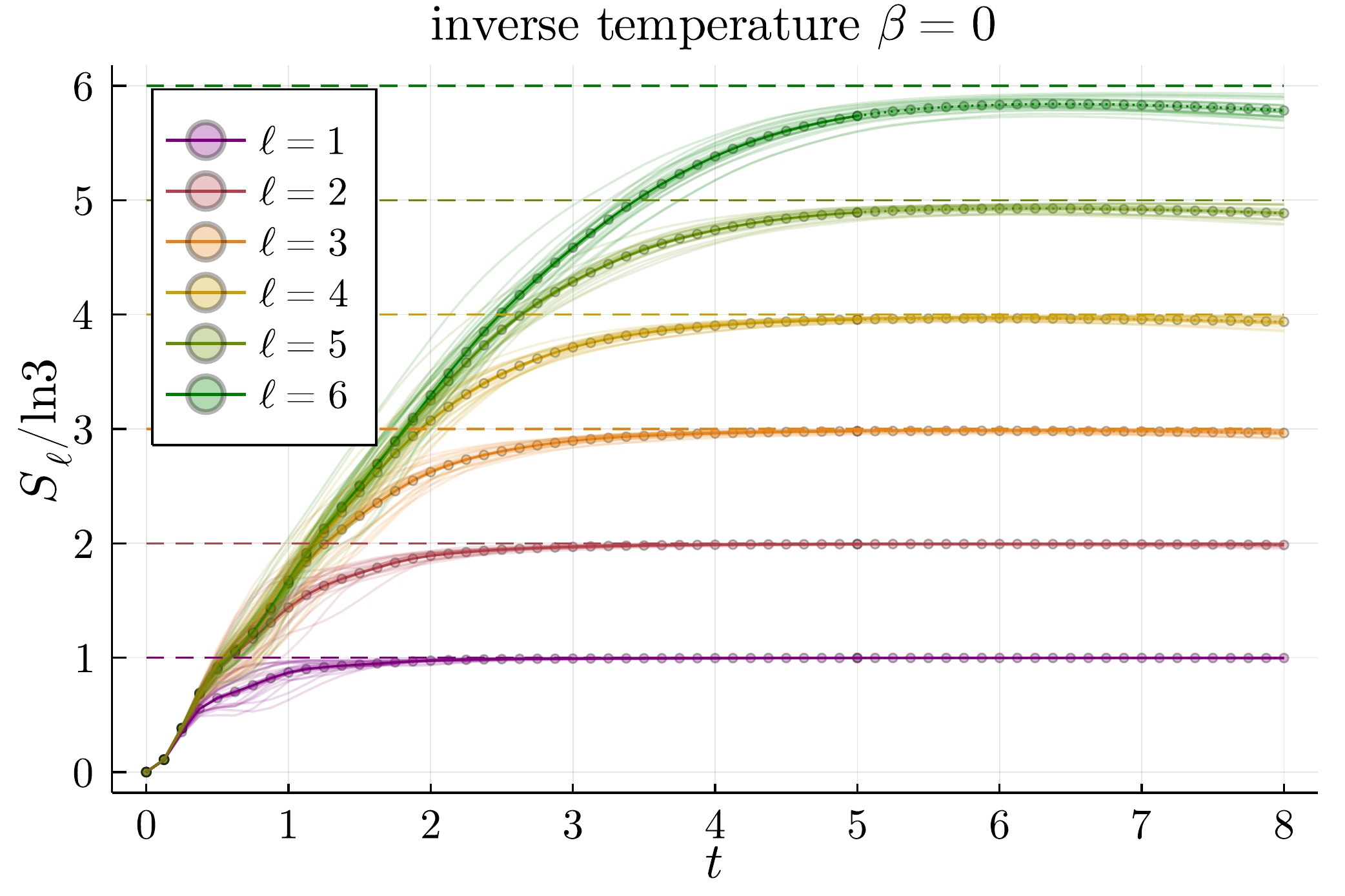}
  \end{minipage}
  \\
   \begin{minipage}{0.23\textwidth}
   \includegraphics[width=\linewidth]{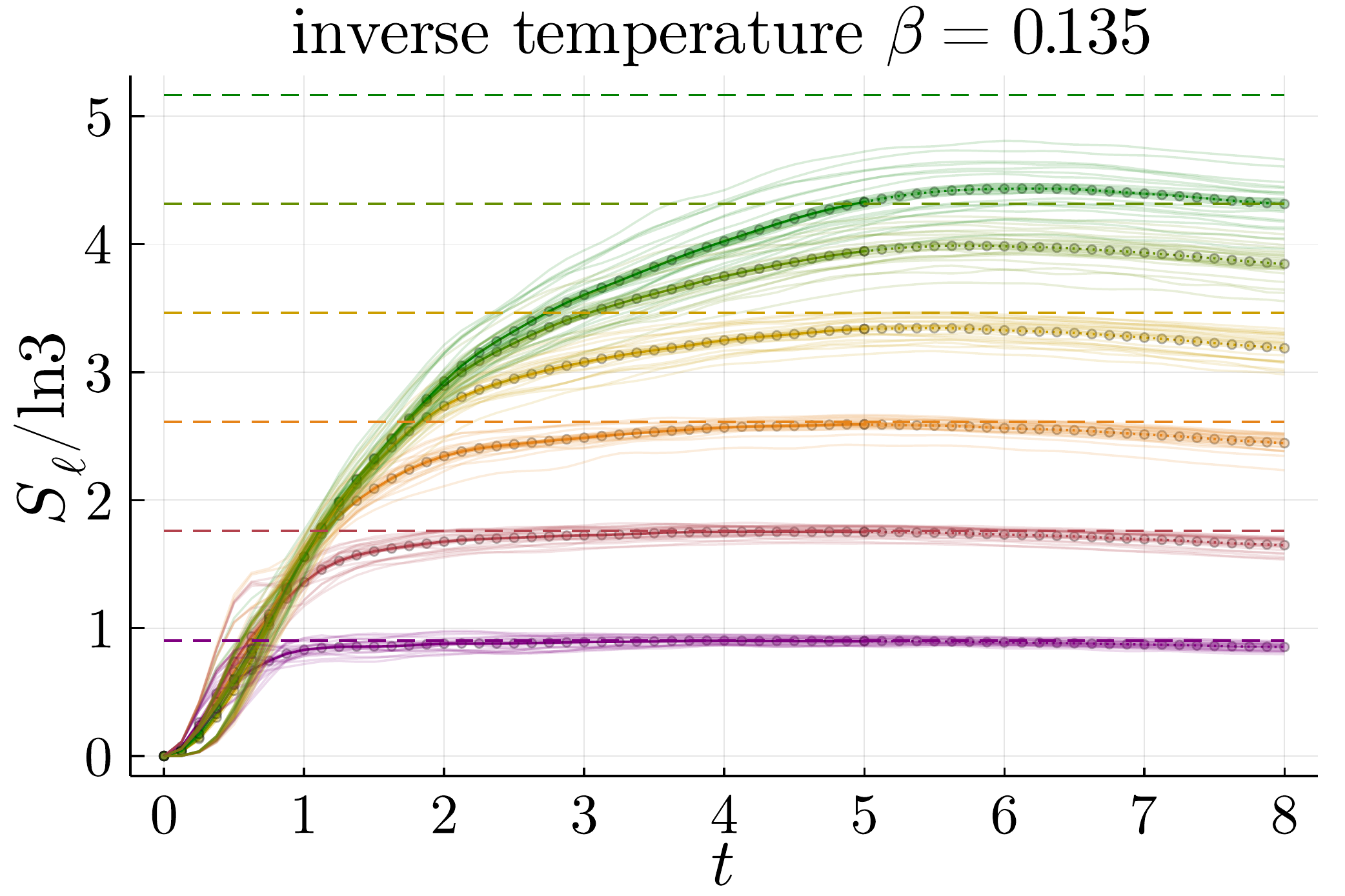}
  \end{minipage}
    \begin{minipage}{0.23\textwidth}
   \includegraphics[width=\linewidth]{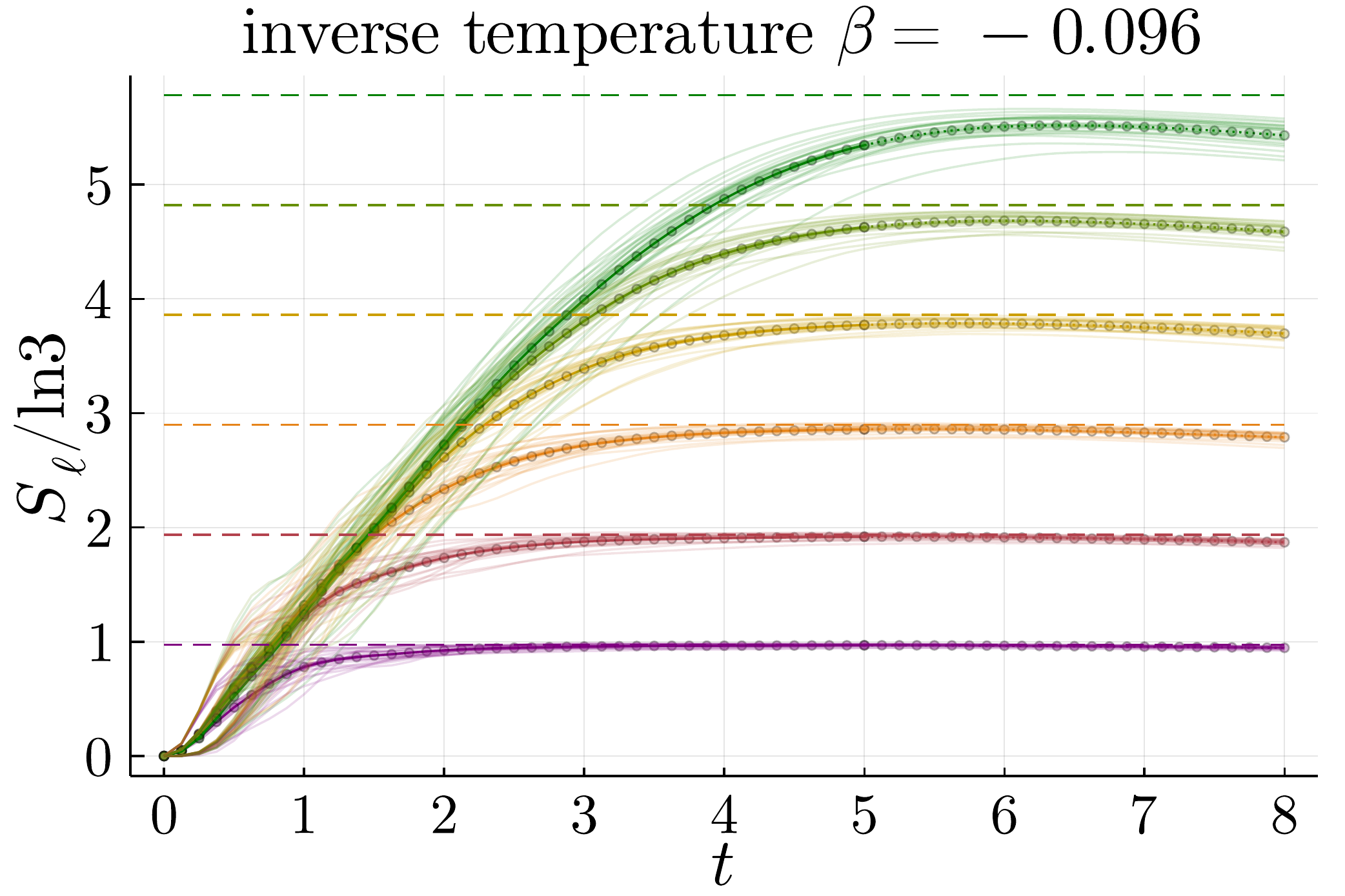}
  \end{minipage}
  \\
    \begin{minipage}{0.23\textwidth}
   \includegraphics[width=\linewidth]{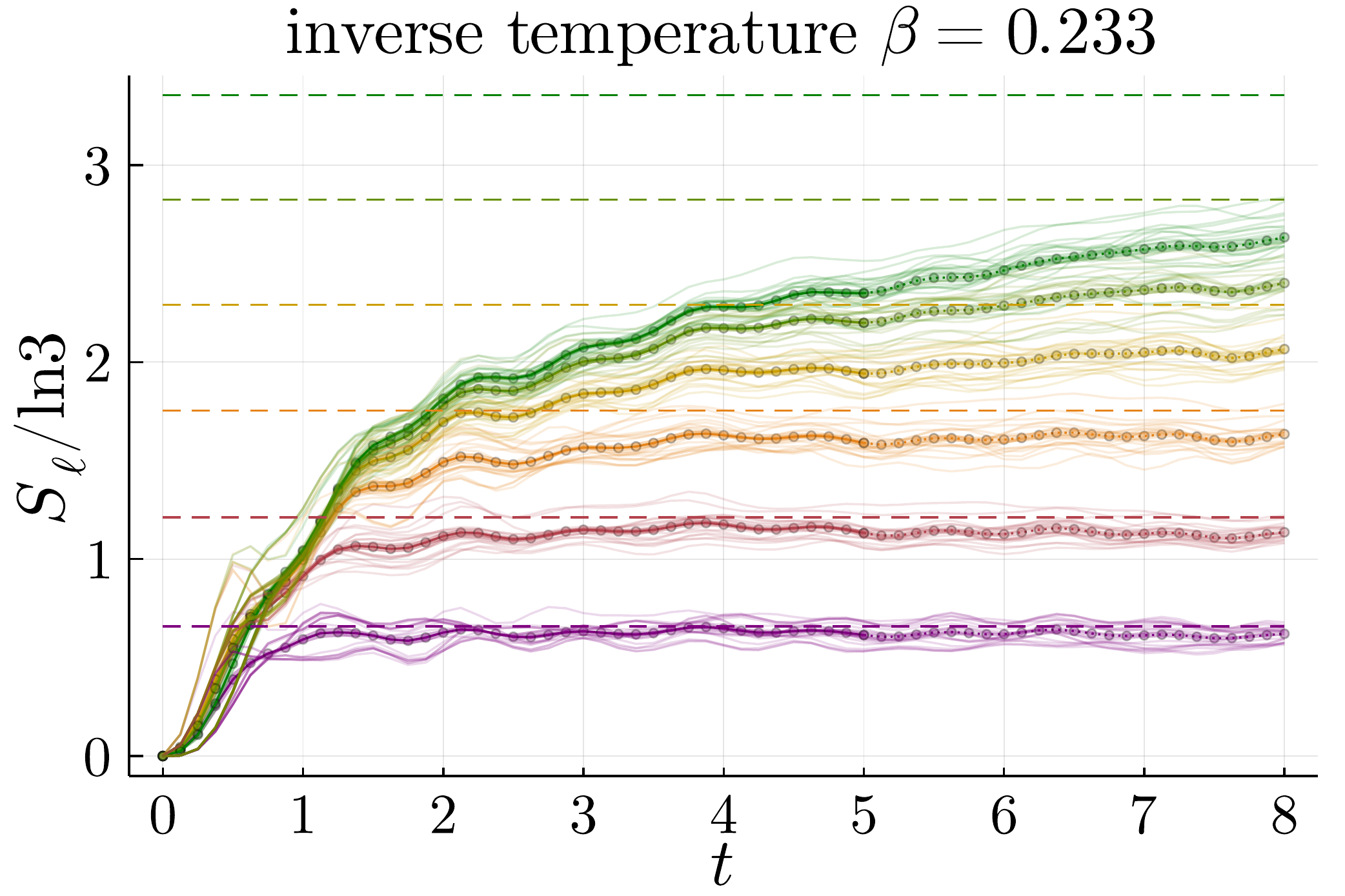}
  \end{minipage}
    \begin{minipage}{0.23\textwidth}
   \includegraphics[width=\linewidth]{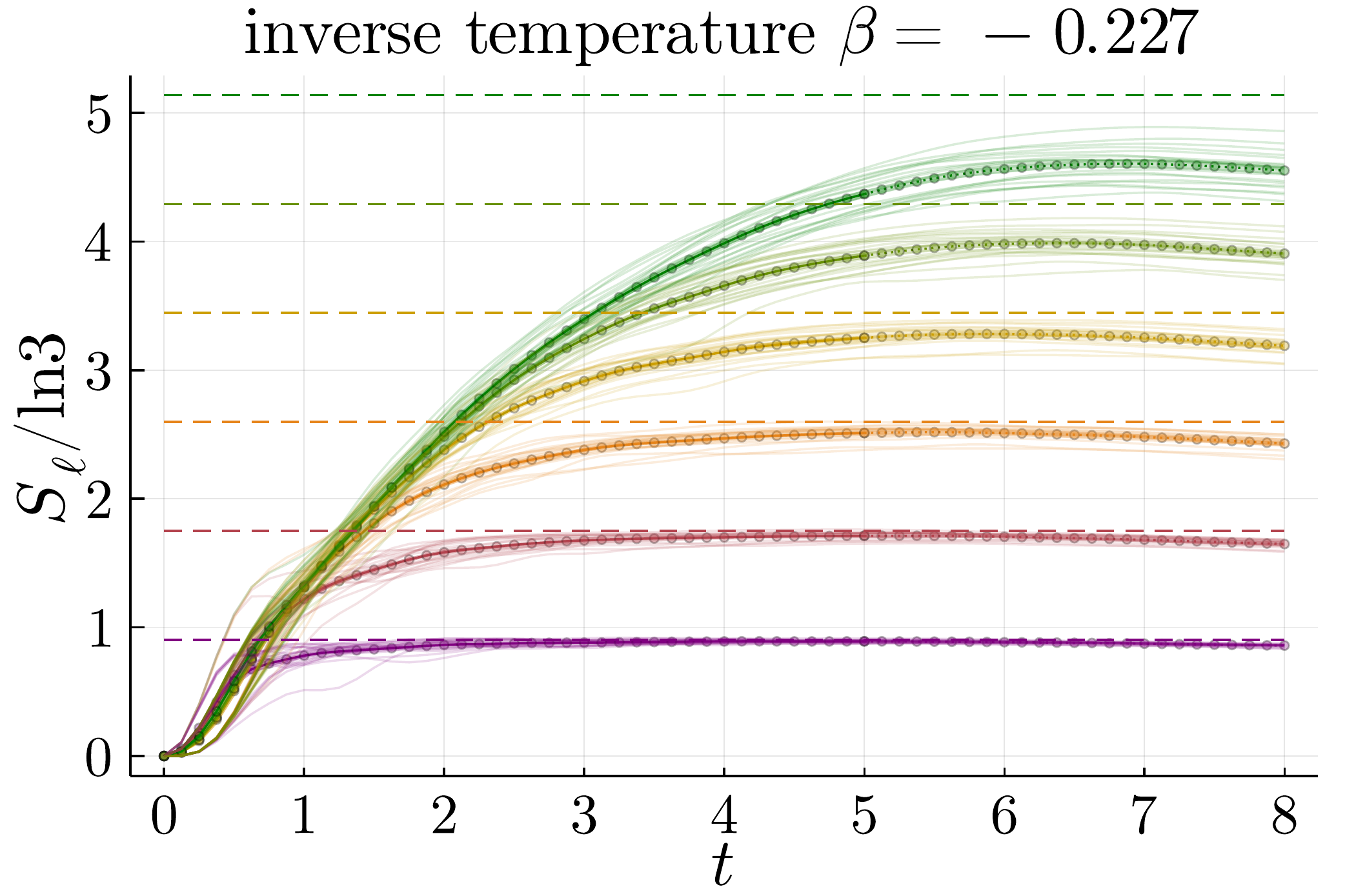}
  \end{minipage}
  \caption{50 qutrit MPS, subsystem entropy measured in trits, infinite and finite temperature initial states, $J=h_x=h_z=1$.
  }
  \label{fig:subent-fintemp}
  \end{figure}

 We often make reference to the entropy deficit $\Delta$ of our subsystems, which is defined as the maximal entropy of the subsystem minus the second R\'enyi entropy of that subsystem. 
 In Fig.~\ref{fig:subent-fintemp} we plot the average R\'enyi entropy of the central subsystems of our 50 qutrit MPS over the course of our TEBD evolution which is used to determine the entropy deficit.
 We also plot in dashed lines the subsystem entropies of Gibbs states, which subsystems should converge to in the process of thermalization.
 Note that these thermal entropies are not the maximal subsystem entropies, so the asymptotic entropy deficit is still nonzero even after subsystems have converged to thermal values of the entropy.
 The infinite temperature case is an exception, since the thermal entropies are maximal and the entropy deficit is zero.
 
 We see that small subsystems quickly converge to their thermal values and stabilize at this entropy.
 However, at later times we begin to see decreases in subsystem entropies which arise from truncation errors in the MPS evolution and are not reflective of accurate dynamics of the subsystem entropies.
 Unfortunately, some of the larger subsystems fail to reach their thermal entropy values before these truncation errors become relevant and start to decrease subsystem entropy.

\section{Thermalization, typicality, and the eigenstate thermalization hypothesis: intuition}\label{thermalization-intuition}
Imagine evolving a tensor product state $\ket{\psi}$ by a Hamiltonian $H$.
Since $\ket{\psi}$ is the eigenstate of some local Hamiltonian,
it has energy uncertainty
\begin{equation}
  \label{eq:quench-var}
  \left[\braket{\psi | H^2 |\psi } - \braket{\psi| H | \psi}^2\right]^{1/2} = \Delta\sqrt{L}\;, 
\end{equation}
where $\Delta$ is some $O(1)$ constant with dimensions energy.
The time-evolved state is
\[ \ket{\psi(t)} = \sum_j e^{-iE_jt}\braket{E_j|\psi} \ket E_j \]
only has weight near the initial energy $E = \braket{\psi| H | \psi}$ of $\ket \psi$.
The phase factors $e^{-iE_j t}$ break the delicate conspiracy between overlaps $\braket{E_j|\psi}$ that results in the state $\psi$ at the initial time $t = 0$,
and we can think of the state as randomly chosen from a distribution on energy eigenstates with width given by \eqref{eq:quench-var}.
That distribution, in turn, is similar to a microcanonical distribution with the same width.

{
This cartoon assumes that the eigenstates have no structure.
If they do, the overlaps $\braket{E_j|\psi}$ will be weighted towards eigenstates that resemble the initial state,
leading to a failure of thermalization.

Suppose the system has a classical limit.
Berry's conjecture \cite{berry_regular_1977,srednicki_chaos_1994} is that the eigenstates are appropriately structureless if the corresponding classical system is chaotic.
(More precisely the energy eigenfunctions are random in such a way that the Wigner function, averaged over a small phase space volume, matches the microcanonical ensemble.)

But Berry's conjecture does not contemplate the discrete Wigner functions we work with!
In order to take the average over a small phase space volume, we would have to work with large onsite Hilbert space dimension $d$.
More broadly the small-$d$ model does not have an obvious classical limit, chaotic or integrable.

If the system does not have a classical limit, as ours does not,
Berry's conjecture becomes the eigenstate thermalization hypothesis (ETH) \cite{deutsch_quantum_1991}%
---that local expectation values in eigenstates match thermal expectation values.
One expects the system to thermalize if its Hamiltonian satisfies the ETH,
and not otherwise.
NB the ETH is a hypothesis, not a conjecture like Berry's conjecture:
a particular Hamiltonian may or may not satisfy the ETH,
so it must be checked (generally numerically.)
The message of our Fig.~\ref{fig:gap-ratio} is that our Hamiltonian at our parameters does satisfy the ETH.

For a thorough and accessible review see \onlinecite{dalessio_quantum_2016},
especially Sec.~2.4 on the semiclassical limit and Berry's conjecture
and Sec.~4.1-4.2 on the eigenstate thermalization hypothesis.
\onlinecite{richter_semiclassical_2022} offers a review of recent work on quantum chaos in the semiclassical limit from a very different point of view.

}

\section{Gibbs state mana}\label{s:gibbs}
  \begin{figure}
  \centering
  \includegraphics[width=\linewidth]{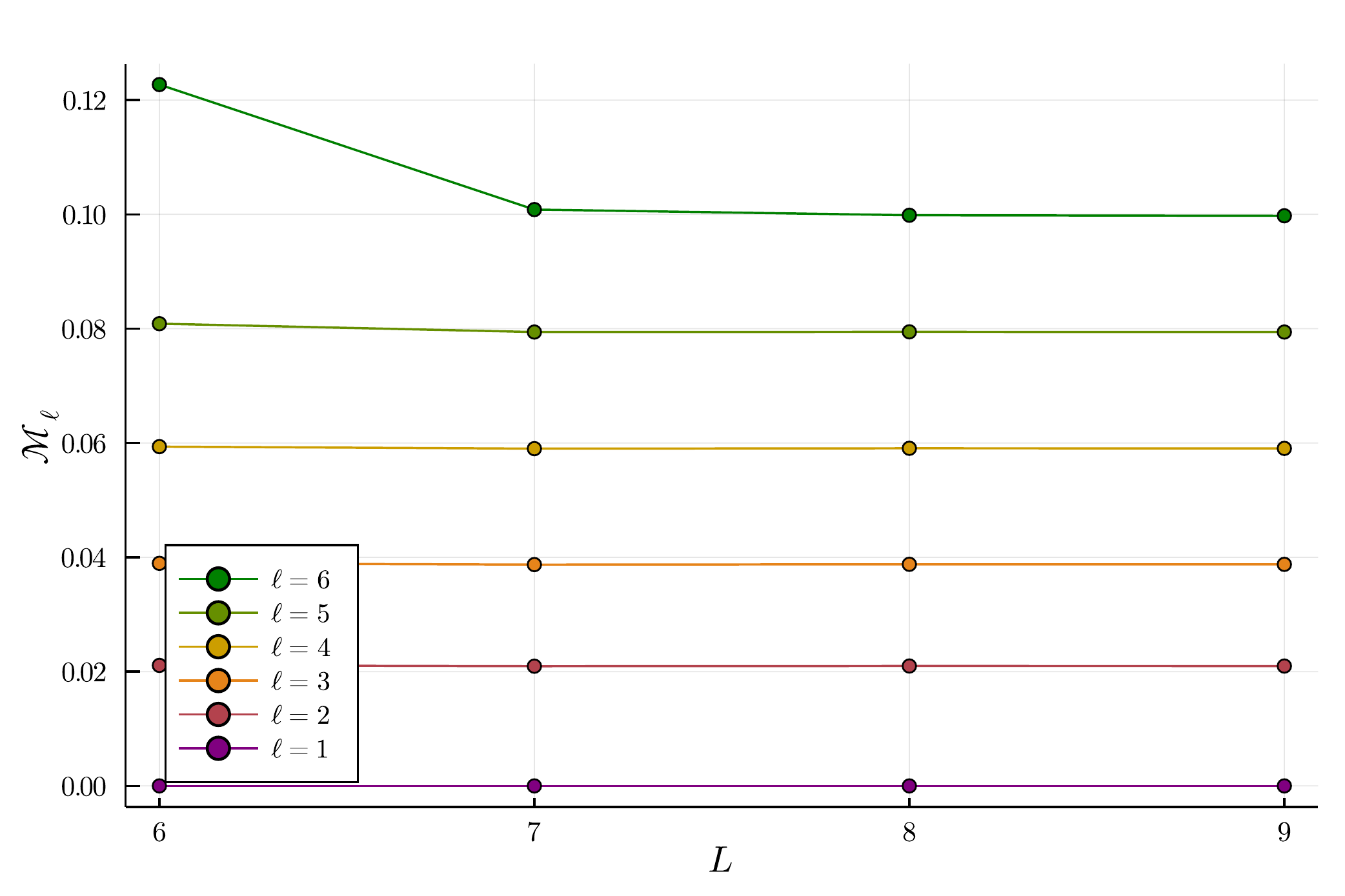}
 \caption{\textbf{Subsystem mana of Gibbs states} with inverse temperature $\beta = -0.25$ for different system sizes. The subsystem mana converges rapidly with system size which allows us to use these small system estimates for our large scale MPS simulations.}
 \label{fig:submana-conv}
\end{figure}

We use exact diagonalization on a system of up to nine sites with periodic boundary conditions to determine the thermal properties of subsystems for the larger qutrit chain.
We use the energy densities of these systems to map initial state energies to temperatures.
Fig.~\ref{fig:submana-conv}

We would like to compare the asymptotic mana of subsystems that have thermalized with nonzero initial energy density to the expected value of thermal subsystem mana. 
From these plots we can extract the subsystem mana for a given inverse temperature as well as the energy density for a given inverse temperature. 
From these we can find the subsystem mana we expect for an initial state of a given energy density.

\section{Bounding the difference in mana between nearby states}\label{s:mana-bnd}

\begin{prop}\label{prop:close-wn}
  Suppose two density matrices $\rho, \sigma$ on a Hilbert space of dimension $d$ are nearby in trace distance:
  \[ \|\rho - \sigma\|_1 \le \eta\;.\]
  Then 
  \begin{align*}
    \| \rho - \sigma \|_W &\le d^2 \eta\quad \text{[linear bound]} \\
    \| \rho - \sigma \|_W &\le \sqrt{d \eta} \quad \text{[square root bound].} 
  \end{align*}
\end{prop}

Applying the reverse triangle inequality to this result will give
\begin{cor}
  \label{cor:wigner-bnd}
  \begin{equation}
    \Big|\|\rho\|_W - \|\sigma\|_W\Big| \le \min(d^2 \eta, \sqrt{d\eta})\;.
  \end{equation}
\end{cor}
and subadditivity of the logarithm will give
\begin{cor}
  \label{cor:mana-bnd}
  \begin{equation}
    \ln \|\rho\|_W \le \ln \|\sigma\|_W + \min\left(2\ln d + \ln \eta, \frac 1 2 \ln d + \frac 1 2 \ln \delta \right)\;.
  \end{equation}
\end{cor}

Now to prove \ref{prop:close-wn}.
\begin{proof}
  Start with the linear bound.
  The difference in Wigner functions is
  \begin{align}
    \begin{split}
      |W_{pq}(\rho) - W_{pq}(\sigma)| &= d^{-1} |\tr A_{pq}(\rho - \sigma)| \\
      &\le d^{-1} \|A_{pq}\|_1 \|\rho - \sigma\|_1\;.
    \end{split}
  \end{align}
  Since the phase-space point operators $A_{pq}$ { of \eqref{eq:psp-def}} are unitary, $\|A_{pq}\|_1 = d$, so
  \begin{equation}
    |W_{pq}(\rho) - W_{pq}(\sigma)| \le \eta
  \end{equation}
  and
  \begin{equation}
   \|\rho - \sigma\|_W = \sum_{pq} |W_{pq}(\rho) - W_{pq}(\sigma)| \le d^2 \eta \;.
  \end{equation}
  as desired.
  
  Turn to the square root bound. Write
  \begin{equation}
    \Delta := \rho - \sigma\;;
  \end{equation}
  the hypothesis is that $\|\Delta\|_1 \le \eta$
  Then, using Cauchy-Schwarz,
  \begin{align}
    \| \Delta \|_W &= \sum_{pq} |W_{pq}(\Delta)| \le d \left[\sum_{pq} W_{pq}^2(\Delta)\right]^{1/2}
  \end{align}
  But
  \begin{equation}
   \sum_{pq} W_{pq}^2 = \frac 1 d \|\Delta\|^2_2 \le \frac 1 d \|\Delta\|_1
  \end{equation}
  since  $-1 \preceq \Delta \preceq 1$,
  so
  \begin{equation}
    \| \Delta \|_W \le \sqrt{d\eta}
  \end{equation}
  as desired.
\end{proof}

\section{Mana and canonical typicality}\label{s:mana-typicality}
Let us use the result of Popescu et al.\cite{popescu_entanglement_2006} to be more precise
about how what canonical typicality means for mana.

That result is as follows.
The subspace of interest is the vector space spanned by the eigenvectors within $\Delta \sqrt L / 2$ of $E$;
call it
\[ R = \mathrm{span} \{ \ket{E_j} : |E_j - E| < \Delta \sqrt L / 2 \}\;,\]
and its dimension $d_R$.
The relevant microcanonical ensemble is
\begin{equation} 
  \mathcal E = \frac 1 {d_E} \sum_{|E_j - E| < \Delta \sqrt L / 2}  \ketbra{E_j}{E_j}\;.
\end{equation}
Call the subsystem of interest $S$,
and its Hilbert space dimension $d_S$;
call the rest of the system $B = \bar S$.
The microcanonical ensemble $\mathcal E$ traces down to
\begin{equation}
  \Omega_S = \tr_{B} \mathcal E
\end{equation}
on $S$\;;
the effective accessible dimension on $B$ is
\begin{equation}
  d_E^{\eff} = \frac 1 {\tr \Omega_B^2}\;,
\end{equation}
where $\Omega_B = \tr_S \mathcal E$.
Finally, write $\mu$ for the Haar measure on the vector (sub)space $R$.
Then
\begin{equation}
  \label{eq:popescu-measure}
  \mu\Big( \ket \phi \in R \ : \ \|\rho_S(\phi) - \Omega_S\|_1 \ge \eta\Big) \le \eta' 
\end{equation}
with
\begin{align}
  \label{eq:popescu-params}
  \begin{split}
    \eta &= \epsilon + \frac 1 2 3^{L/2}\sqrt{d_{E}^{\eff}} \\
    \eta' &= 4e^{-Cd_R^2\epsilon}\;.
  \end{split}
\end{align}

Now use \eqref{eq:popescu-measure} and \eqref{eq:popescu-params} together with Cor.~\ref{cor:wigner-bnd}
to bound the difference in Wigner norms between $\rho_S(\phi)$ and $\ket{\Omega_S}$
Write
\begin{align*}
  \mathcal W_{\phi} &= \ln \|\rho_S(\phi)\|_W  \\
  \mathcal W_{\Omega} &= \ln \|\Omega_S\|_W  
\end{align*}
for the subsystem mana of the randomly chosen state $\ket \phi$ and the microcanonical ensemble $\Omega$,
respectively. 
Applying Cor.~\ref{cor:wigner-bnd} from App.~\ref{s:mana-bnd}, we find that
\begin{align}
  \mu\Big( \phi \in R\ :\ \mathcal W_\phi - \mathcal W_\rho \ge \delta\Big) \le \eta'
\end{align}
with
\begin{align}
  \delta = \min(d_S^2 \eta, \sqrt{d\eta})\;,
\end{align}
$\eta$ as in Eq.~\eqref{eq:popescu-params}.

To get some intuition for what this result means,
let us look for the $\eta$ (and hence $\delta$) corresponding to $\eta' = 1/2$:
that is, (an upper bound on) the median discrepancy in Wigner norm.
This $\eta' = 1/2$ gives
\[ \epsilon = \frac {\ln 8}{C d_R^2}\;\]
for
\begin{align*}
  \eta &= \epsilon + \frac 1 2 \sqrt{\frac{d_S}{d_E^{\eff}}} \\
  &\le \frac {\ln 8}{C d_R^2} + \frac {d_S}{2\sqrt{d_R}}
\end{align*}
using that [bound on $d_E^{\eff}$ from Popescu].
The second term dominates the first, so
\[ \eta \lesssim  \frac {d_S}{2\sqrt{d_R}} \;.\]
(Were $\eta'$ sufficiently small this would not be the case.)
Now
\begin{align}
  \delta \approx \frac{d_S}{\sqrt{2\sqrt{d_R}}} \sim d_S d_R^{-1/4}\;.
\end{align}

\section{Krylov subspace evolution} \label{app:krylov}

We want to corroborate our stories about thermalization and subsystem mana in small enough systems where more exact methods than TEBD evolution of MPS states can be done for long times. 
We also need the system to be large enough to see thermalization of the smallest subsystems and finite size correction emerging for larger subsystems, which is challenging using exact diagonalization of a chain of qutrit sites. 
So, we turn to Krylov subspace methods to time evolve states of our Potts model for intermediate system sizes and high-precision long time dynamics. 
The Potts model is the same as in  Eq. \ref{eq:ham}, but with periodic boundary conditions and the coefficients $J=h_x=h_z=1$. 
We evolve the product stabilizer initial states for an $L=11$ qutrit chain up to time $t=100$ and measure the mana and entropy of subsystems less than half the system size. 
 
\begin{figure}
  \centering
     \begin{minipage}{0.48\textwidth}
  \includegraphics[width=\linewidth]{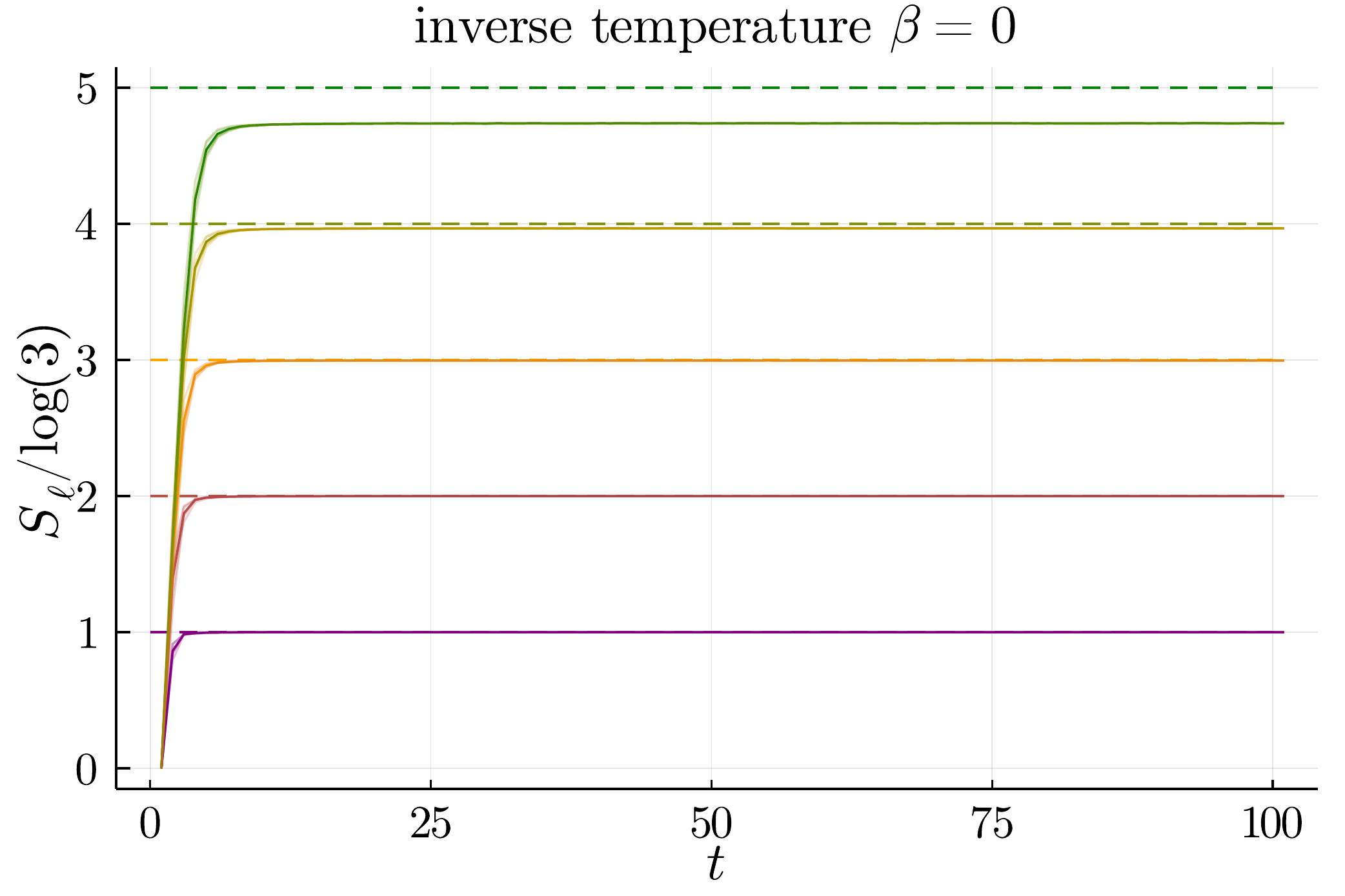}
 \end{minipage}
 \\
  \begin{minipage}{0.23\textwidth}
  \includegraphics[width=\linewidth]{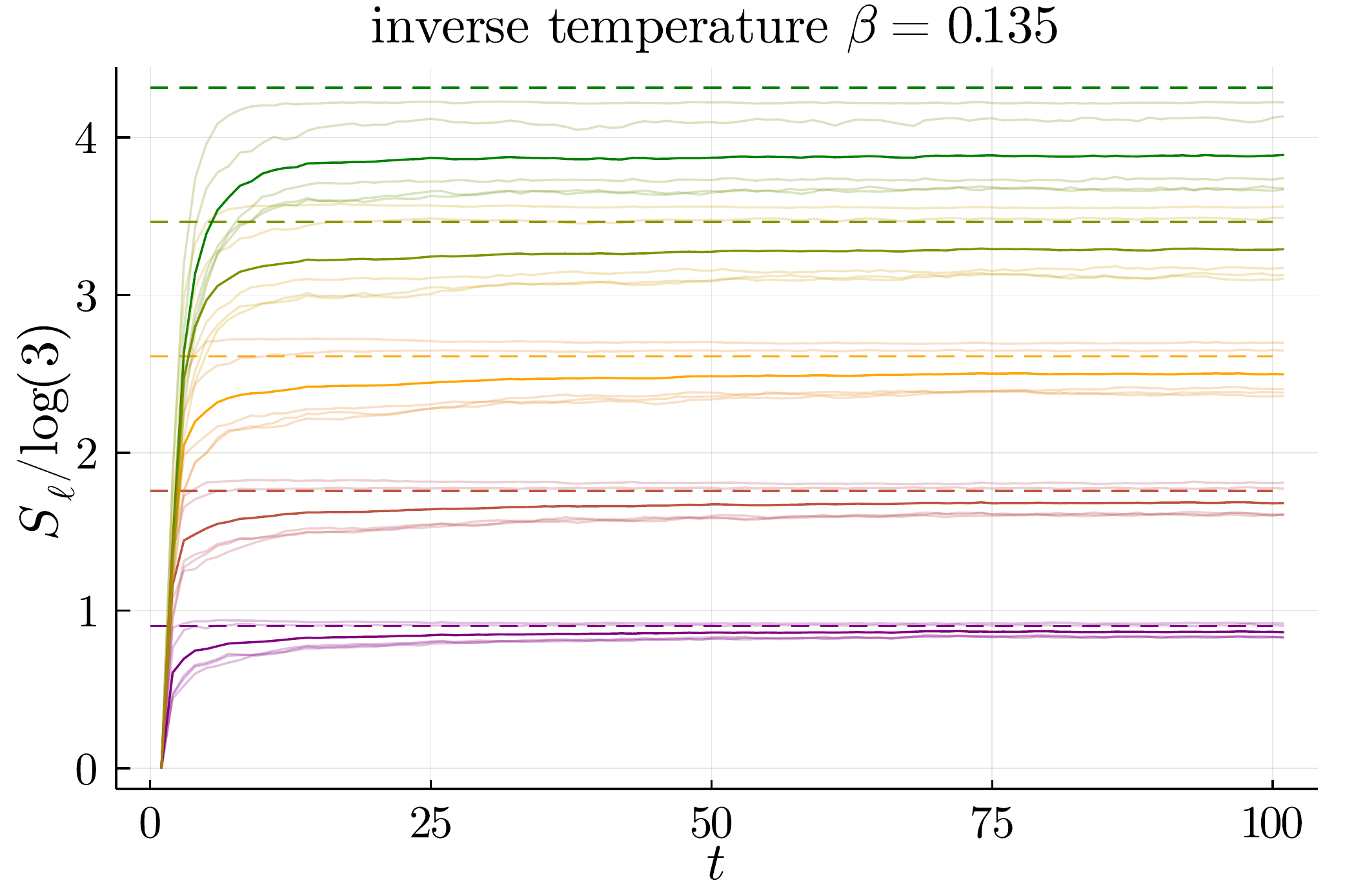}
 \end{minipage}
   \begin{minipage}{0.23\textwidth}
  \includegraphics[width=\linewidth]{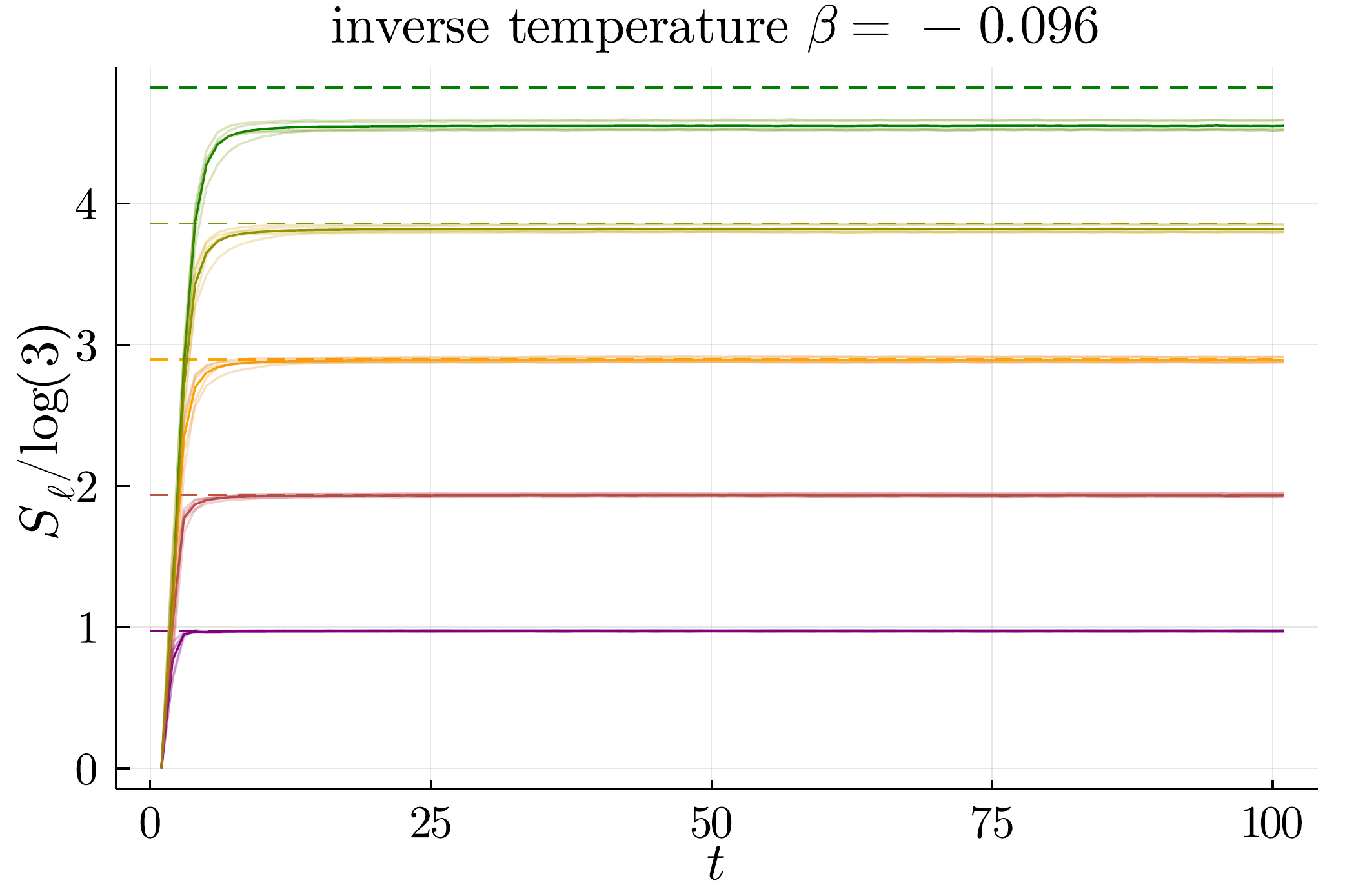}
 \end{minipage}
 \\
   \begin{minipage}{0.23\textwidth}
  \includegraphics[width=\linewidth]{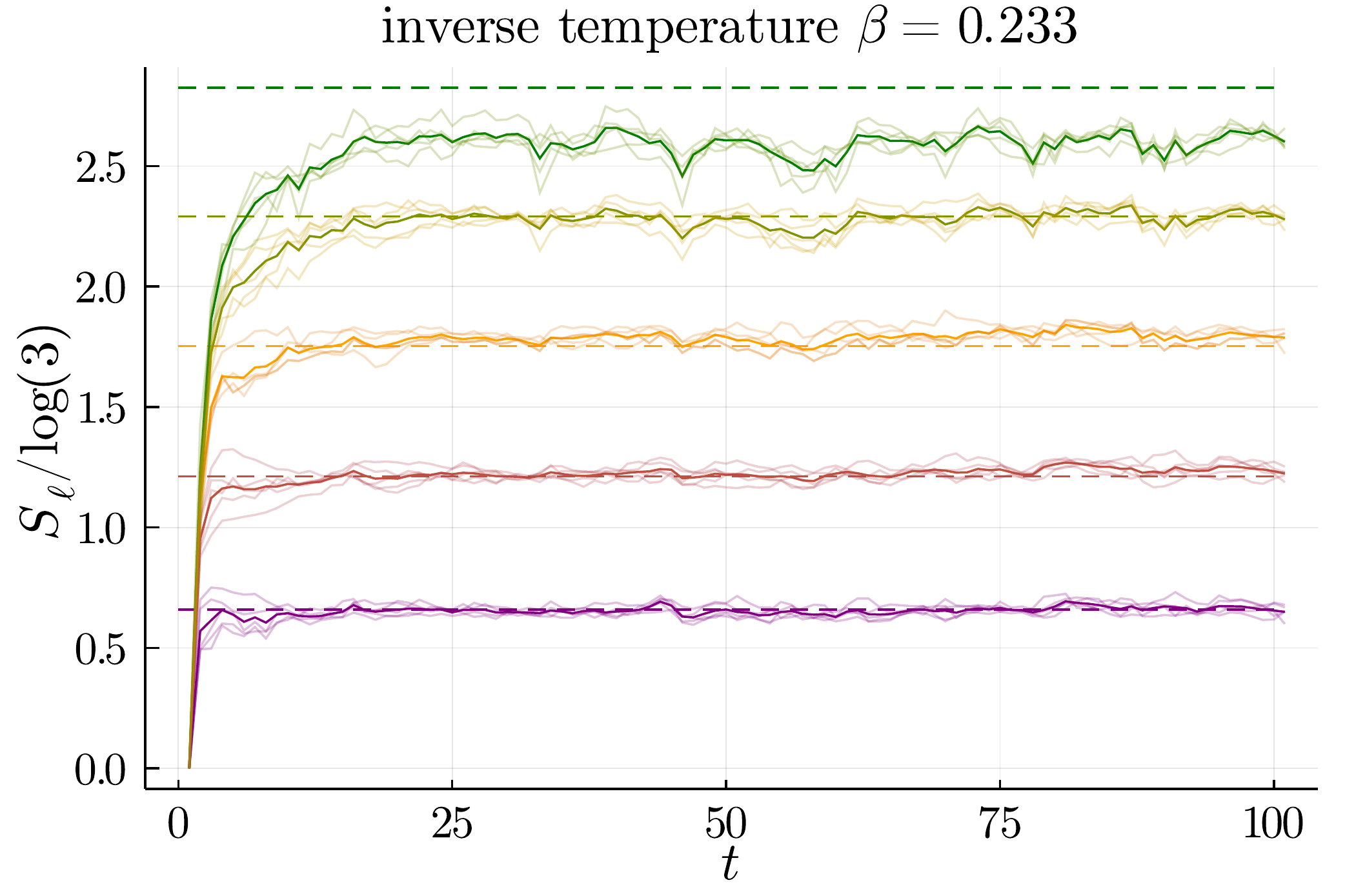}
 \end{minipage}
   \begin{minipage}{0.23\textwidth}
  \includegraphics[width=\linewidth]{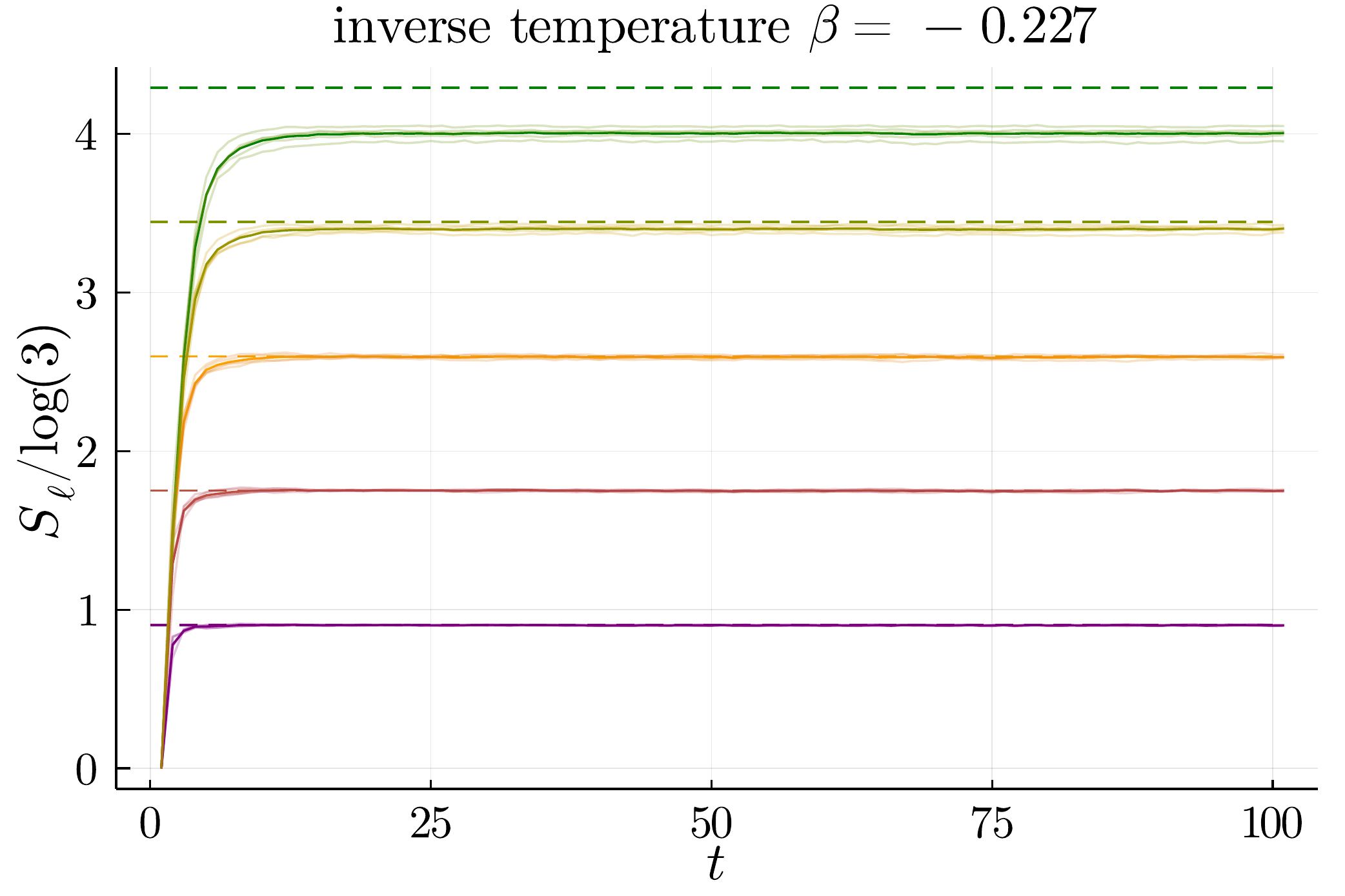}
 \end{minipage}
 \caption{11 qutrit Krylov evolution, subsystem entropy in trits up to $t=100$, thermal subsystem values of 8-qutrit system dashed.
 }
 \label{fig:krylov-submana-fintemp}
 \end{figure}

The  time evolved state which has thermalized is compared to the Gibbs state values obtained through exact diagonalization of an 8 qutrit system. 
The subsystem entropies asymptotically approach the thermal values, with deficits visible in four and five qutrit systems due to finite size effects. See Fig. \ref{fig:krylov-submana-fintemp}.

The entropy of the $\beta = 0.135$ states are lower and have larger standard deviation between different initial states than other temperatures. 
The entropies of the $\beta = 0.233$ states have fluctuations which are the same across the initial state, which have more correlation due to their relatively lower statistics from energy constraints. These fluctuations are maintained in averaging and result in a low standard deviation.

\begin{figure}
  \centering
  \begin{minipage}{0.23\textwidth}
  \includegraphics[width=\linewidth]{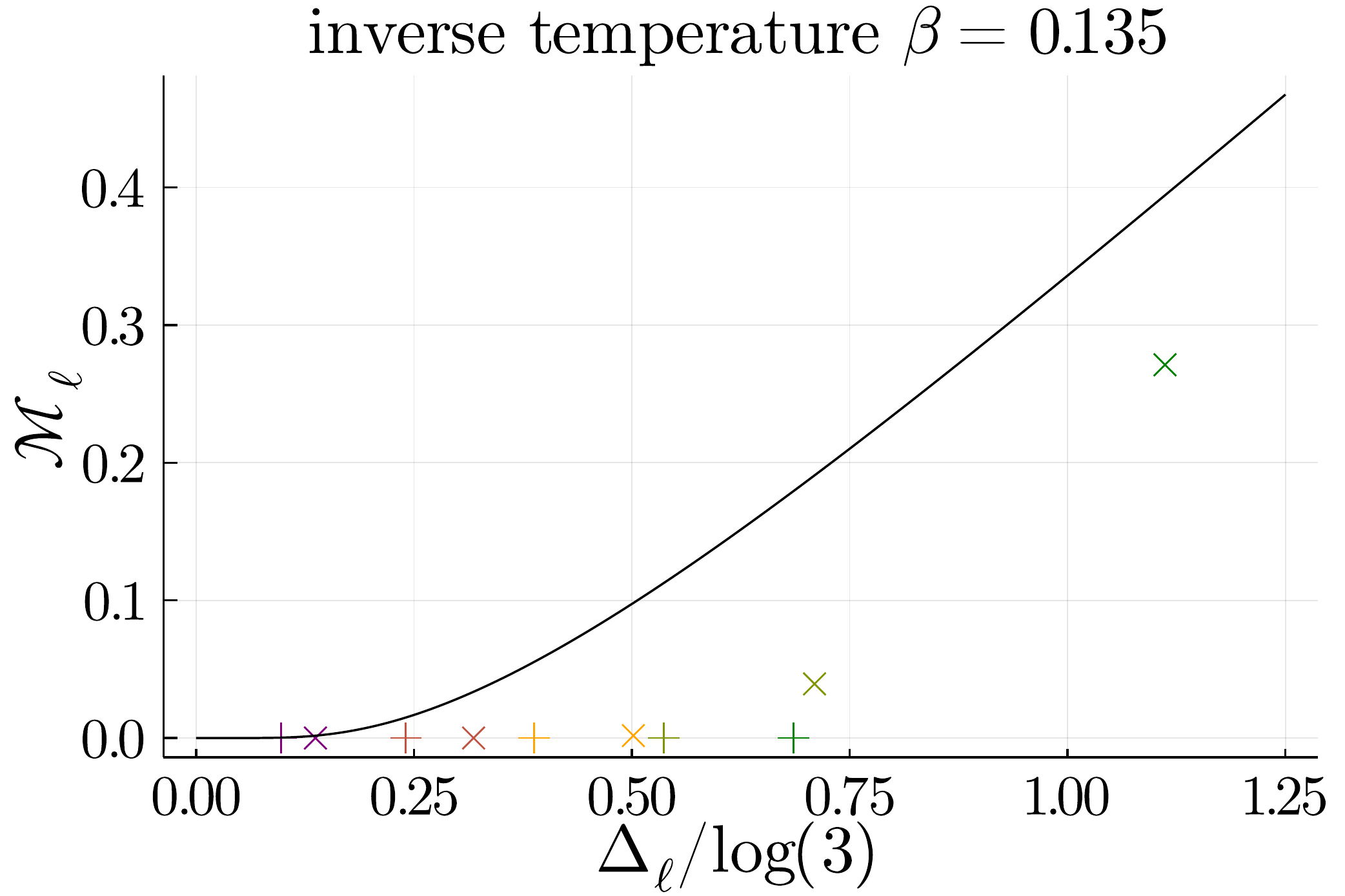}
 \end{minipage}
   \begin{minipage}{0.23\textwidth}
  \includegraphics[width=\linewidth]{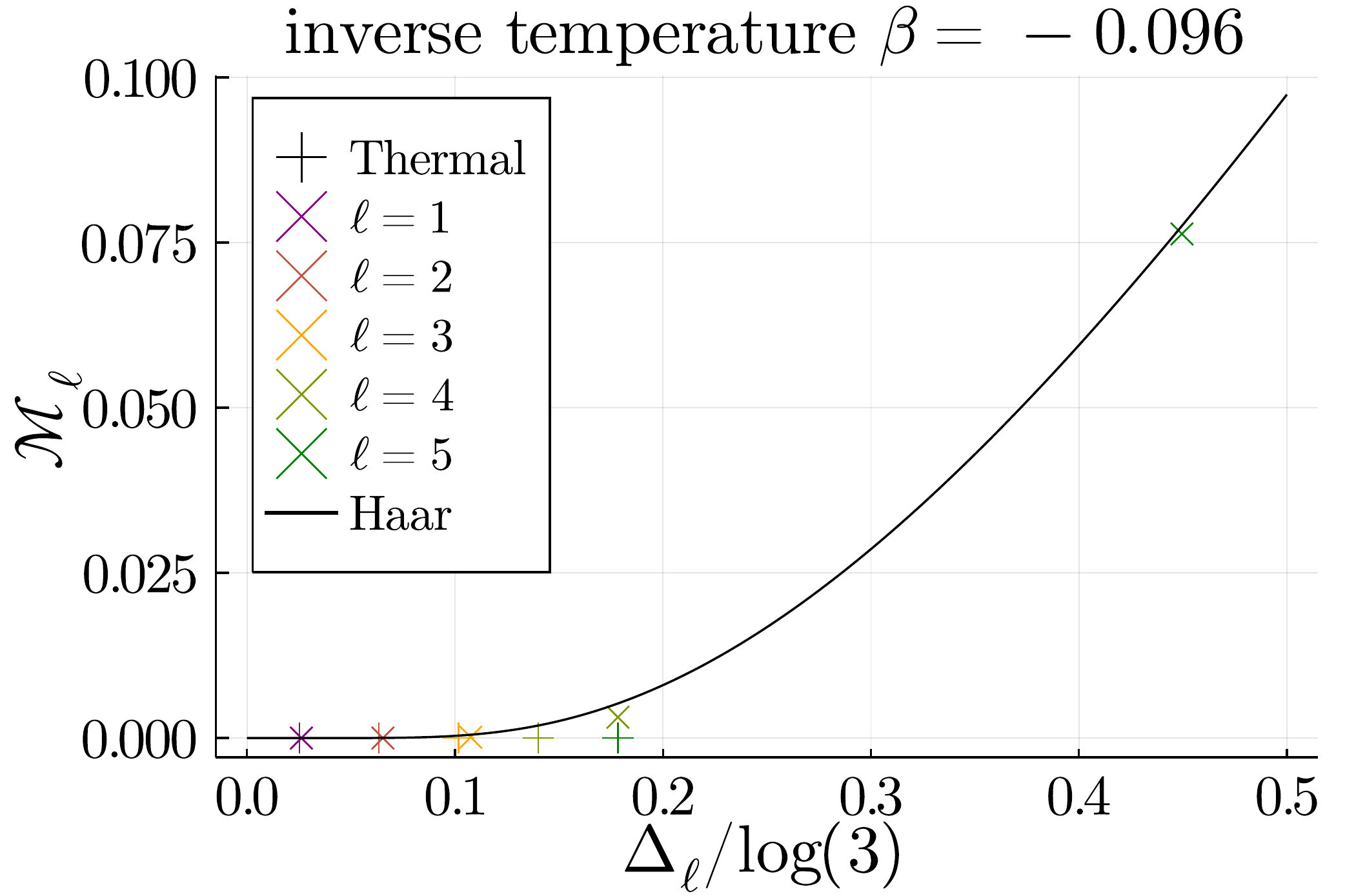}
 \end{minipage}
 \\
   \begin{minipage}{0.23\textwidth}
  \includegraphics[width=\linewidth]{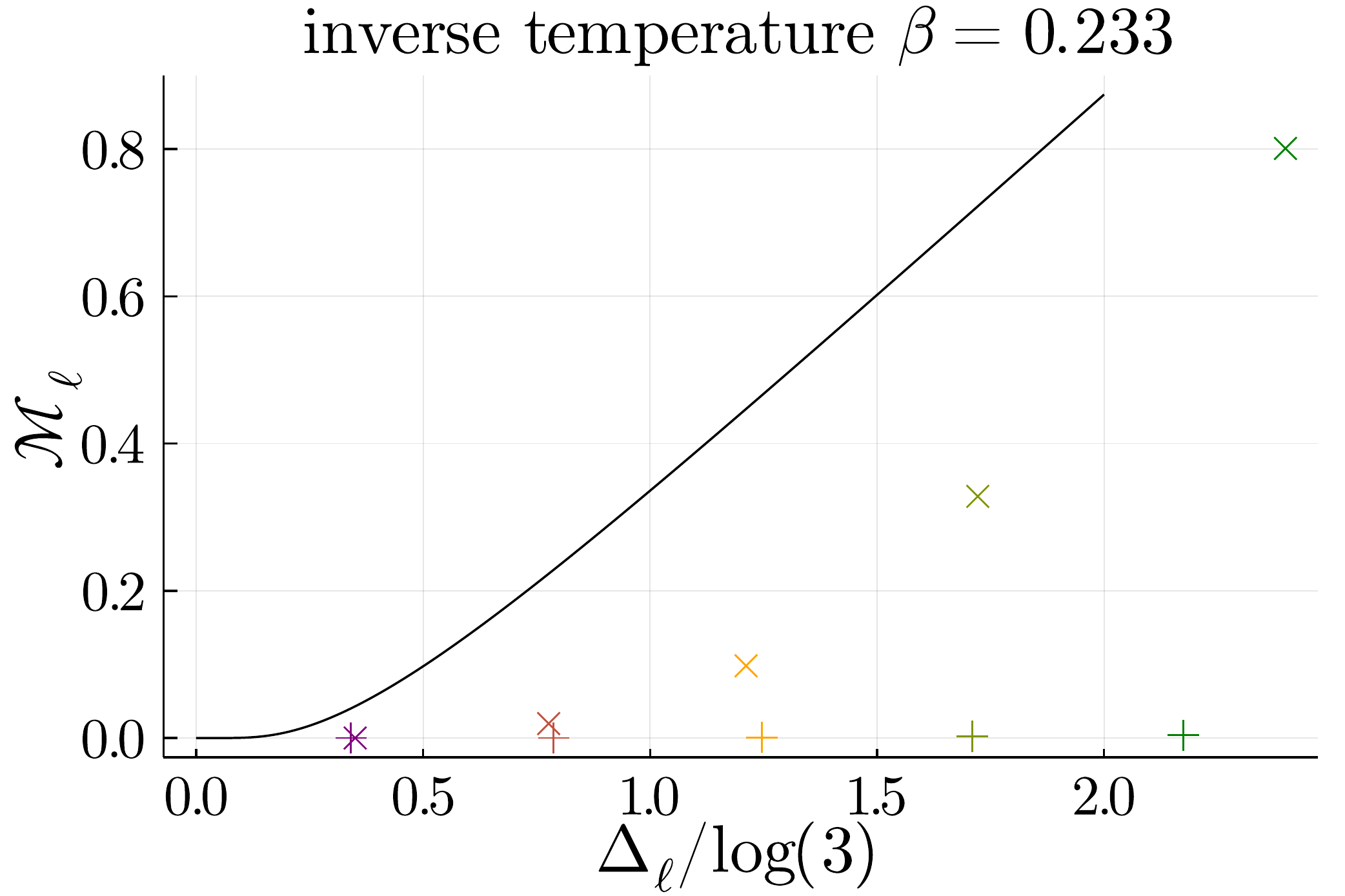}
 \end{minipage}
   \begin{minipage}{0.23\textwidth}
  \includegraphics[width=\linewidth]{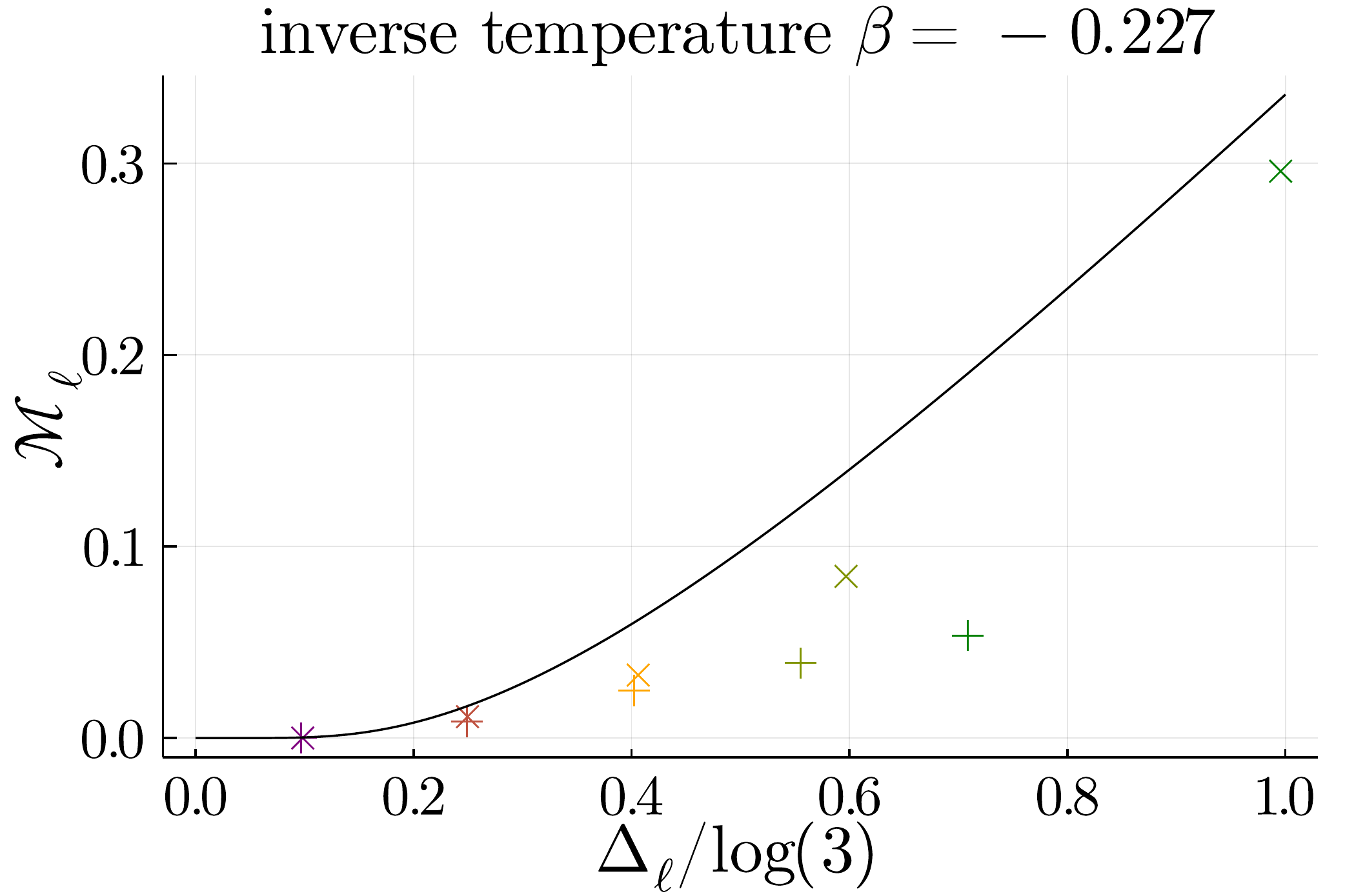}
 \end{minipage}
 \caption{11 qutrit Krylov evolution at $t=100$ compared with finite temperature subsystems of an 8 qutrit chain. Entropy deficit in trits versus subsystem mana, with the relationship for Haar random states given by the solid black line. }
 \label{fig:krylov-submana-fintemp-deficit}
 \end{figure}

 For subsystem mana, subsystems smaller than four which have zero thermal mana match, with larger subsystems having finite size corrections.
 However, for $\beta = 0.233$ and $\beta = -0.227$, which have small but nonzero mana in most subsystems, have significantly more mana than thermal values even in smaller subsystems. See Fig. \ref{fig:krylov-submana-fintemp-deficit} for comparison of late time subsystem mana and entropy deficit with thermal subsystem values.

What we also notice for the Gibbs states is that for $-0.12 \leq \beta \leq 0.2$ all subsystems have zero mana, thus the Gibbs state itself may be inside the stabilizer hull. 
For our zero energy initial states, most subsystems are maximally mixed with finite size effects reducing the entropy only for the largest subsystems.
 The maximally mixed state is in a sense at the center of the stabilizer hull and far from any states with mana, so all subsystems of the late time states for zero energy initial states have zero mana except for the $\ell=5$ subsystem where finite size effects are seen.

Our Gibbs states with $\beta = 0.135, -0.96$ are also inside the stabilizer hull, but closer to the boundary. 
The late time subsystems for these intiial states also have zero mana for most subsystems, though the $\ell=4$ subsystems noticeably have mana where the same sized subsystems for zero energy initial states did not. 
This could still be from finite size effects, but being closer to the stabilizer hull boundary the introduction of mana does not require as large of a deviation from the true Gibbs state subsystems. 

For  Gibbs states with $\beta = 0.233, -0.237$ subsystems do have a small amount of mana and lie just outsize the stabilizer hull.
 We see however an excess of mana in subsystems over the Gibbs state values even for small subsystems. 
These states, although close to the Gibbs state subsystems in some sense, have enough 'wiggle room' in a small neighborhood to amount significantly more mana.

Our subsystem mana estimates for thermal states were carried out by numerically finding the exact Gibbs state for small systems and then computing the mana of various sized subsystems. 
We see in Fig.~\ref{fig:submana-conv} that the subsystem mana converge rapidly with overall system size and so we can accurately use these values as estimates of the subsystem mana for large system thermal states. 
Moreover, the finite size effects we do see increase the subsystem mana, so it is unlikely that these values would underestimate the mana of thermal subsystems.
This shows further that the large excess of mana seen in long time simulations is due to canonical typicality effects of time evolved pure states.

\end{document}